\RequirePackage{fix-cm}
\documentclass[twocolumn,natbib]{svjour3}          
\smartqed  
\usepackage{graphicx}
\graphicspath{{./figures/}{./}}
\newcommand{\eq}[1]{(\ref{#1})} 
\newcommand{\mbf}[1]{\mbox{\boldmath$#1$\unboldmath}} 
\newcommand{\dder}[2]{\frac{\mathrm{d} #1}{\mathrm{d} #2}} 
\usepackage{color}
\journalname{Statistics and Computing}
\begin{document}

\title{Density Estimation with Distribution Element Trees\thanks{
The author is grateful to Marco Weibel for his help during the preparation of this manuscript. Very valuable feedback from an associate editor and two reviewers and helpful input from Oliver Brenner and Florian M\"uller are gratefully acknowledged. Moreover, the author acknowledges helpful comments from Nina Roth and feedback on the initial version of this manuscript from Patrick Jenny, both from ETH Z\"urich. The author has been financially supported by ETH Z\"urich.}
}

\titlerunning{Distribution Element Trees}        

\author{Daniel W.\ Meyer}


\institute{Daniel W.\ Meyer \at
              Institute of Fluid Dynamics \\
              ETH Z\"urich \\
              Tel.: +41-44-633-9273\\
              Fax: +41-44-632-1147\\
              \email{meyerda@ethz.ch}           
}

\date{Received: 06/11/2016 / Accepted: 02/05/2017}

\maketitle

\begin{abstract}
The estimation of probability densities based on available data is a central task in many statistical applications. Especially in the case of large ensembles with many samples or high-dimensional sample spaces, computationally efficient methods are needed. We propose a new method that is based on a decomposition of the unknown distribution in terms of so-called distribution elements (DEs). These elements enable an adaptive and hierarchical discretization of the sample space with small or large elements in regions with smoothly or highly variable densities, respectively. The novel refinement strategy that we propose is based on statistical goodness-of-fit and pair-wise (as an approximation to mutual) independence tests that evaluate the local approximation of the distribution in terms of DEs. The capabilities of our new method are inspected based on several examples of different dimensionality and successfully compared with other state-of-the-art density estimators.
\end{abstract}

\section{Introduction}\label{secIntro}

In this work, we propose a new method for estimating a probability density $p(\mathbf{x})$ of the random variable vector~$\mathbf{X}$ at position $\mathbf{x} = (x_1,\ldots,x_d)^\top$ of the bounded, $d$-dimensional probability space $\Omega$ based on a given ensemble of $n$ samples $\mathbf{x}_1,\ldots,\mathbf{x}_n$. Unlike other methods, our approach is applicable for large and/or high-dimensional datasets. Density estimation methods are essentially categorized into parametric and non-parametric methods \citep[e.g.,][]{Haerdle:2004a,Scott:2015a}. While we focus in this work on the development of a non-parametric approach, our method can be viewed as a hybrid between the two categories \citep{Yang:2008a}. Therefore, before we focus on non-parametric methods, we start the following literature review by briefly explaining the concept of parametric density estimation.

In parametric density estimation, a parametric density model $p(\mathbf{x}|\mbf{\theta})$ is given in analytical form. With the parameters~$\mbf{\theta} = (\theta_1,\theta_2,\ldots)$ estimated from the available data, we arrive at a complete density characterization. This characterization can be evaluated with statistical tests or more precisely composite goodness-of-fit tests such as for example the $\chi^2$ test of \citet{Pearson:1900a}. Here, for a given significance level~$\alpha$, we either do not reject the null hypothesis $p(\mathbf{x}) = p(\mathbf{x}|\mbf{\theta})$ and reject the alternative $p(\mathbf{x}) \neq p(\mathbf{x}|\mbf{\theta})$ or vice versa. For small ensembles, these tests lose their power meaning that they become unable to detect alternatives and instead do not reject the null hypothesis \citep[e.g.,][]{Steele:2006a}. Parametric density estimation is of limited generality, since a predefined density model is needed.

Non-parametric density estimation is more general as no parametric density model is required. A widely used representative from this category of density estimators is the histogram \citep[section~3]{Fix:1951a,Scott:2015a}. Here, the probability space is typically discretized into equally-spaced bins and the probability density within one bin is set proportional to the number of samples in that bin. To determine the bin width~$h$, the mean integrated square error (MISE), or more precisely its asymptotic approximation referred to as AMISE, is minimized for cases with known $p(\mathbf{x})$; leading for example with a Gaussian $p(\mathbf{x})$ to an~$h$ given by the so-called normal reference rule \citep[equation~(2.26)]{Haerdle:2004a}. Histograms are conceptually simple, but have certain drawbacks. First of all, they are relatively inaccurate and with the overall number of bins growing exponentially with the number of dimensions~$d$, conventional histograms become prohibitive in cases with large~$d$ \citep[section~3.4.1]{Scott:2015a}. More economical and possibly accurate variants with adaptive bin widths are available \citep[e.g.,][]{Kogure:1987a}. These variants attempt to better resolve regions with large density variation with finer bins, while using larger bins in more uniform areas. Adaptive histograms rely on estimates of gradients of $p(\mathbf{x})$ or the use of percentile meshes \citep[section~3.2.8]{Scott:2015a}.

One possible generalization of histograms are cubic log-splines \citep{Kooperberg:1991a}. Here, the one-dimensional probability space is partitioned into bins like in a histogram, but within a bin, a parametric model in the form of a cubic polynomial or spline is applied. The bin bounds or knots are placed subject to a rule that was derived based on experience in fitting log-spline models \citep[section~5.1]{Kooperberg:1991a}. The coupled polynomial coefficients or spline parameters in turn are determined numerically with a Newton--Raphson method by maximizing the likelihood of the log-spline estimator. The overall number of knots is either calculated by a predefined rule or sequential knot-deletion. A related approach that is applied in multi-variate settings are so-called polynomial histograms. Here, the density in a bin varies according to a linear, quadratic, or higher order polynomial \citep{Scott:1997a,Jing:2012a}. Unlike in the log-spline method, the polynomial coefficients are determined locally based on statistical moments estimated within individual bins (conditional moments). The use of higher order polynomials enables the use of larger bins globally while maintaining the same MISE compared to conventional histograms \citep[table~1]{Jing:2012a}. This is because to some extent the density variation is accounted for already at the level of an individual bin. Especially in cases where $d$ is large, this reduces computational costs. So far, equidistant and prescribed non-equidistant bin grids were considered \citep{Jing:2012a,Zaunders:2016a}.

Besides histograms, a second important category of non-parametric methods is kernel density estimation (KDE) \citep{Rosenblatt:1956a,Sheather:2004a}. Here, samples are not grouped into bins, but are equipped with kernel functions, e.g., triangular, Gaussian, etc.\ \citep[table~3.1]{Silverman:1998a}. Similarly like the bin width in histograms, kernels have a certain support or bandwidth~$h$. Optimal global bandwidths were determined with AMISE analysis based on second order derivatives of $p(\mathbf{x})$ \citep[e.g.,][equations~(6.18) and~(6.50)]{Scott:2015a}. Adaptive more accurate methods that reduce the bandwidth in dense areas and use wider kernels in sparse regions have been documented \citep[e.g.,][]{Loftsgaarden:1965a,Achilleos:2012a}, but are not completely satisfactory yet. For example \citet[section~6.8]{Scott:2015a} summarized quite recently: `Adaptive methods hold much promise, but usually introduce many new parameters that are difficult to estimate, and frequently introduce artifacts of the sample (rather than the underlying density).' Certain efficient spectral KDE implementations rely on bandwidths that are constant within different directions~$x_i$ in probability space \citep[e.g.,][section~3.5]{OBrien:2016a,Silverman:1998a}. The highly-cited KDE method of \citet{Botev:2010a} on the other hand is adaptive, is available in the form of an efficient spectral implementation \citep{Botev:2007a}, and reduces boundary bias effects of existing KDE approaches. Recently, a new KDE method for bounded domains that eliminates boundary bias issues was presented in the context of functional data analysis by \citet{Petersen:2016a}. A high-dimensional implementation of the method of \citet{Botev:2010a} was recently made available, where the cost per density query scales as an exponential fraction of the number of samples \citep{Botev:2016a}. Alternative KDE methods were summarized in \citep{Park:1990a,Park:1992a,Cao:1994a,Jones:1996a,Silverman:1998a,Haerdle:2004a,Scott:2015a}.

A density estimator for exponential families---which is a broad class of densities---that has superior convergence properties compared to KDE was presented by \citet{Sriperumbudur:2013a}. This estimator is based on minimizing the Fisher divergence and requires the solution of an $(n d + 1)\times (n d + 1)$ linear system. Most interestingly, the advantage to KDE grows with increasing dimensionality~$d$ as was shown numerically.

Density estimation based on mixture distributions can be viewed as a generalization of KDE, where the unknown distribution is expressed like in KDE as a superposition of probability densities \citep[e.g.,][]{Wang:2015a}. These densities are referred to as mixture models and each mixture model is associated with an ensemble subset containing---unlike in KDE---multiple samples. In this context, the so-called Dirichlet process mixture models, that go back to \citep{Ferguson:1973a}, have received renewed attention after progress was made in the numerical estimation of mixture parameters \citep[e.g.,][]{Neal:2000a}.

To arrive at an efficient adaptive method, tree-based approaches have been proposed more recently \citep[e.g.,][]{Ram:2011a,Wong:2010a,Jiang:2016a}. These approaches start from the probability space $C = \Omega$, which is typically assumed to be a $d$-dimensional hypercube. The root cube~$\Omega$, or cuboid to be more precise, is recursively subdivided into smaller cuboids, e.g., $C_1$ and~$C_2$ with $C = C_1 \bigcup C_2$, based on suitable conditions. More specifically, \citet{Ram:2011a} discussed density estimation trees or shorter density trees that are derived from decision trees \citep{Breiman:1984a} and are constructed based on optimal split operations of cuboids at tree nodes. These splits are optimal in the sense that they maximally reduce the integral square error (ISE), i.e., $\mbox{ISE}(C) > \mbox{ISE}(C_1) + \mbox{ISE}(C_2)$. The optimum is found from all possible splits in each dimension. The large number of possibilities renders the method expensive for large datasets \citep[equation~(9)]{Ram:2011a}. A preset lowest threshold is set to stop the splitting. The splitting process is followed by a tree pruning and cross-validation step. The resulting tree or more precisely its leafs provide a histogram with adaptive bin widths. The tree structure enables a fast density estimation at a cost proportional to the tree depth.

\citet{Wong:2010a} and \citet{Jiang:2016a}, on the other hand, have introduced and numerically implemented, respectively, the optional P\'olya tree (OPT) method. In this approach, cuboids $C$ are partitioned and uniform cuboid probability densities $p(\mathbf{x}|C_i)$ (or $q(\mathbf{x}|C_i)$ in their work) are assigned according to probabilistic processes involving Bernoulli and Dirichlet random numbers, respectively. The random partitioning process relies on selection and stopping probabilities, $\lambda_i(C|\mathbf{x})$ and $\rho(C|\mathbf{x})$, respectively, that are calculated based on a recursive expression \citep[equation~(2.1) in][]{Jiang:2016a} that necessitates partitioning down to subregions with either zero or one sample. This leads to a close to exponential growth in computing time as a function of the number of dimensions \citep[figure~1]{Jiang:2016a}. As a remedy, in naive inexact OPT (NI-OPT) partitioning limits were introduced like smallest number of points in subregions or smallest size of subregions to arrive at manageable trees. Moreover, limited-lookahead OPT (LL-OPT) introduces two additional tuning parameters that control the tree depth for the recursive calculation and thus help to reduce memory requirements and computing times of OPT and NI-OPT. The density estimates $p(\mathbf{x})$ that result from OPT implementations are piecewise uniform within subregions~$C$ or bins and are adaptive within individual dimensions.

To relate back to our initial discussion on parametric methods, \citet{Ma:2011a} outline an OPT method for goodness-of-fit testing of large datasets against a given base distribution $p_0$ (or $q_0$ in their work). Here, instead of uniform cuboid probability densities, $p_0$ is used, but otherwise the OPT methodology of \citet{Wong:2010a} applies. As a measure of the overall goodness-of-fit, the integral stopping probability $\rho(\Omega)$ was proposed \citep[section~3]{Ma:2011a}.

In the present contribution, we develop a new non-parametric density estimator that is adaptive and cost efficient. In view of the so-called curse of dimensionality, that is diminishing MISE convergence rates for increasing dimensionality \citep[e.g.,][equation~(3.67)]{Scott:2015a}, methods that enable the treatment of large datasets at small computational costs become important. While our method adopts certain features from the previously introduced polynomial histograms and tree-based techniques, our new method is, however, conceptually and algorithmically simpler and computationally more efficient compared to these approaches.

Our development starts by recognizing that a histogram is essentially a collection of disjoint piecewise uniform distributions. In a histogram with equally-sized bins, depending on the true density distribution, a uniform approximation may be accurate in certain bins, while inaccurate in others. Like in the context of parametric methods, we could apply a statistical test to evaluate the goodness-of-fit of the data in individual bins against a uniform null hypothesis. Thus a natural recursive way of constructing a histogram emerges: $\Omega$ is defined as a root hypercuboid that encloses the available data. With a suitable goodness-of-fit test, it is tested whether the uniform null hypothesis is adopted at a given significance level based on the data in the cuboid. If not, the root cuboid is split and the testing and splitting is recursively repeated for all cuboids resulting from this and subsequent splits. Eventually the test will adopt the uniform hypothesis in all resulting subcuboids. This is because the power of the test diminishes as the number of samples in each subcuboid becomes smaller during consecutive splitting. Finally an adaptive histogram results, where in each bin the uniform distribution is supported by a positive outcome of a goodness-of-fit test. This approach is not entirely non-parametric as a significance level has to be prescribed. The nested spatial arrangement of cuboids and subcuboids, that is organized in a tree structure, enables the fast query of density estimates locally.

To further reduce the number of bins, a next step is to apply instead of uniform or constant bin-densities linear or higher-order densities similar to polynomial histograms. The corresponding polynomial coefficients are derived from the bin data and a composite goodness-of-fit test is applied. In this sense, the proposed method is a hybrid between parametric and non-parametric density estimation. With the density inside a bin being a polynomial of a certain order, that approximates the density distribution locally, we refer to it as a distribution element (DE). Since the resulting density estimate is given in the form of a tree with DEs at its leafs, we refer to our approach as DE tree (DET) density estimator. In the following section~\ref{secFormulation}, we provide details about the formulation of the DET estimator and in section~\ref{secApplications} present comparative applications. Concluding remarks are provided in section~\ref{secConclusions}.

\section{Formulation}\label{secFormulation}

We aim at accurately estimating the probability density $p(\mathbf{x})$ at different positions $\mathbf{x} = (x_1,\ldots,x_d)^\top \in \Omega$ based on a given ensemble including~$n$ samples $\mathbf{x}_1,\ldots,\mathbf{x}_n$. The probability space $\Omega$ is defined by a hypercuboid, i.e.,
\begin{displaymath}
\Omega = \prod_{i = 1}^d [x_{i,l},x_{i,u}],
\end{displaymath}
with lower and upper bounds $x_{i,l}$ and $x_{i,u}$, respectively, of components~$x_i$ such that $\mathbf{x}_j \in \Omega \;\forall\; j = 1,2,\ldots,n$. This condition is for example met if
\begin{displaymath}
x_{i,l} = \min_{j = 1}^n x_{i,j} \mbox{ and } x_{i,u} = \max_{j = 1}^n x_{i,j},
\end{displaymath}
where $x_{i,j}$ refers to the $i$th component of sample $\mathbf{x}_j$.\footnote{Fixing the domain bounds based on the data range leads to bounds that are almost certainly too narrow. Accordingly, the resulting density estimates will display a bias toward too high values.} In our method, $\Omega$ is split recursively into $m$ smaller hypercuboids or simply cuboids
\begin{displaymath}
C_k = \prod_{i = 1}^d [x_{i,l}^k,x_{i,u}^k] \subset \Omega
\end{displaymath}
with $k = 1,2,\ldots,m$. The cuboids $C_k$ are disjoint and satisfy $\Omega = \bigcup_{k = 1}^m C_k$. A cuboid $C_k$ comprises together with a local density $p_k(\mathbf{x})$ the $k$th DE. Based on all DEs, the DET density estimator is then given by
\begin{equation}\label{eqDETPDF}
p(\mathbf{x}) = \sum_{k = 1}^m p_k(\mathbf{x})\;\forall\;\mathbf{x}\in\Omega.
\end{equation}
In the following section~\ref{subsecDE}, we define $p_k(\mathbf{x})$ and in section~\ref{subsecDET}, we introduce the recursive splitting method that leads to the DET density estimator.

\subsection{Distribution Elements}\label{subsecDE}

Similar to a bin in a histogram, a DE can be viewed as the least complex building block or atom of a density distribution estimate or DET. Therefore, we chose a simple analytical form that is suitable to approximate the density in a small subregion $C_k$ of $\Omega$. We define the probability density of DE~$k$ as
\begin{equation}\label{eqDEPDF}
p_k(\mathbf{x}) = \left\{\begin{array}{ll}
\displaystyle\frac{n(C_k)}{n}\prod_{i = 1}^d p[x_i|\mbf{\theta}_i(C_k)] & \forall\;\mathbf{x}\in C_k \\
0 & \mbox{otherwise,}
\end{array}\right.
\end{equation}
where $p[x_i|\mbf{\theta}_i(C_k)]$ are marginal densities of components $x_i$ with local parameter vectors~$\mbf{\theta}_i$ and $n(C_k)/n$ is the fraction of all samples~$n$ that reside in~$C_k$. Therefore in each DE, the random variables $X_i$ are approximated as statistically independent. This has important implications, since it enables us to break the exponential growth of bins in terms of~$d$ as will become clear in the next section.

Insertion of expression~\eq{eqDEPDF} in equation~\eq{eqDETPDF} and integration over the entire probability space $\Omega$ reveals that, since $\sum_{k = 1}^m n(C_k) = n$, the DET estimator integrates to one and therefore satisfies the normalization condition of a probability density function (PDF). Moreover, if $p[x_i|\mbf{\theta}_i(C_k)]\ge 0\;\forall\;x_i\in[x_{i,l}^k,x_{i,u}^k]$, the DET estimator is non-negative for $\mathbf{x}\in\Omega$ and therefore is a PDF.

With the uniform marginal density
\begin{equation}\label{eqDEConstant}
p[x_i|\mbf{\theta}_i(C_k)] = \frac{1}{x_{i,u}^k-x_{i,l}^k},
\end{equation}
which defines a so-called constant DE, we obtain from equation~\eq{eqDEPDF}
\begin{displaymath}
p_k(\mathbf{x}) = \frac{n(C_k)}{n}\frac{1}{\prod_{i = 1}^d (x_{i,u}^k-x_{i,l}^k)},
\end{displaymath}
which is the familiar density in a histogram bin \citep[e.g.,][section~3.4]{Scott:2015a}. Similarly, the density of a linear DE is given as
\begin{equation}\label{eqDELinear}
p[x_i|\mbf{\theta}_i(C_k)] = \frac{\left(\frac{x_i-x_{i,l}^k}{x_{i,u}^k-x_{i,l}^k} - \frac{1}{2}\right)\theta_{i,1}(C_k) + 1}{x_{i,u}^k-x_{i,l}^k},
\end{equation}
where $\theta_{i,1}(C_k)$ is a slope parameter. We notice that a linear DE with $\theta_{i,1} = 0$ is equivalent to a constant DE. To estimate the slope based on the data points inside~$C_k$, we apply the following minimum mean square error (MMSE) estimator
\begin{equation}\label{eqSlope}
\theta_{i,1}(C_k) = \frac{n(C_k)s_i(C_k)^3}{n(C_k)s_i(C_k)^2 + 144\langle X_i^{\prime 2}|C_k\rangle}
\end{equation}
with $s_i(C_k) = 6(2\langle X_i|C_k\rangle - 1)$. Here, the quantities $\langle X_i|C_k\rangle$ and $\langle X_i^{\prime 2}|C_k\rangle$ are the mean and variance estimates based on component $x_i$ of the samples contained in cuboid~$C_k$. A derivation of the MMSE slope estimator~\eq{eqSlope} is included in the appendix. To make sure that $p[x_i|\mbf{\theta}_i(C_k)] \ge 0$, we clip $\theta_{i,1}(C_k)$ such that $\theta_{i,1}(C_k) \in [-2,2]$. In this work, we focus on DEs with uniform and linear marginal PDFs, but higher order DEs are conceivable as well.

It is pointed out that the polynomials~\eq{eqDEPDF} are differently constructed compared to the ones applied by \citet[equations~(5) and~(6)]{Jing:2012a}. Our DE classification is based on the marginal densities $p[x_i|\mbf{\theta}_i(C_k)]$, while \citeauthor{Jing:2012a} classify the expanded polynomial $p_k(\mathbf{x})$ and allow for statistical dependence among the components $x_i$.

\subsection{Distribution Element Tree}\label{subsecDET}

For the construction of the DET density estimator~\eq{eqDETPDF}, a constant or linear DE is assigned to the probability space~$\Omega$ or root cuboid depending on the required order. To verify for the root DE or any subsequent DE~$k$ whether the data included in~$\Omega$ or~$C_k$ are compatible with DE density~\eq{eqDEPDF}, we apply goodness-of-fit tests for the marginal distributions and test independence of the joint distribution. This testing approach is directly implied by the structure of the DE density~\eq{eqDEPDF}. (a) If one or several of these tests fail, meaning that the null hypothesis given by DE density~\eq{eqDEPDF} is rejected, the distribution of the data within $C_k$ is most likely more complex (and there is sufficient data within DE~$k$ for rejecting). Consequently, DE $k$ is considered interim and a split along one or two probability space directions is conducted. Next, the testing and splitting process continues for each of the resulting two or four DEs. On the other hand, (b) if  DE~$k$ passes all tests or contains no data, the splitting process stops, DE~$k$ is final, and becomes one of the $m$ DEs that are part of the DET density estimator~\eq{eqDETPDF}. During the outlined estimator construction process, a tree is emerging with its root given by the DE on $\Omega$, branches to interim DEs resulting from DE splits, and final DEs at the leafs of the tree. More details about the testing and splitting processes follow in the next few paragraphs.

\subsubsection{Splitting}

If a split of DE $k$ on cuboid~$C_k$ along dimension~$x_i$ is conducted, cuboid~$C_k$ is split either into subcuboids with equal volume (equal size split) or subcuboids with approximately equal number of samples (equal score split). If splits in directions $x_{i}$ and $x_{j}$ are to be conducted, we first split along direction~$x_{i}$ and then split each of the two resulting DEs along $x_{j}$.

\subsubsection{Goodness-of-Fit Testing}

To verify whether the data in DE~$k$ are compatible with density~\eq{eqDEPDF}, we apply a two-stage testing sequence. First, we apply goodness-of-fit tests in each of the $d$ directions $x_i$ to verify whether the marginal distributions of the data are compatible with the null hypothesis given by the marginal DE densities~\eq{eqDEConstant} or~\eq{eqDELinear}, depending on the DE order. If in one or several directions the null hypothesis is rejected, we split along the rejecting direction with the smallest $p$-value. In this work, we focus on Pearson's $\chi^2$ goodness-of-fit test \citep{Pearson:1900a} with significance level $\alpha_g$.

This test was originally developed for categorical data, but is often used for continuous variables as well with the continuous data grouped into classes \citep[e.g.,][]{Mann:1942a}. We define the classes such that the number of samples in
\begin{equation}\label{eqNClassesGoF}
n_c = \min\left[\frac{n}{5}, 4\sqrt[5]{\frac{2(n-1)^2}{c^2}}\right]
\end{equation}
classes, with $c = \sqrt{2}\,\mbox{erfcinv}(2\alpha_g)$, is approximately equal. Here, erfcinv is the inverse complementary error function. In expression~\eq{eqNClassesGoF}, the first contribution $n/5$ is the rule of thumb, stating that each class should contain at least five samples \citep[section~7]{Cochran:1952a}, and the second contribution was proposed by \citet[theorem~1]{Mann:1942a} and maximizes the power of the $\chi^2$ test under certain conditions. With constant and linear DEs, the $\chi^2$ test is conducted with $n_c-1$ and $n_c-2$ degrees of freedom, respectively (composite test).

To investigate the influence of the goodness-of-fit test on the performance of the DET method, we apply in section~\ref{subsecDepTestParams} in addition to the previously mentioned $\chi^2$ goodness-of-fit test a Kolmogorov--Smirnov (KS) test \citep{Smirnov:1948a}. Even though the KS test is a non-parametric test in the sense that no discrete classes have to be prescribed, we favor the $\chi^2$ test for reasons of computational efficiency and convenience in cases where large samples and composite tests are involved, respectively.

\subsubsection{Independence Testing}

Second, if $d > 1$ and all tests concerning the marginal distributions are passed without any split, we verify whether the components of the data in DE~$k$ are pairwise independent and thus approximately compatible with DE distribution~\eq{eqDETPDF}. We point out that this approach is approximate, since (mutually) independent random variables, as represented by distribution~\eq{eqDETPDF}, are pairwise independent, but for $d > 2$ pairwise independence does generally not imply mutual independence \citep[p.~184]{Papoulis:1991a}. If not stated otherwise, we apply Pearson's $\chi^2$ independence test \citep{Pearson:1900a} with significance level $\alpha_d$ for pairs of components~$x_{i}$ and~$x_{j}$ with $i = 1,2,\ldots,d-1$ and $j = i+1,i+2,\ldots,d$. We use contingency tables with classes such that the samples are equally distributed among $\sqrt{n_c}$ classes in the $x_{i}$- and $x_{j}$-direction \citep[section~3.1]{Bagnato:2012a}. Moreover, a $\chi^2$ distribution with $(\sqrt{n_c}-1)^2$ degrees of freedom is applied. 

To inspect the role of the independence test statistic on the performance of the DET method, we apply in section~\ref{subsecDepTestParams} for comparison an independence test based on Kendall's $\tau$ \citep{Kendall:1938a}. However, for reasons of computational efficiency in high-dimensional cases, we favor the $\chi^2$ independence test.

If for one or several component pairs the independence null hypothesis is rejected, DE splitting is done along the components of the rejecting pair with the smallest $p$-value. Within component pairs, the component with smallest goodness-of-fit $p$-value is split first.

\subsubsection{Computational Cost}

The outlined approximate testing sequence avoids the initiation of simultaneous splits in more than two directions and therefore, the DET estimator is not subject to the exponential growth of bins in terms of~$d$. At each DE in the tree, (a) $d$ goodness-of-fit tests are conducted and if $d > 1$ and no split was induced by these tests, (b) of the order of $d^2/2$ pair-wise independence tests follow. This testing sequence is applied at each tree node, that is each interim and final DE. In the case of score-based splitting, an upper bound for the number of interim and final DEs, that determines the DET construction cost, exists. If each final DE contains just one sample, the maximal possible number of final DEs $m = n$. Moreover, if $n$ can be expressed as an integer-valued power of two, the number of all interim and final DEs is equal to
\begin{equation}\label{eqNumbDEs}
1 + 2 + 4 + 8 + \ldots + n/2 + n = 2n - 1.
\end{equation}
Here, the sum starts with~1 representing the root DE and after multiple generations of binary splits ends with $n$ final DEs. If $n$ cannot be written as an integer power of two, the next larger integer that is a power of two is $n^\prime = 2^{\mathrm{ceil}[\log_2(n)]}$, which satisfies $n < n^\prime < 2n$. Then based on result~\eq{eqNumbDEs}, an upper bound for the number of all interim and final DEs is given by $2n^\prime - 1 < 4n - 1$. To summarize, when applying equal score splits, the computational effort for the DET construction scales in the worst case linearly with the number of samples~$n$.

To evaluate the density estimate~\eq{eqDETPDF} at a certain location~$\mathbf{x}$, i.e., $p(\mathbf{x})$, the sum~\eq{eqDETPDF} involving~$m$ final DEs could be evaluated. A more efficient approach, however, exploits the tree structure of the DET estimator: The only non-zero term in sum~\eq{eqDETPDF} can be quickly identified by starting at the root DE and by sequentially identifying at the DET forks the interim DE~$l$ where $\mathbf{x}\in C_l$. After of the order of~$n_t$ decision operations, where $n_t$ is the number of DE splits or tree depth, the leaf or final DE~$k$ is reached that contains point~$\mathbf{x}$. Typically $n_t \ll m < n$, which renders the DET estimator computationally efficient for density queries. In the next few sections, we document the capabilities of the DET density estimator for a range of test cases.

\section{Density Estimation with the DET Method}\label{secApplications}

For the following computations, if not mentioned otherwise, significance levels $\alpha_g = \alpha_d = 0.001$ were applied. All computations were carried out on a state-of-the-art laptop computer with a 2.8GHz Intel Core~i7 processor. We inspect the accuracy of the DET estimator for a diverse set of one-, two-, four-, and seven-dimensional cases in the following three sections. To this end, we estimate the MISE defined as
\begin{equation}\label{eqMise}
\left\langle\int_\Omega [\hat{p}(\mathbf{x}) - p(\mathbf{x})]^2\,\mathrm{d}\mathbf{x}\right\rangle,
\end{equation}
where $\hat{p}(\mathbf{x})$ with $\mathbf{x}\in\Omega$, given for example by the DET method~\eq{eqDETPDF}, is an estimate of the exact PDF $p(\mathbf{x})$ based on an ensemble of samples and angular brackets represent the expectation with respect to that ensemble. To put our results into perspective with state-of-the-art density estimation, we compare against the adaptive KDE method of \citet{Botev:2010a}, the density-estimation-tree or in short density-tree method of \citet{Ram:2011a}, the LL-OPT estimator of \citet{Jiang:2016a}, and finally a conventional histogram, where the bin width in each dimension was determined based on the normal reference rule \citep[equation~(3.66)]{Scott:2015a}.

To numerically evaluate the integration over probability space $\Omega$ in the MISE~\eq{eqMise}, we applied for $d = 1$ and~2 trapezoidal rules with equidistant Cartesian grids. As an exception to this, an exponentially stretched grid was applied for the one-dimensional gamma PDF example (following in section~\ref{subsubsec1dC6}) to better resolve the region of high probability near the origin. To assert the accuracy of the resulting MISE, adaptive quadrature methods as outlined by \citet{Shampine:2008a,Shampine:2008b} were applied as well, leading to virtually the same results as with trapezoidal rules. In connection with the DET estimator, here the integration region~$\Omega$ was decomposed among all DEs and adaptive quadrature was applied in each element individually. For the four- and seven-dimensional cases, quadrature or trapezoidal rules become too expensive and we resorted to Monte Carlo (MC) integration instead. To this end, we rewrite the MISE~\eq{eqMise} as
\begin{eqnarray}\label{eqMiseMC}
& & \left\langle\int_\Omega \left[\frac{\hat{p}(\mathbf{x})^2}{p(\mathbf{x})} - 2\hat{p}(\mathbf{x}) + p(\mathbf{x})\right] p(\mathbf{x})\,\mathrm{d}\mathbf{x}\right\rangle \nonumber \\
& & = \left\langle\lim_{n\to\infty}\frac{1}{n}\sum_{j = 1}^n \left[\frac{\hat{p}(\mathbf{x}_j)^2}{p(\mathbf{x}_j)} - 2\hat{p}(\mathbf{x}_j) + p(\mathbf{x}_j)\right] \right\rangle,
\end{eqnarray}
with samples~$\mathbf{x}_j$ distributed as prescribed by PDF $p(\mathbf{x})$.

Computationally efficient spectral KDE implementations for $d = 1$ and~2, that operate on equidistant Cartesian grids, are available \citep{Botev:2007a}. Since these implementations become prohibitive for $d > 2$, an alternative was recently provided by \citet{Botev:2016a}. However, here similar to conventional KDE the computational effort per density evaluation depends on~$n$. A density-tree implementation is available via the MLPACK library \citep{Curtin:2013a}. For the LL-OPT computations, the implementation cited by \citet[section~7]{Jiang:2016a} was applied. We used the outlined implementations with prescribed standard parameter values.

In the following one- and two-dimensional examples, MISE ensemble averages $\langle\,\ldots\,\rangle$ were approximated based on 50 samples. Moreover, for the DET method, based on these samples, average number of DEs $\langle m\rangle$ and tree depths $\langle n_t\rangle$ are reported. Here, $n_t$ for one DET sample corresponds to the maximal depth within the tree.

We start our assessment of the DET method, by comparing in the following sections~\ref{subsec1d} and~\ref{subsec2d} the different DET variants to the established KDE method of \citet{Botev:2010a}. In a next set of comparisons contained in section~\ref{subsec1d2d}, we focus on linear DETs with equal-size splits and compare with density trees, LL-OPTs, and histograms.

\subsection{One-Dimensional Examples}\label{subsec1d}

If not stated otherwise, for the following one-dimensional adaptive KDE, a grid with $2^{14} = 16384$ nodes or cosine modes was deployed \citep{Botev:2007a}. The following one-dimensional examples involving normal mixture densities were taken from \citep{Marron:1992a}.

\subsubsection{Kurtotic Unimodal PDF}

\begin{figure*}
\unitlength\textwidth
\begin{picture}(1,0.62)
\put(0.0,0.31){\includegraphics[width=0.5\textwidth]{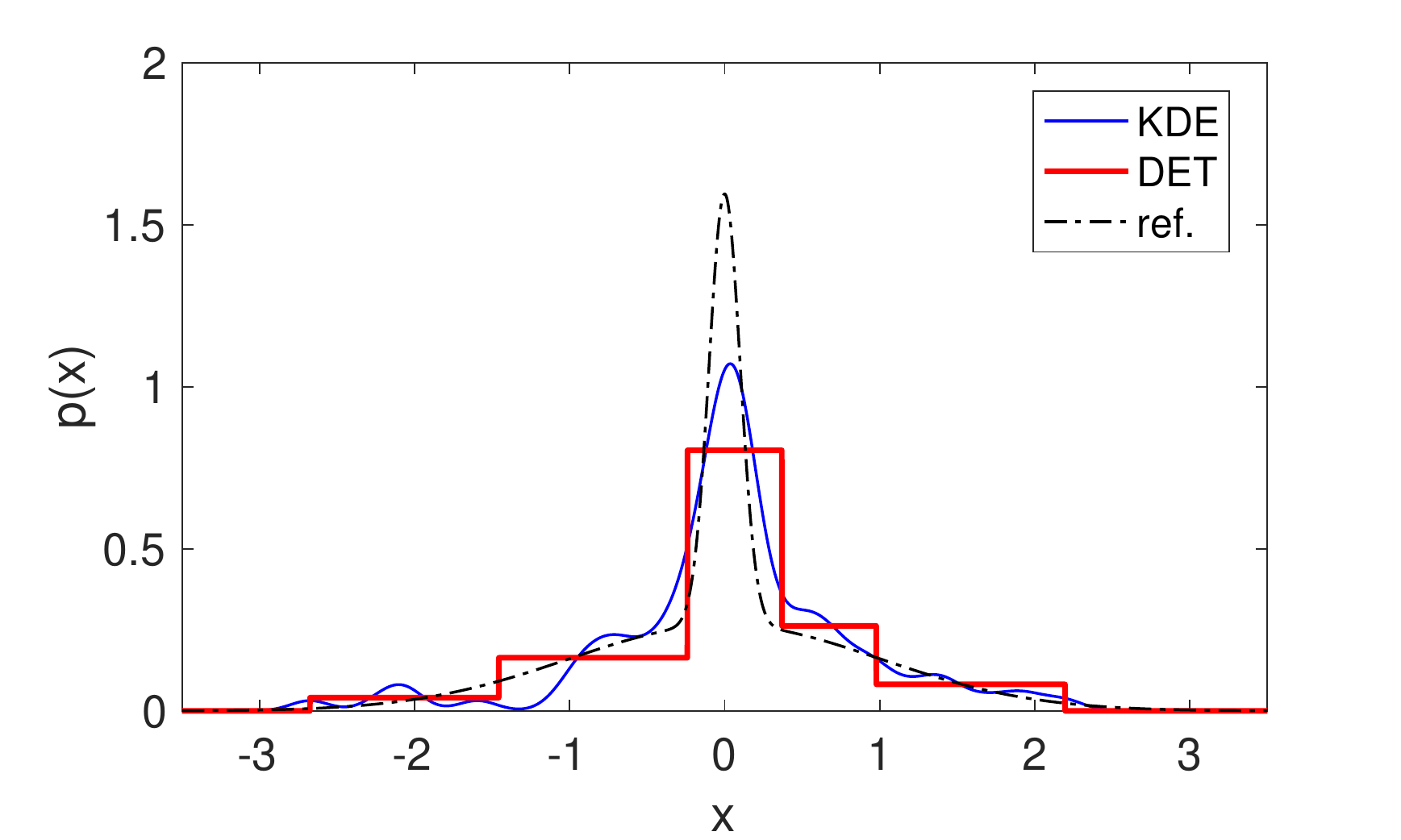}}
\put(0.01,0.61){\makebox(0,0){(a1)}}
\put(0.5,0.31){\includegraphics[width=0.5\textwidth]{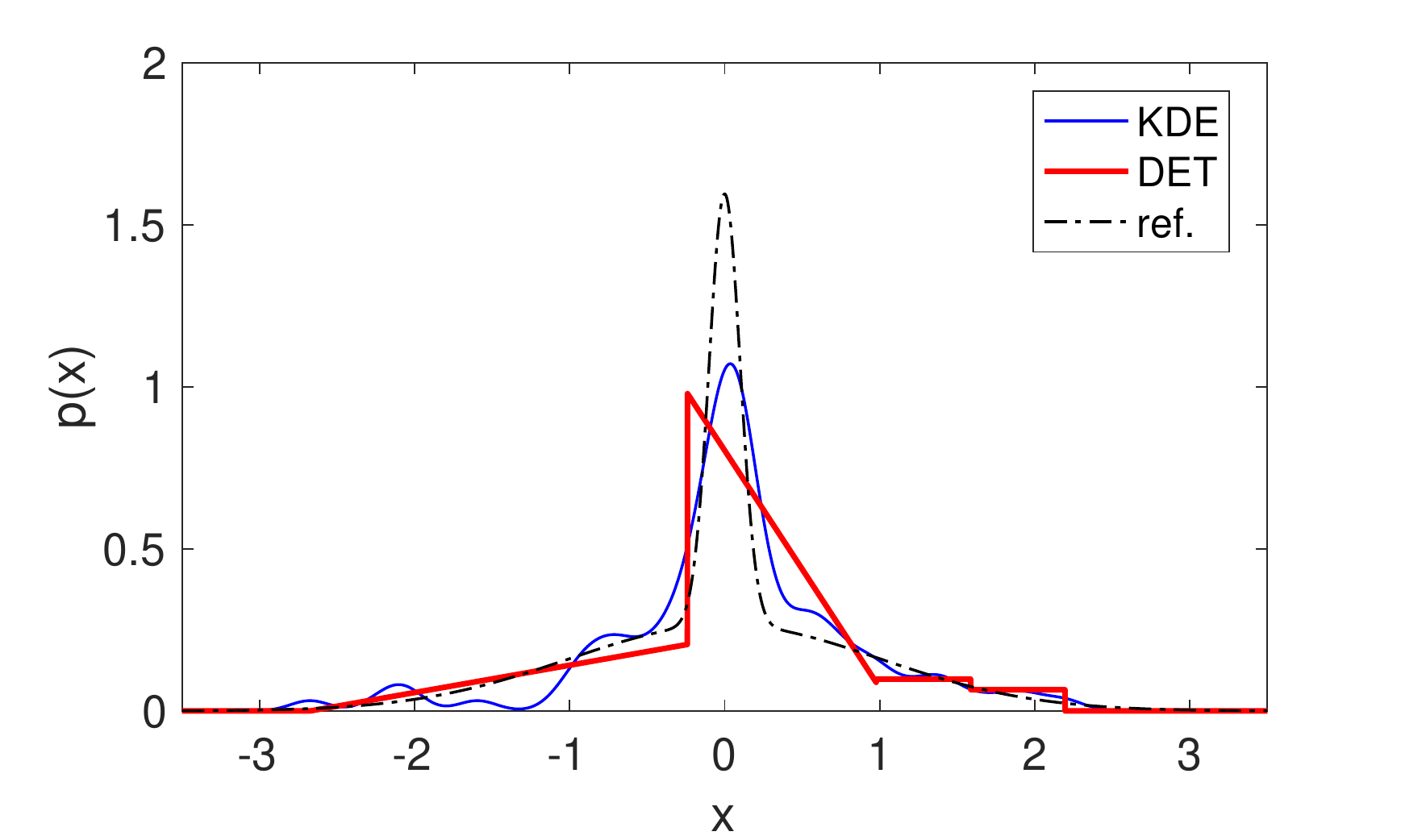}}
\put(0.51,0.61){\makebox(0,0){(b1)}}
\put(0.0,0.0){\includegraphics[width=0.5\textwidth]{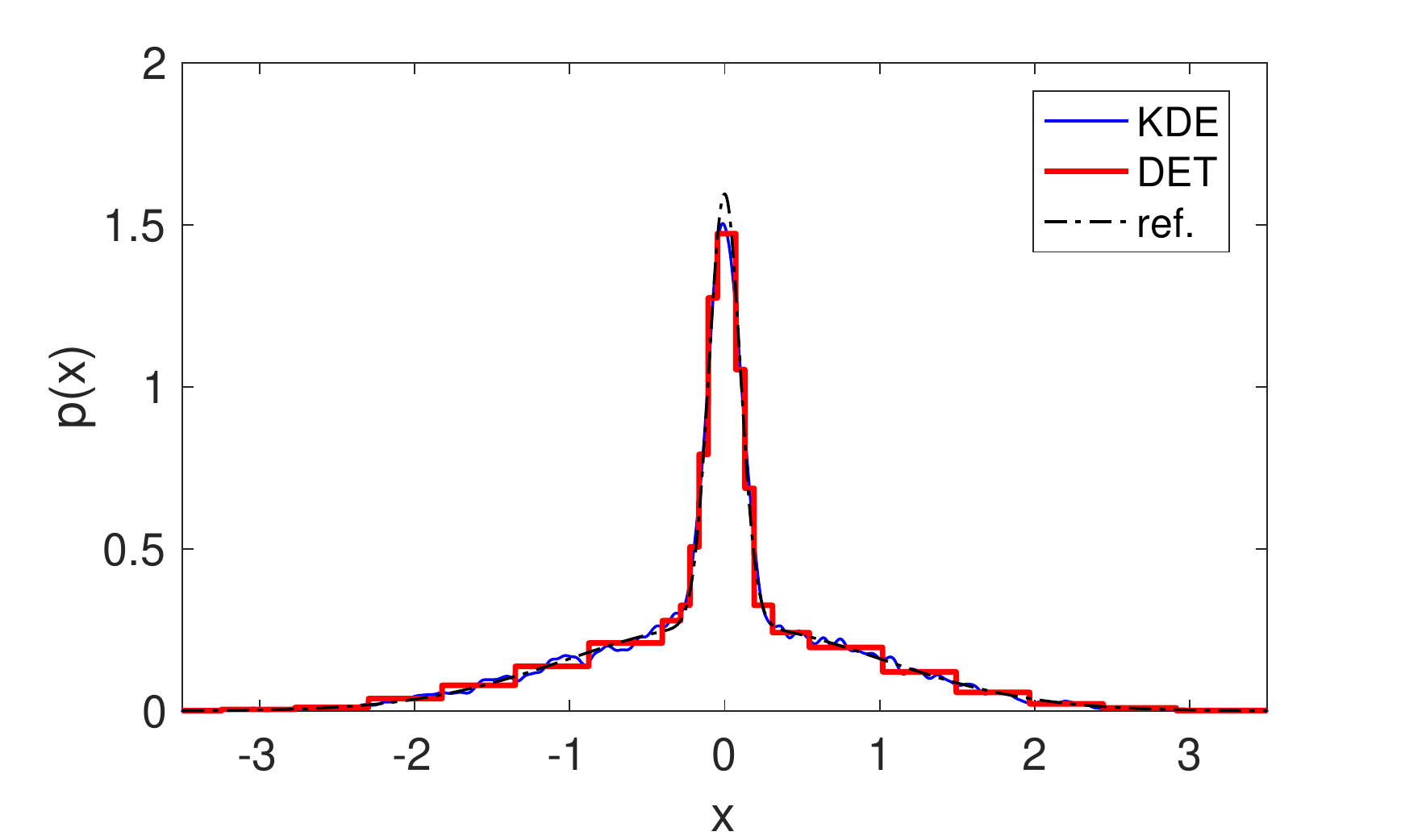}}
\put(0.01,0.3){\makebox(0,0){(a2)}}
\put(0.5,0.0){\includegraphics[width=0.5\textwidth]{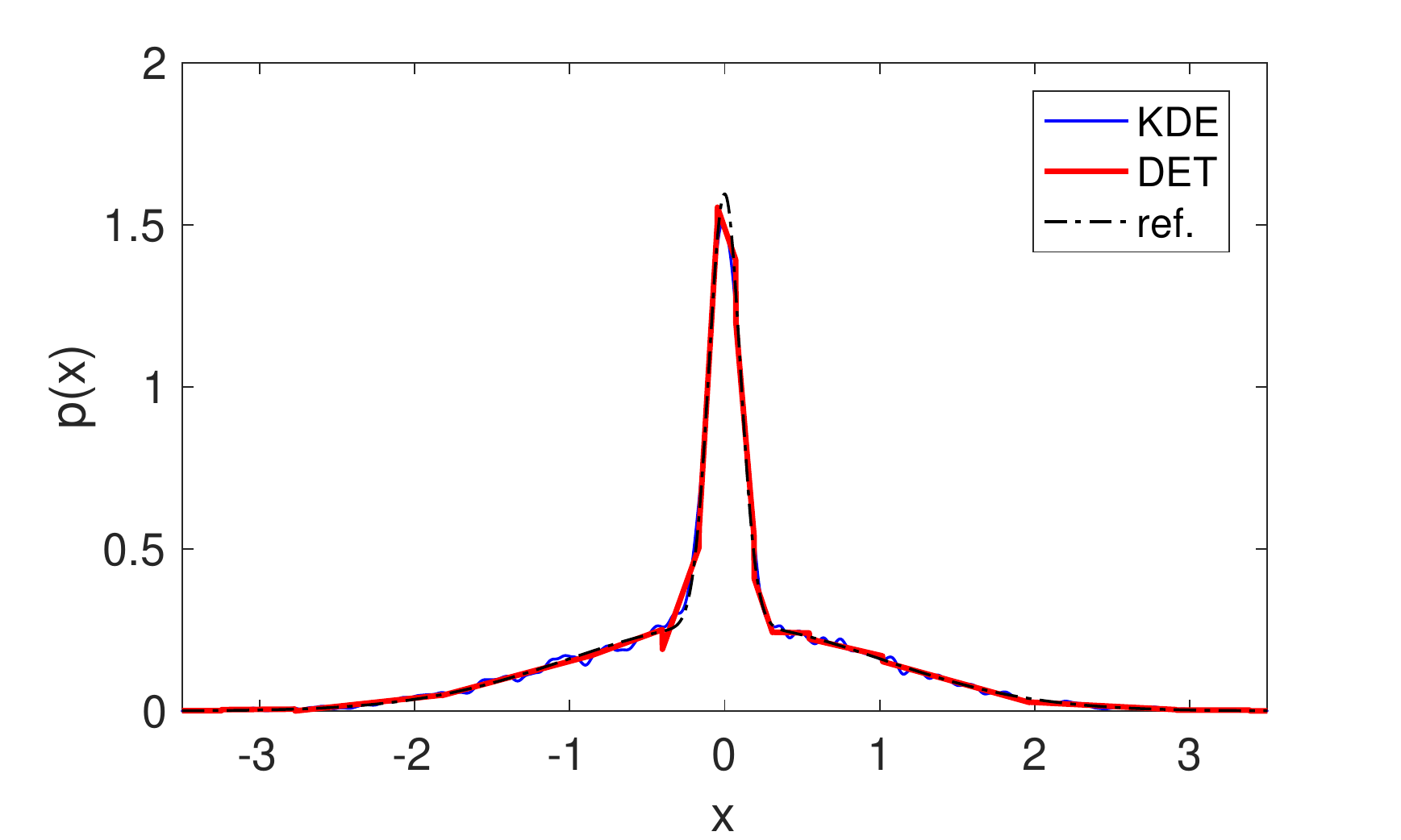}}
\put(0.51,0.3){\makebox(0,0){(b2)}}
\end{picture}
\caption{PDF estimates resulting from (blue thin solid) adaptive KDE and (red thick solid) the size-split DET method with particle ensembles including (1) $n = 100$ and (2) $10^4$ samples are compared with (black dash dot) the reference PDF~\eq{eq1dC2PDF}. In panels (a) and (b), DET estimates with constant and linear elements are depicted, respectively.}\label{fig1dC2PDF}
\end{figure*}
\begin{figure*}
\unitlength\textwidth
\begin{picture}(1,0.75)
\put(0.0,0.375){\includegraphics[width=0.5\textwidth]{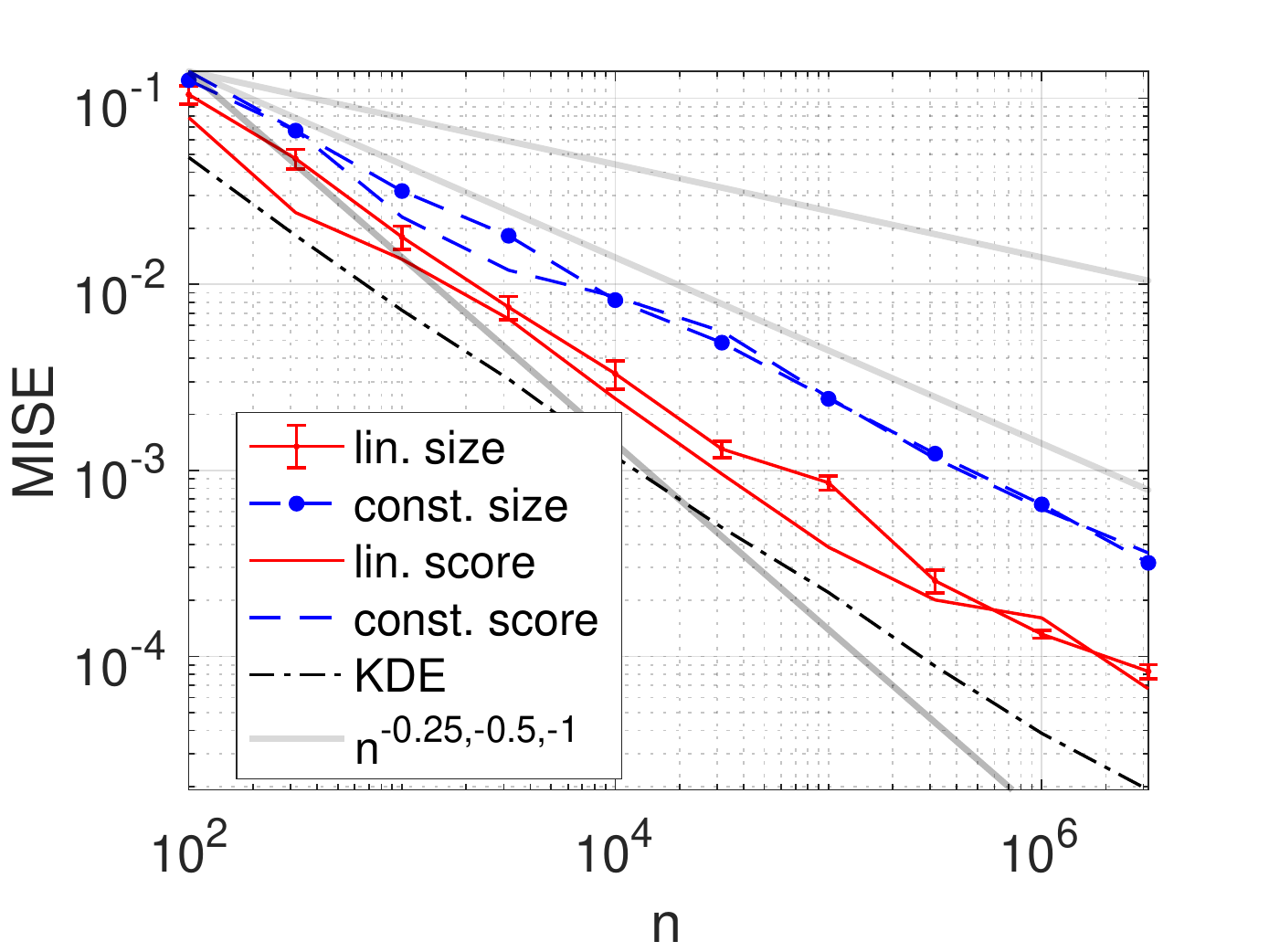}}
\put(0.01,0.74){\makebox(0,0){(a)}}
\put(0.5,0.375){\includegraphics[width=0.5\textwidth]{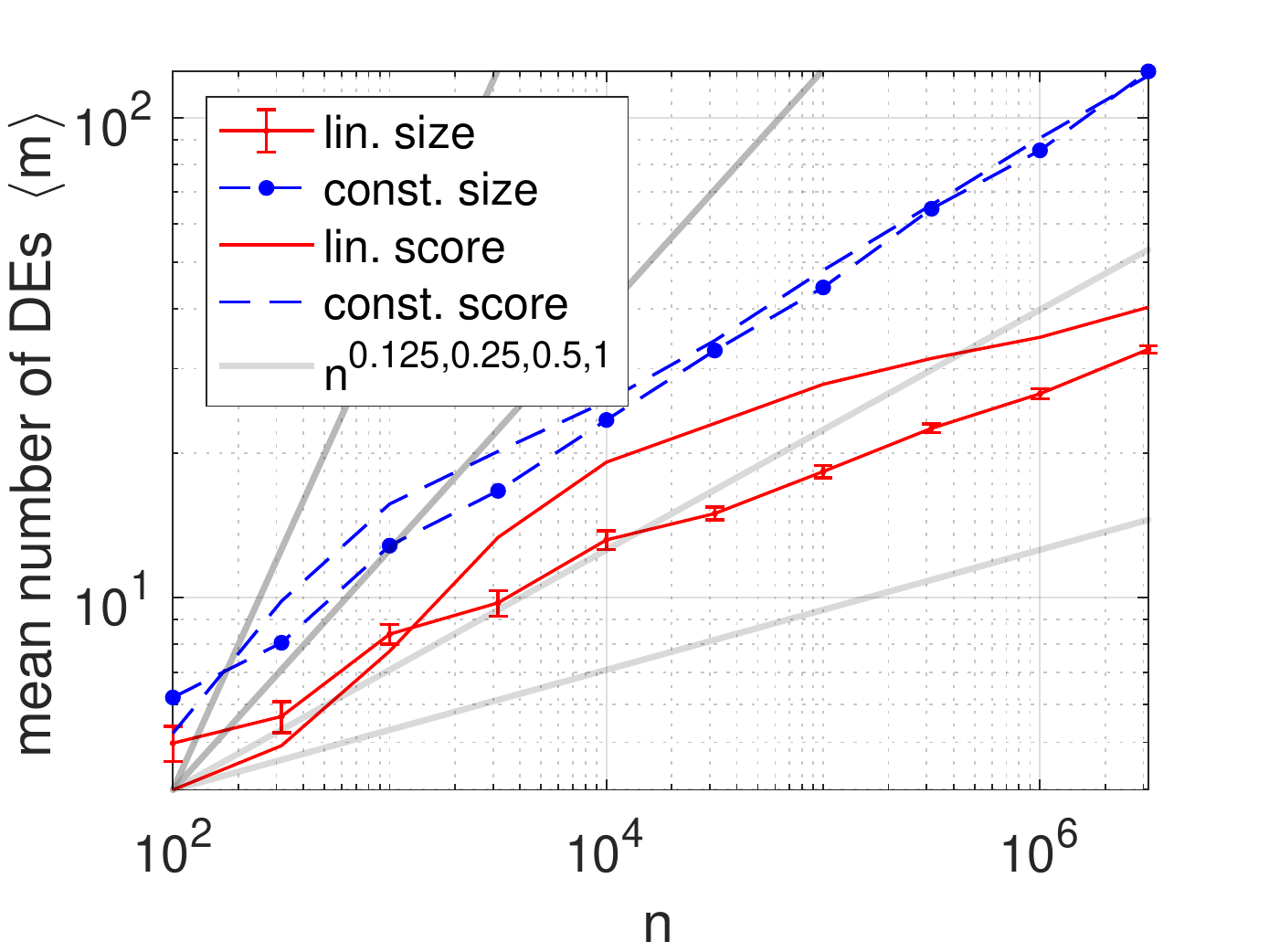}}
\put(0.51,0.74){\makebox(0,0){(b)}}
\put(0.0,0.0){\includegraphics[width=0.5\textwidth]{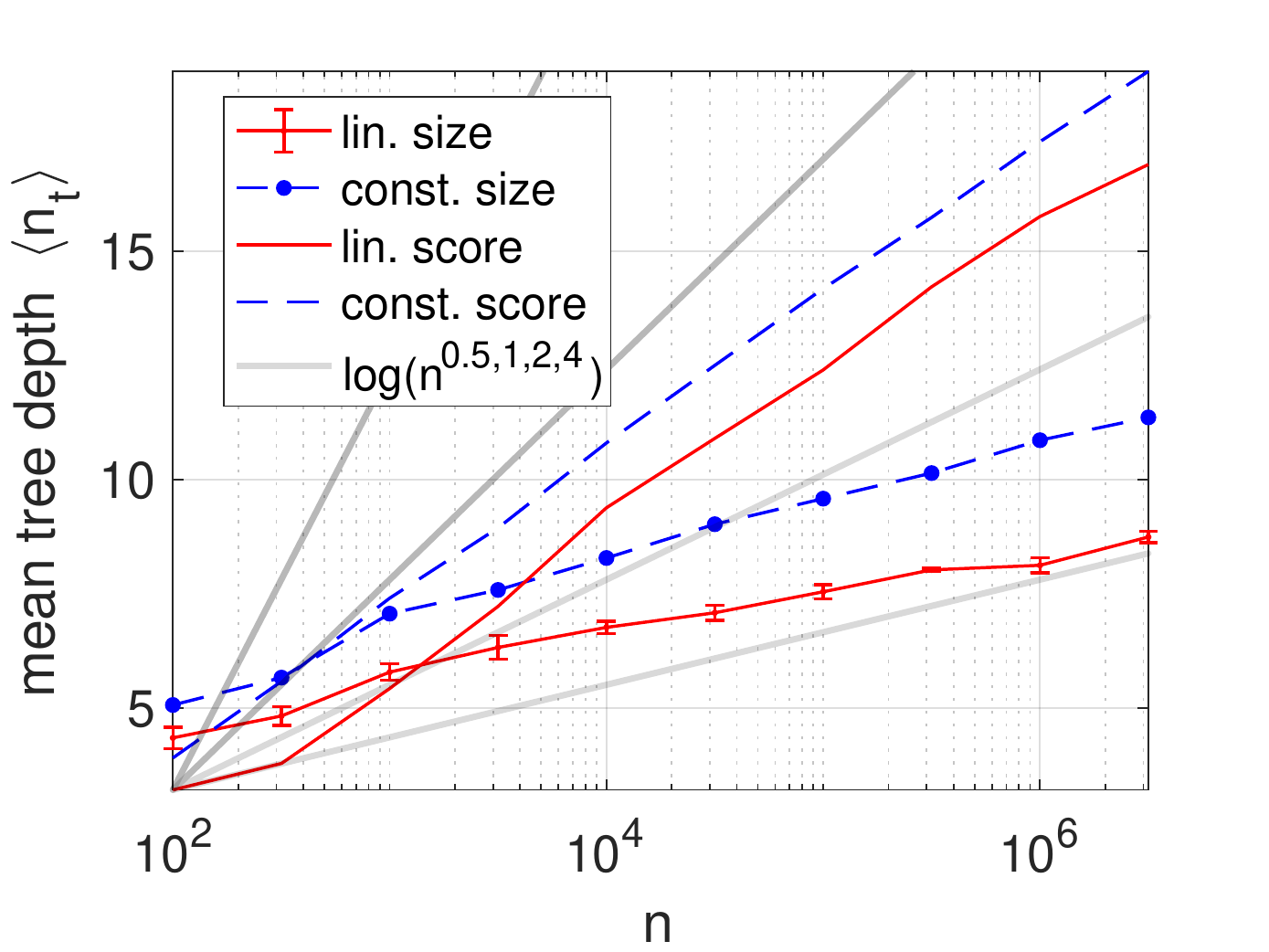}}
\put(0.01,0.365){\makebox(0,0){(c)}}
\put(0.5,0.0){\includegraphics[width=0.5\textwidth]{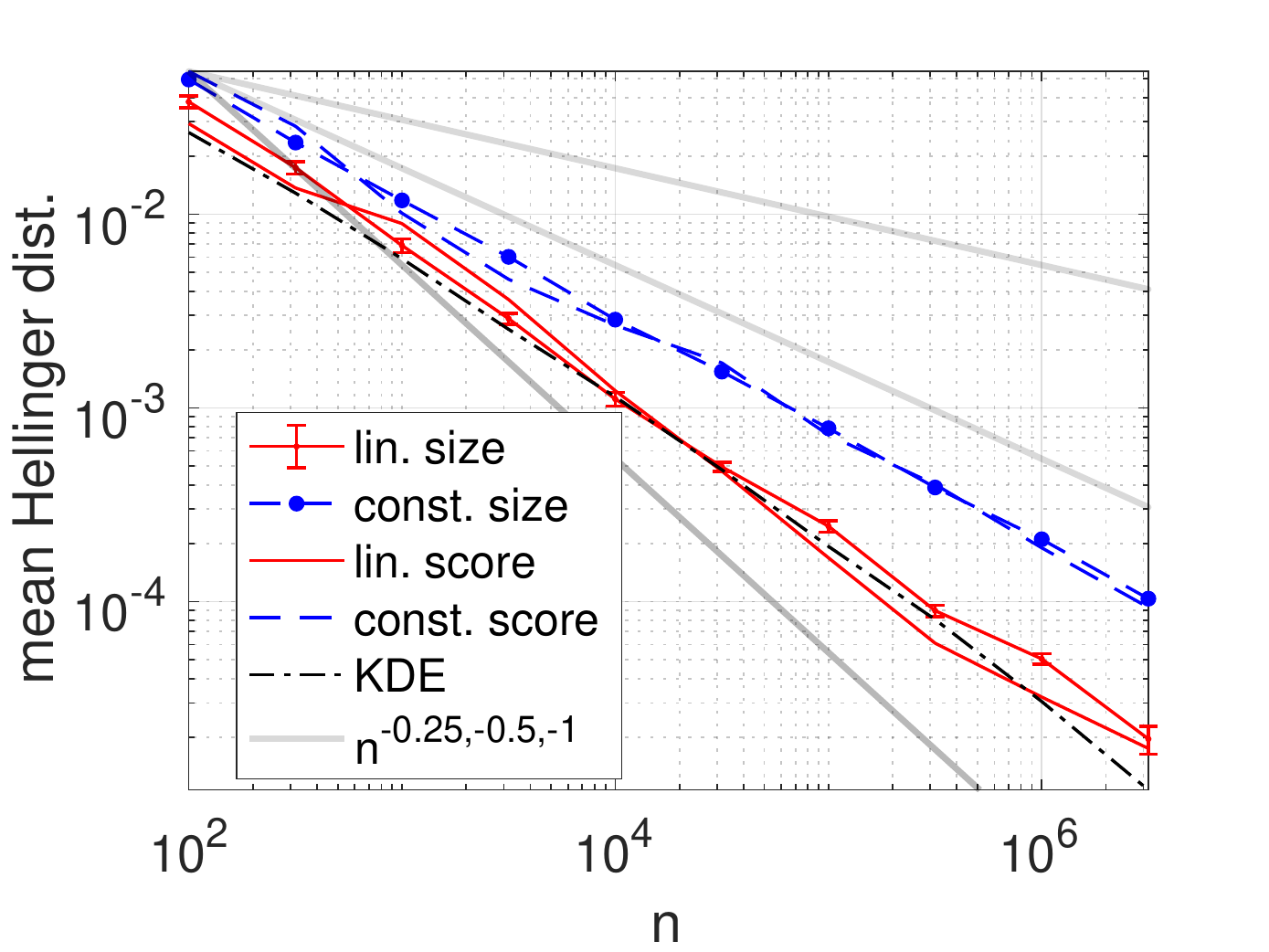}}
\put(0.51,0.365){\makebox(0,0){(d)}}
\end{picture}
\caption{Estimation of kurtotic unimodal PDF~\eq{eq1dC2PDF}. Evolutions of (a) the MISE and (d) the mean Hellinger distance as a function of the number of samples~$n$ for DET estimators (red solid and blue dashed) with equal size (symbols) and score splits (no symbols) are plotted. The MISE and mean Hellinger distance resulting from the adaptive KDE is included as well (black dash dot). For the DET estimators the resulting mean numbers of DEs~$\langle m\rangle$ and mean tree depths~$\langle n_t\rangle$ are given in panels (b) and (c), respectively. Power law and logarithmic scalings with exponents indicated in the figure legends are depicted (gray thick solid). Exemplary 95\% confidence intervals are provided for the linear DET estimator with size splits.}\label{fig1dC2MISE}
\end{figure*}
First, we inspect the performance of the DET estimator for the kurtotic unimodal distribution no.~4 in their paper, i.e.,
\begin{equation}\label{eq1dC2PDF}
p(x) = {\textstyle\frac{2}{3}}N(0,1) + {\textstyle\frac{1}{3}N(0,1/10)},
\end{equation}
where $N(\mu,\sigma)$ is the normal distribution with mean~$\mu$ and standard deviation~$\sigma$. Exemplary estimates resulting from the DET method and KDE are depicted together with expression~\eq{eq1dC2PDF} in \figurename{}~\ref{fig1dC2PDF}. The DET method adaptively allocates finer bins in the region where part $N(0,1/10)$ dominates and resorts to larger DEs to represent $N(0,1)$. The corresponding MISEs are depicted in \figurename{}~\ref{fig1dC2MISE}(a) as a function of the number of samples~$n$. A power law decay with exponents between $-1/2$ and -1 is observed for all DET estimators and the adaptive KDE. The linear DET methods and adaptive KDE share approximately the same decay rate, with KDE having a smaller MISE than the DET variants.

Besides the inferior MISE decay rate, constant DEs require more DEs as is seen in \figurename{}~\ref{fig1dC2MISE}(b). Here, the mean number of constant and linear DEs grows approximately with $n^{1/4}$ and $n^{1/8}$ for large~$n$. The number of DEs $\langle m\rangle$ determines the computational cost associated with the DET construction. These growth rates are independent of the splitting method. While the DET splitting method has little effect on $\langle m\rangle$ and the MISE, the tree growth, measured by $\langle n_t\rangle$, is smaller for size- vs.\ score-based splitting (see \figurename{}~\ref{fig1dC2MISE}(c)). However, in both cases, $\langle n_t\rangle$ increases logarithmically with~$n$ to a good approximation.

With $m$ and $n_t$ being discrete variables and the DE splitting being a discrete process, the DET datasets in \figurename{}~\ref{fig1dC2MISE} are not as smooth as for example the MISE decay in the adaptive KDE. Deviations from regular scalings in the present, and more so in the next examples, are especially apparent for small~$n$, where the DETs are comprised of few splitting levels and DEs. In \figurename{}~\ref{fig1dC2MISE}, exemplary 95\% confidence intervals (given a Gaussian likelihood model) illustrate that the ensembles of density estimator realizations used in this work are sufficiently large to keep statistical uncertainties small.

\subsubsection{Outlier PDF}

\begin{figure*}
\unitlength\textwidth
\begin{picture}(1,0.34)
\put(0.0,0.0){\includegraphics[width=0.56\textwidth]{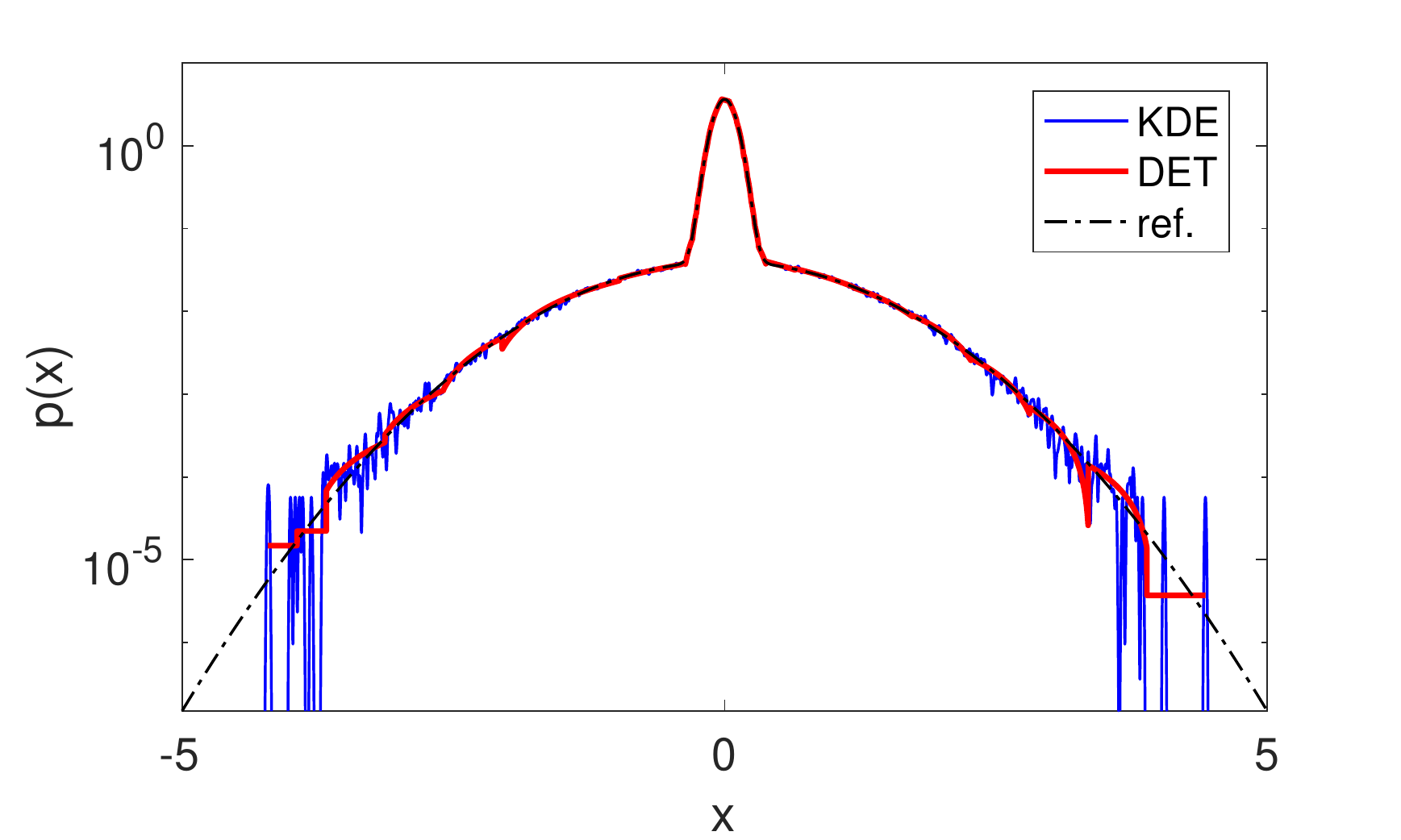}}
\put(0.01,0.33){\makebox(0,0){(a)}}
\put(0.56,0.0){\includegraphics[width=0.44\textwidth]{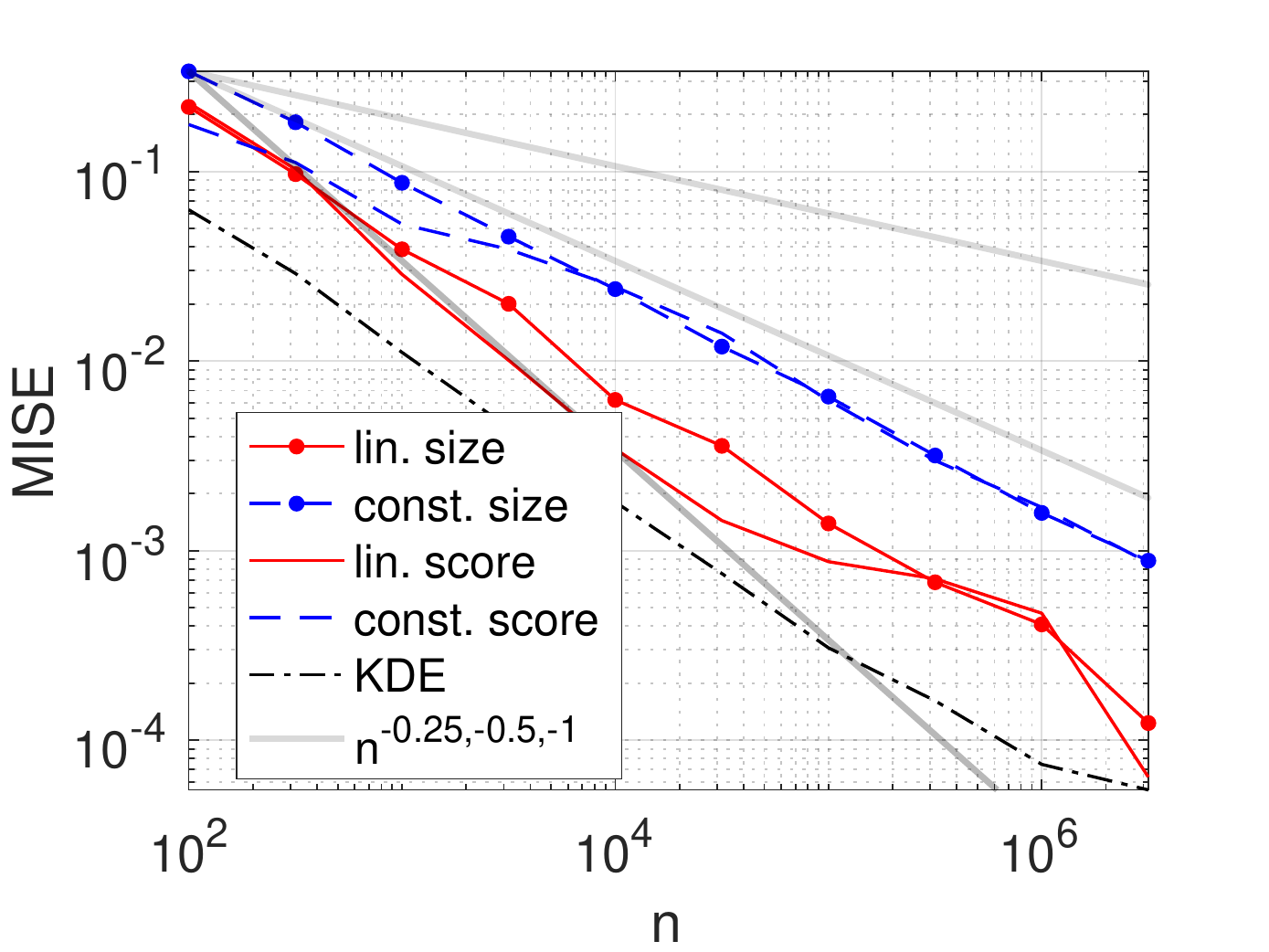}}
\put(0.57,0.33){\makebox(0,0){(b)}}
\end{picture}
\caption{(a) PDF estimates resulting from (blue thin solid) adaptive KDE and (red thick solid) the size-split linear DET method based on a particle ensembles with $n = 10^6$ samples are compared with (black dash dot) the reference PDF~\eq{eq1dC3PDF}. The corresponding MISE evolution as a function of the number of samples~$n$ is shown in panel (b) like in \figurename{}~\ref{fig1dC2MISE}(a).}\label{fig1dC3MISEPDF}
\end{figure*}
In a next step, we study the DET estimator for the outlier distribution no.~5 from \citep{Marron:1992a} given by
\begin{equation}\label{eq1dC3PDF}
p(x) = {\textstyle\frac{1}{10}}N(0,1) + {\textstyle\frac{9}{10}N(0,1/10)}.
\end{equation}
The PDF estimates and resulting MISE are plotted in \figurename{}~\ref{fig1dC3MISEPDF}. Except for the slightly worse MISE of the DET variants compared with KDE, all previous observations from the kurtotic unimodal distribution~\eq{eq1dC2PDF} carry over to the present case.

\subsubsection{Asymmetric Claw PDF}

\begin{figure*}
\unitlength\textwidth
\begin{picture}(1,0.68)
\put(0.0,0.34){\includegraphics[width=0.56\textwidth]{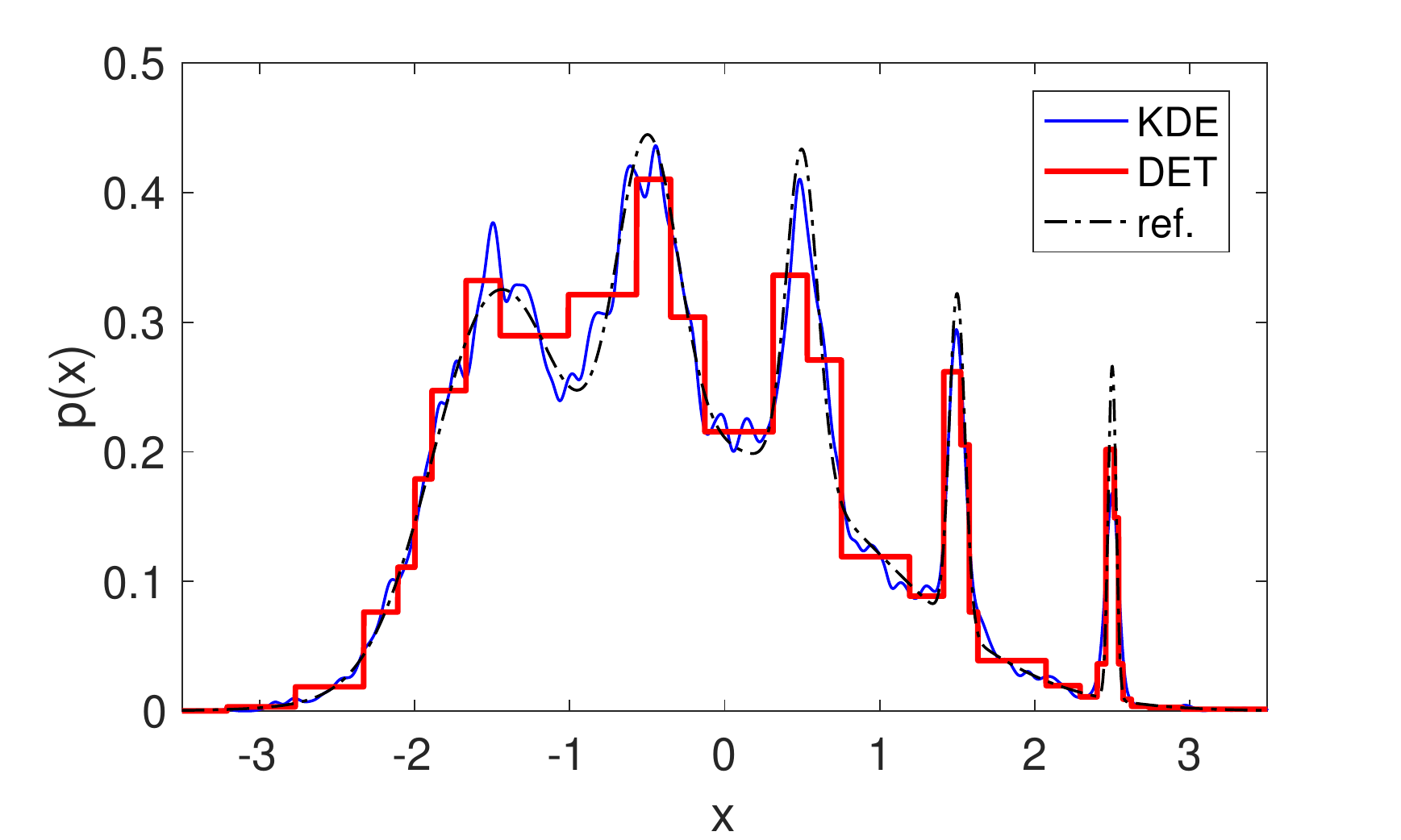}}
\put(0.01,0.67){\makebox(0,0){(a1)}}
\put(0.56,0.34){\includegraphics[width=0.44\textwidth]{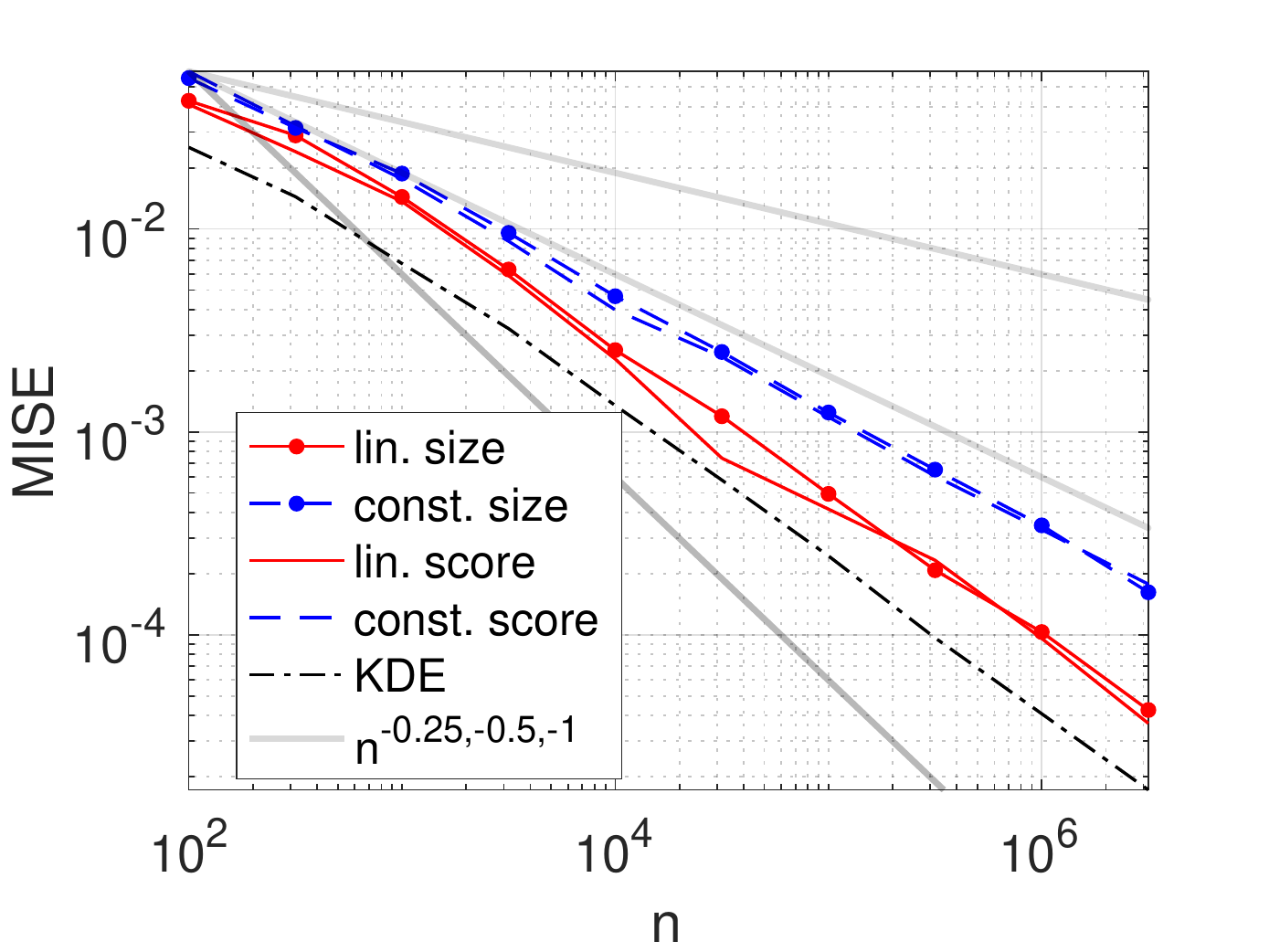}}
\put(0.57,0.67){\makebox(0,0){(b)}}
\put(0.0,0.0){\includegraphics[width=0.56\textwidth]{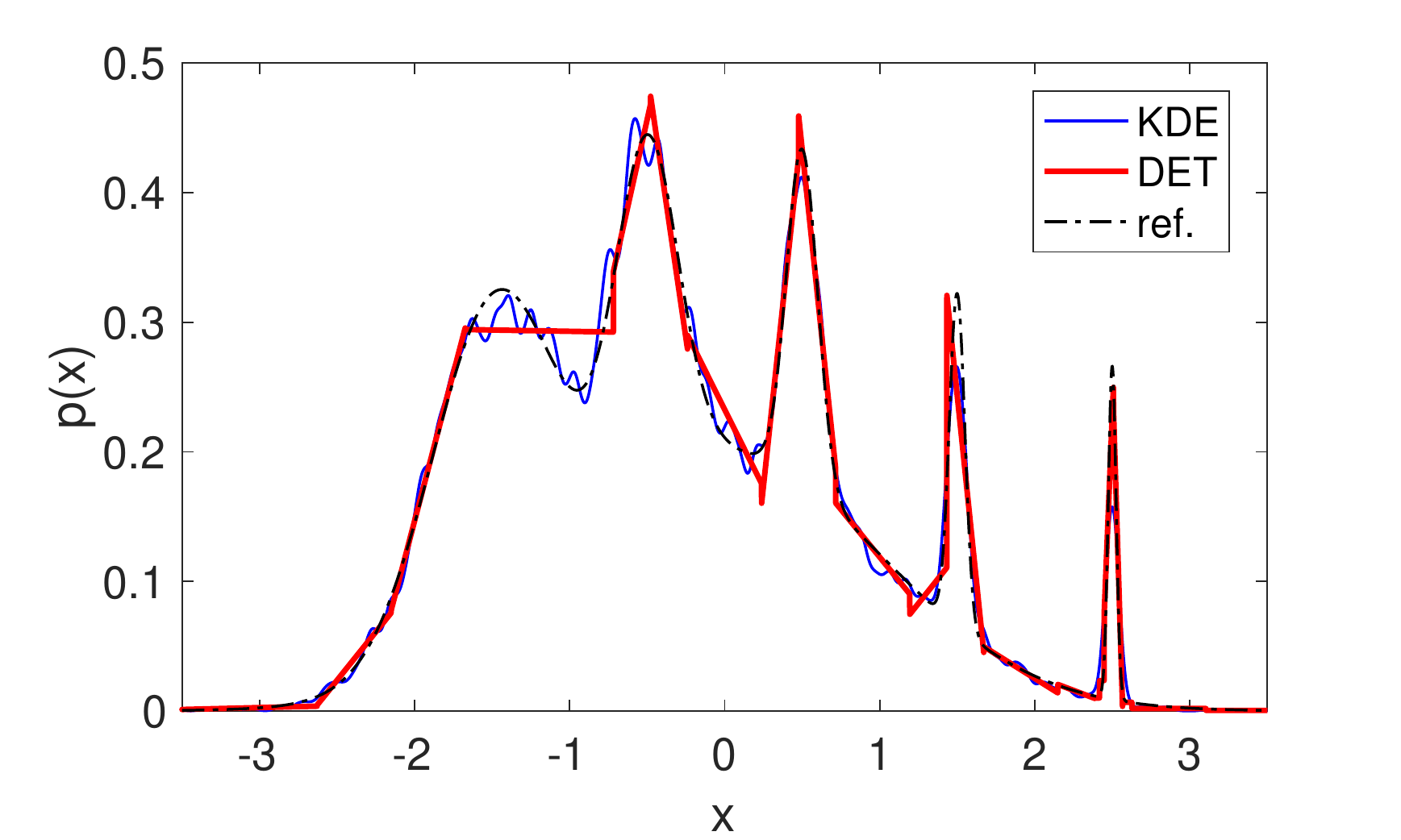}}
\put(0.01,0.33){\makebox(0,0){(a2)}}
\put(0.56,0.0){\includegraphics[width=0.44\textwidth]{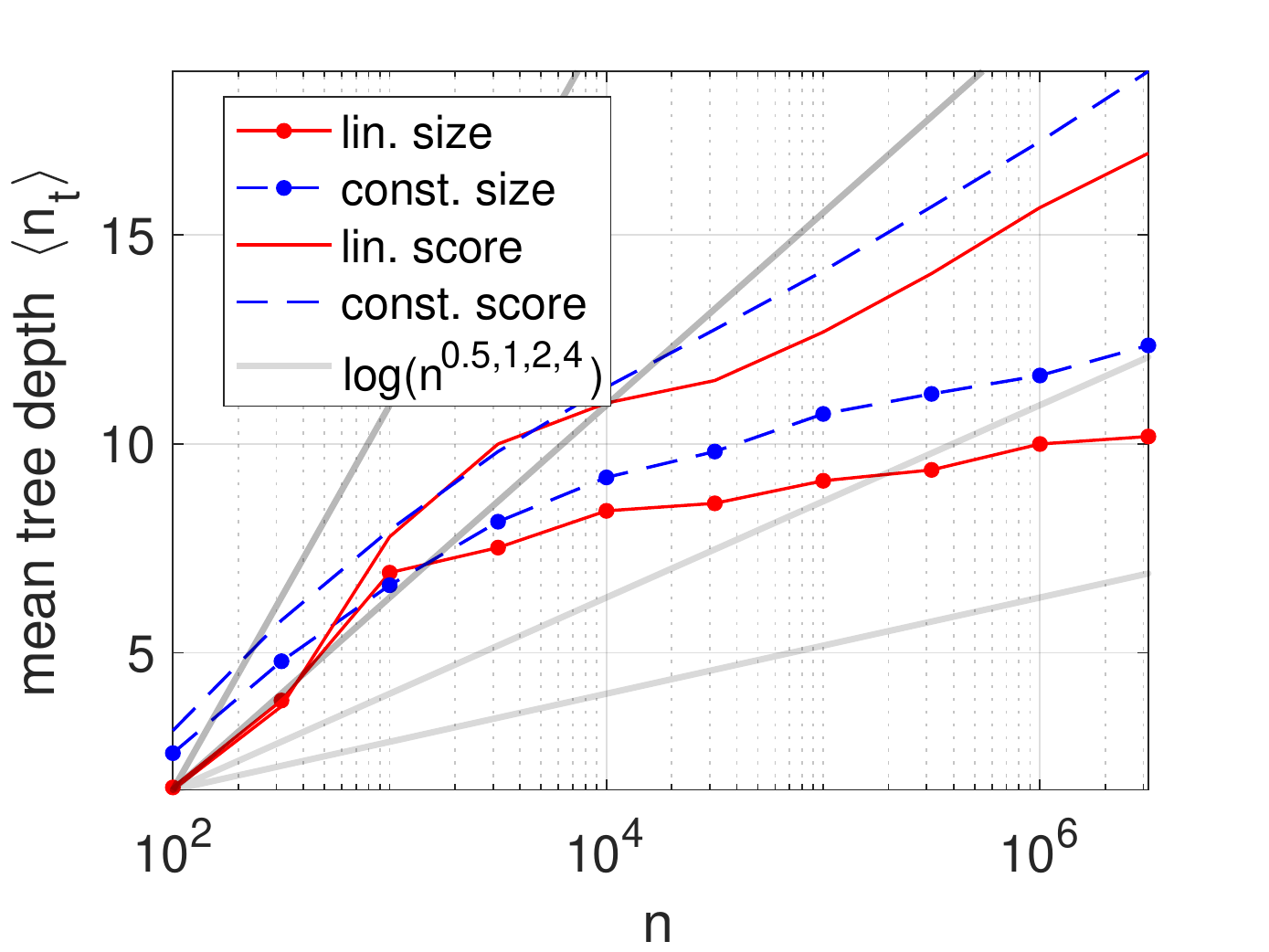}}
\put(0.57,0.33){\makebox(0,0){(c)}}
\end{picture}
\caption{(a) PDF estimates resulting from (blue thin solid) adaptive KDE and (red thick solid) size-split (1) constant and (2) linear DETs based on a particle ensembles with $n = 10^4$ samples are compared with (black dash dot) the reference PDF~\eq{eq1dC1PDF}. The corresponding MISE evolution and mean tree depth as a function of the number of samples~$n$ are shown in panels (b) and (c) like in \figurename{}~\ref{fig1dC2MISE}.}\label{fig1dC1MISEPDF}
\end{figure*}
The third test distribution is the asymmetric claw distribution no.~12 \citep{Marron:1992a}, i.e.,
\begin{equation}\label{eq1dC1PDF}
p(x) = {\textstyle\frac{1}{2}}N(0,1) + \sum_{l = -2}^2 \frac{2^{1-l}}{31} N(l+{\textstyle\frac{1}{2}},2^{-l}/10).
\end{equation}
Estimates of this PDF based on the size-based DET estimators and adaptive KDE are included in \figurename{}~\ref{fig1dC1MISEPDF}(a) and illustrate the adaptive bin width selection of the DET method based on the local sample density. The MISE decay given in panel (b) is slightly closer to the adaptive KDE compared with the previous cases. Again a logarithmic growth of the tree depth as a function of the ensemble size~$n$ is recovered with similar depths like in the previous examples as seen in \figurename{}~\ref{fig1dC2MISE}(c).

\subsubsection{Spiky Uniforms PDF}

\begin{figure*}
\unitlength\textwidth
\begin{picture}(1,0.62)
\put(0.0,0.31){\includegraphics[width=0.5\textwidth]{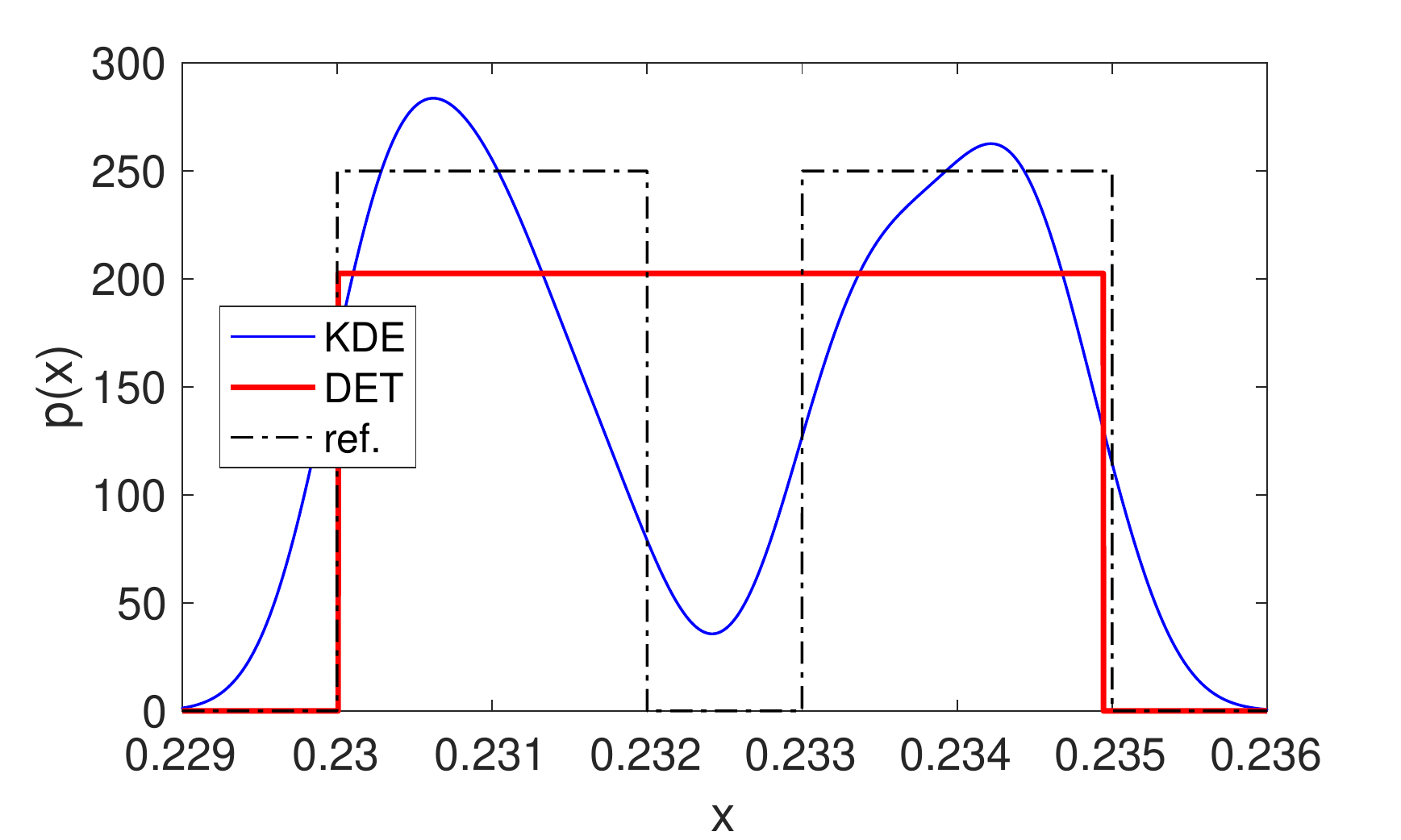}}
\put(0.01,0.61){\makebox(0,0){(a1)}}
\put(0.5,0.31){\includegraphics[width=0.5\textwidth]{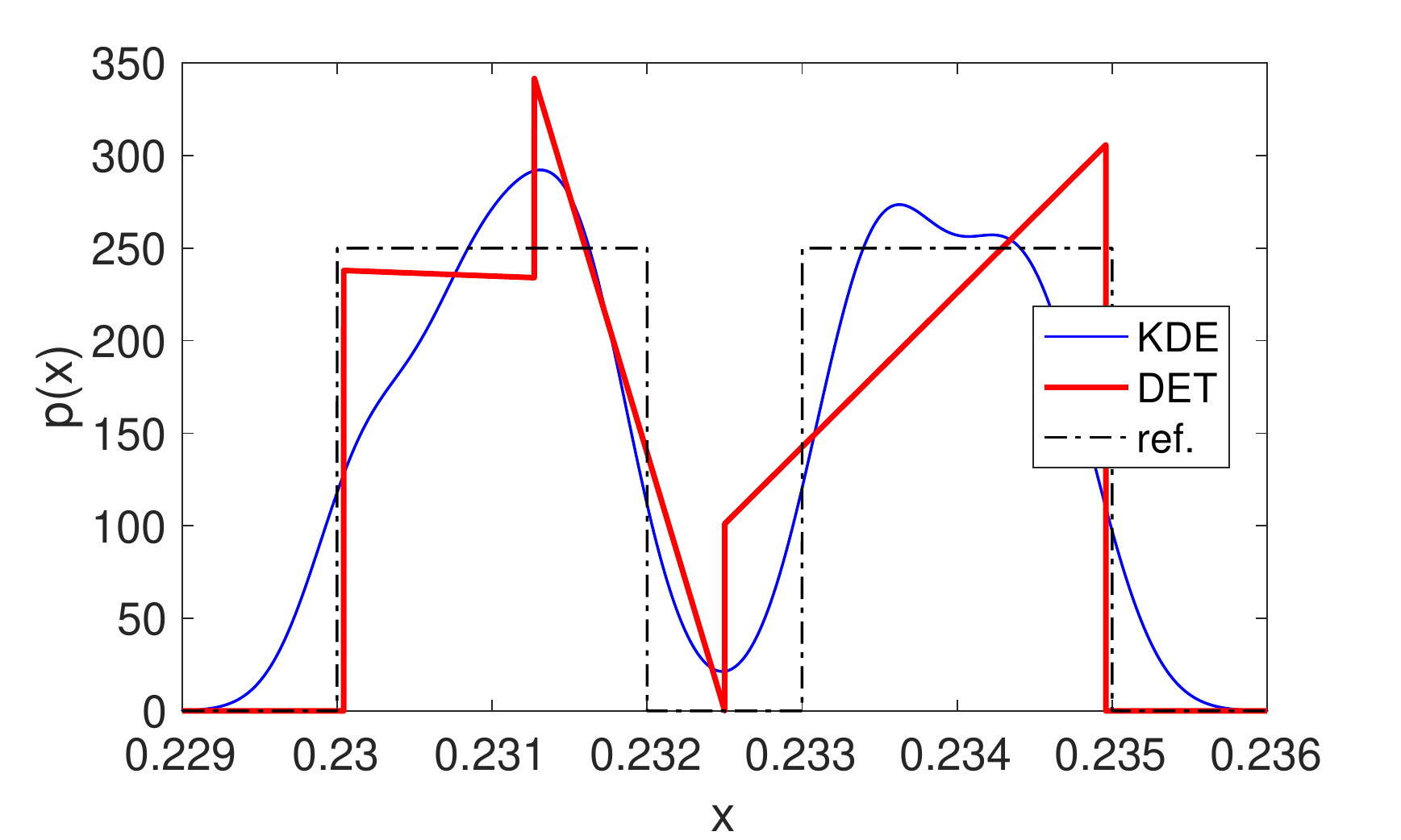}}
\put(0.51,0.61){\makebox(0,0){(b1)}}
\put(0.0,0.0){\includegraphics[width=0.5\textwidth]{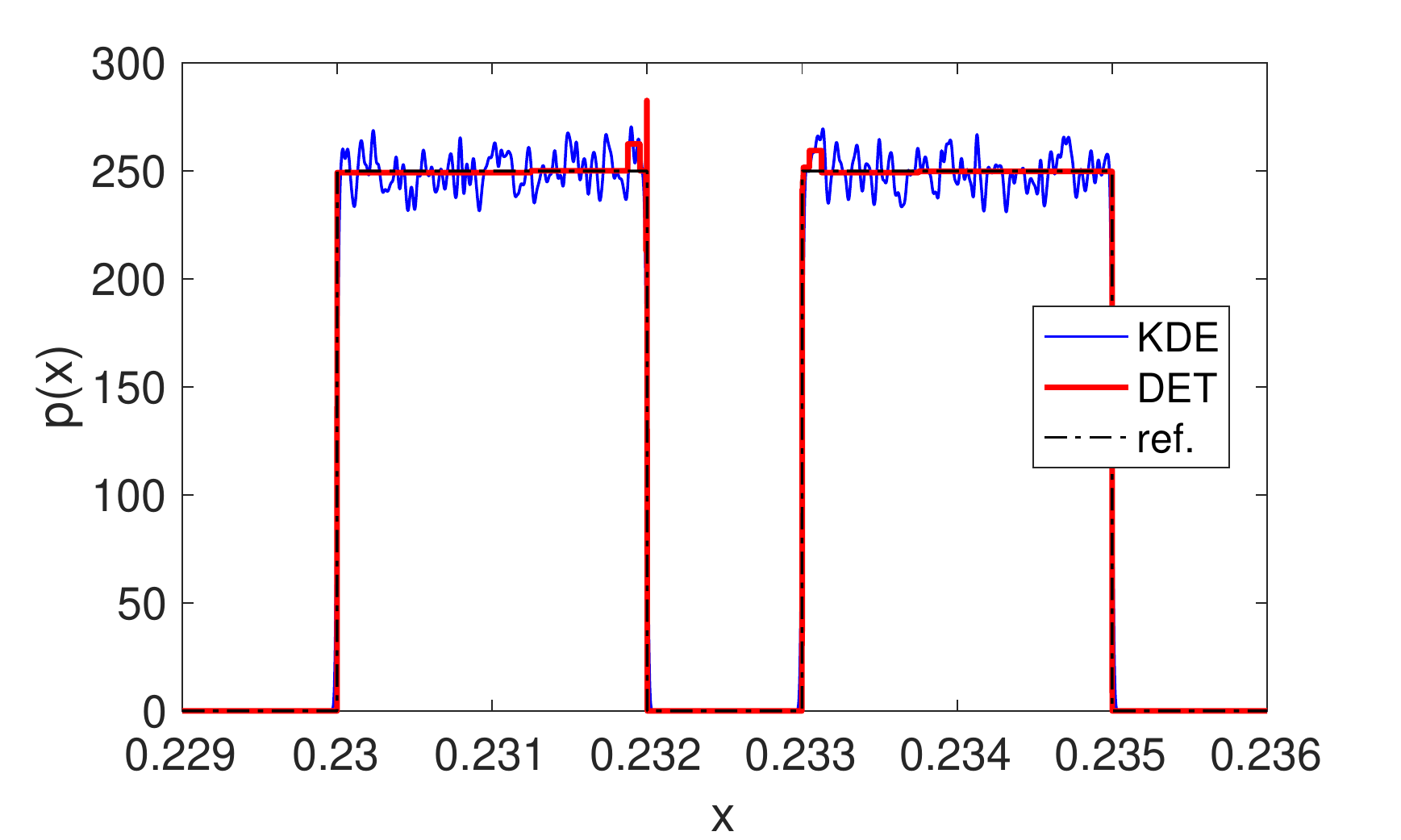}}
\put(0.01,0.3){\makebox(0,0){(a2)}}
\put(0.5,0.0){\includegraphics[width=0.5\textwidth]{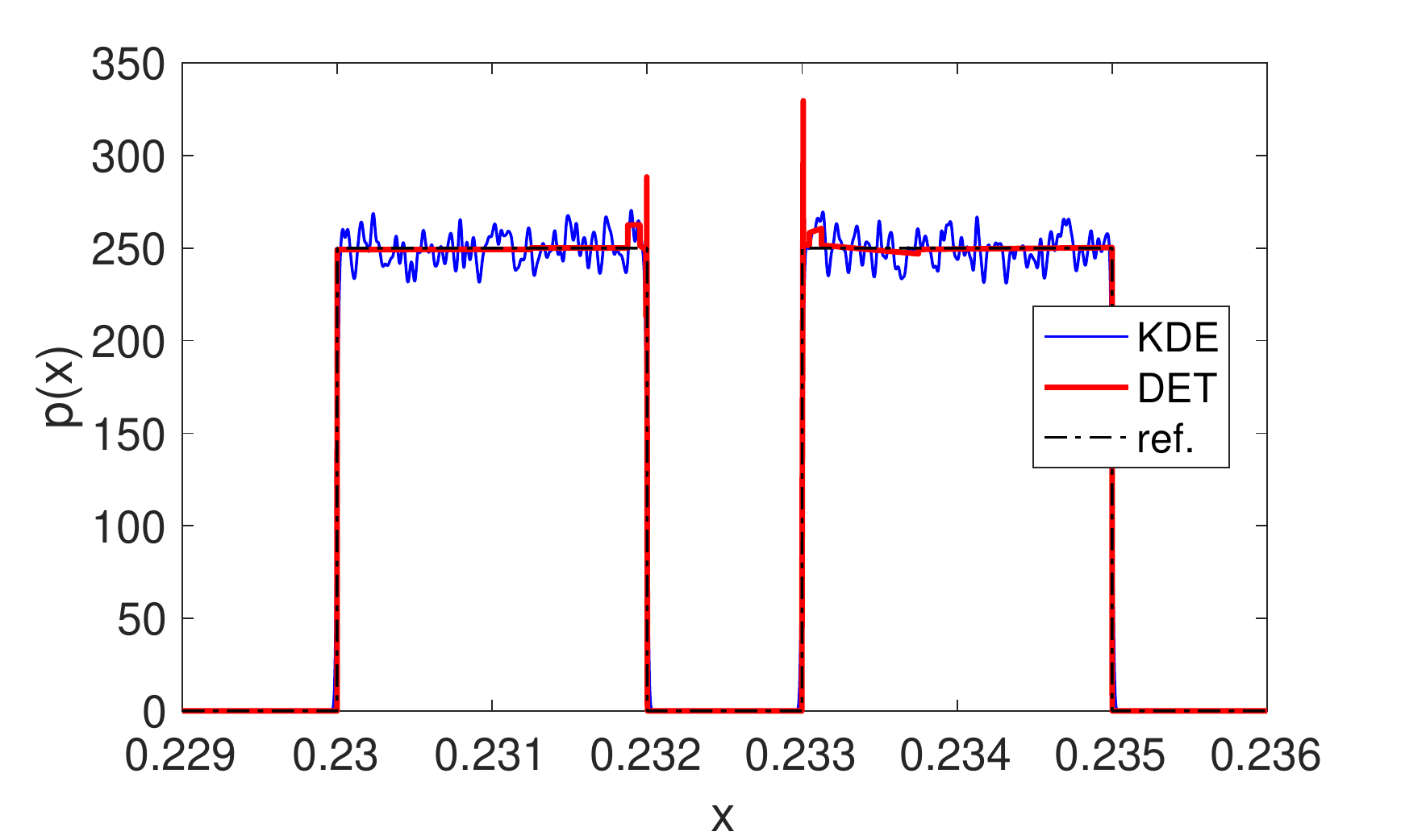}}
\put(0.51,0.3){\makebox(0,0){(b2)}}
\end{picture}
\caption{PDF estimates resulting from (blue thin solid) adaptive KDE and (red thick solid) the size-split DET method with particle ensembles including (1) $n = 100$ and (2) $10^5$ samples are compared with (black dash dot) the reference PDF~\eq{eq1dC4PDF}. In panels (a) and (b), DET estimates with constant and linear elements are depicted, respectively.}\label{fig1dC4PDF}
\end{figure*}
\begin{figure*}
\unitlength\textwidth
\begin{picture}(1,0.375)
\put(0.0,0.0){\includegraphics[width=0.5\textwidth]{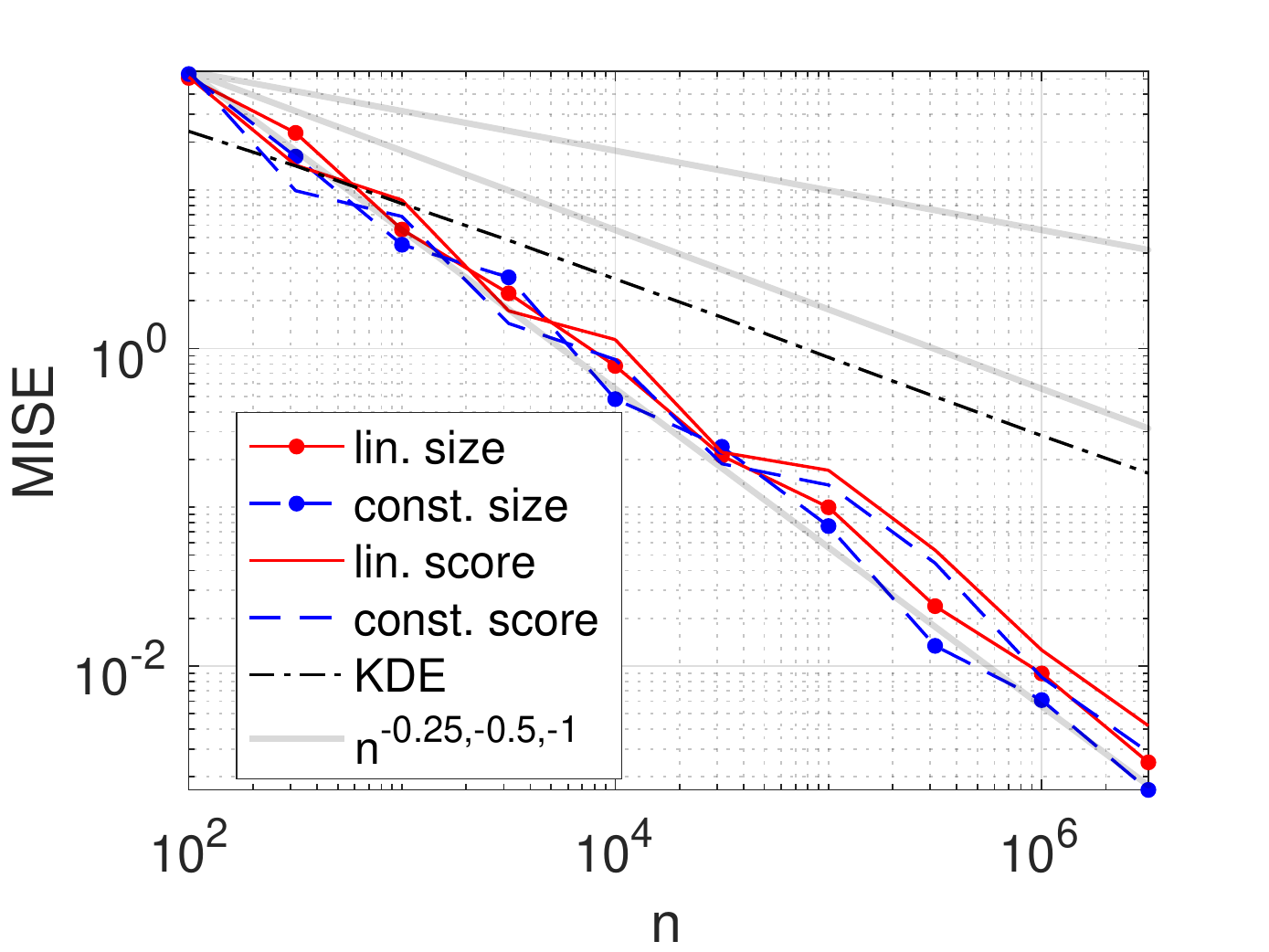}}
\put(0.01,0.365){\makebox(0,0){(a)}}
\put(0.5,0.0){\includegraphics[width=0.5\textwidth]{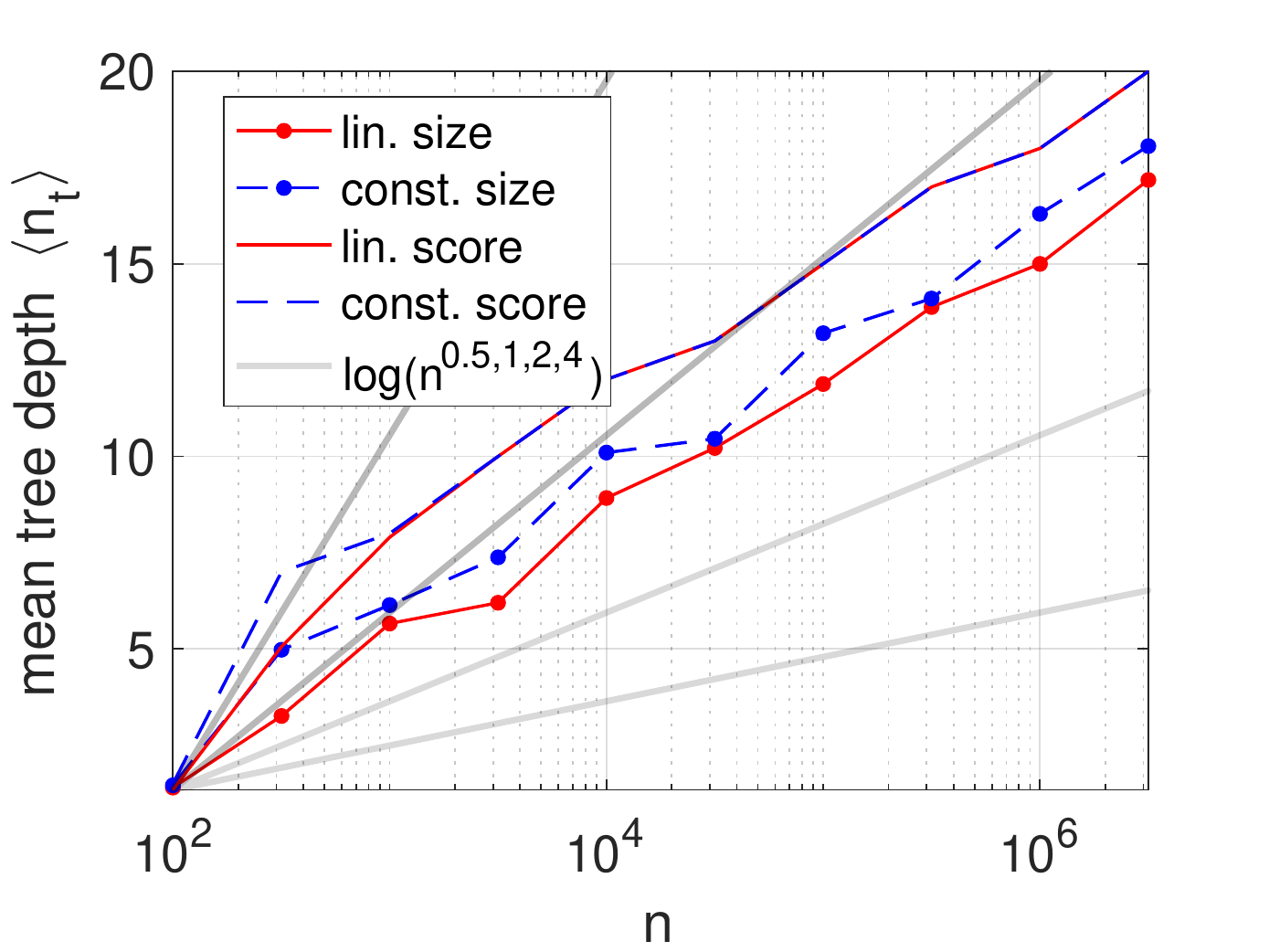}}
\put(0.51,0.365){\makebox(0,0){(b)}}
\end{picture}
\caption{Estimation of spiky uniforms PDF~\eq{eq1dC4PDF}. (a) Evolutions of the MISE as a function of the number of samples~$n$ for DET estimators (red solid and blue dashed) with equal size (symbols) and score splits (no symbols) are plotted. The MISE resulting from the adaptive KDE is included as well (black dash dot). For the DET estimators the resulting mean tree depths~$\langle n_t\rangle$ are given in panel (b). Power law and logarithmic scalings with exponents indicated in the figure legends are depicted (gray thick solid).}\label{fig1dC4MISE}
\end{figure*}
The previous test cases have focused on Gaussian mixtures. The next three examples deal with different PDFs like the spiky uniforms distribution taken from \citep[example~6]{Wong:2010a} and given by
\begin{equation}\label{eq1dC4PDF}
p(x) = {\textstyle\frac{1}{2}}U(0.23,0.232) + {\textstyle\frac{1}{2}U(0.233,0.235)},
\end{equation}
where $U(a,b)$ is a uniform distribution defined on the interval $[a,b]$. DET-based density estimates and adaptive KDE are compared in \figurename{}~\ref{fig1dC4PDF}. We can observe that while the DET methods tend to produce oscillations at the bounds of the uniforms, the adaptive KDE is subject to noise apparent in the constant sections of the uniforms. This difficulty is detected as well by the MISE plotted in \figurename{}~\ref{fig1dC4MISE}(a), where the DET methods converge faster with approximately $1/n$ to the true density compared to adaptive KDE with $1/\sqrt{n}$. The DET methods become more accurate than KDE for ensembles with $n > 10^3$. The mean tree depth of the DET method shown in \figurename{}~\ref{fig1dC4MISE}(b) and the mean number of DEs (not shown) display logarithmic and sublinear growth as observed previously. As an exception to all one-dimensional cases considered, for the spiky uniform distribution, there is no advantage in terms of MISE convergence, number of DEs, and tree depth when using linear vs.\ constant DEs. Given the piecewise constant form of the PDF, this is expected.

\subsubsection{Beta PDF}

\begin{figure*}
\unitlength\textwidth
\begin{picture}(1,0.62)
\put(0.0,0.31){\includegraphics[width=0.5\textwidth]{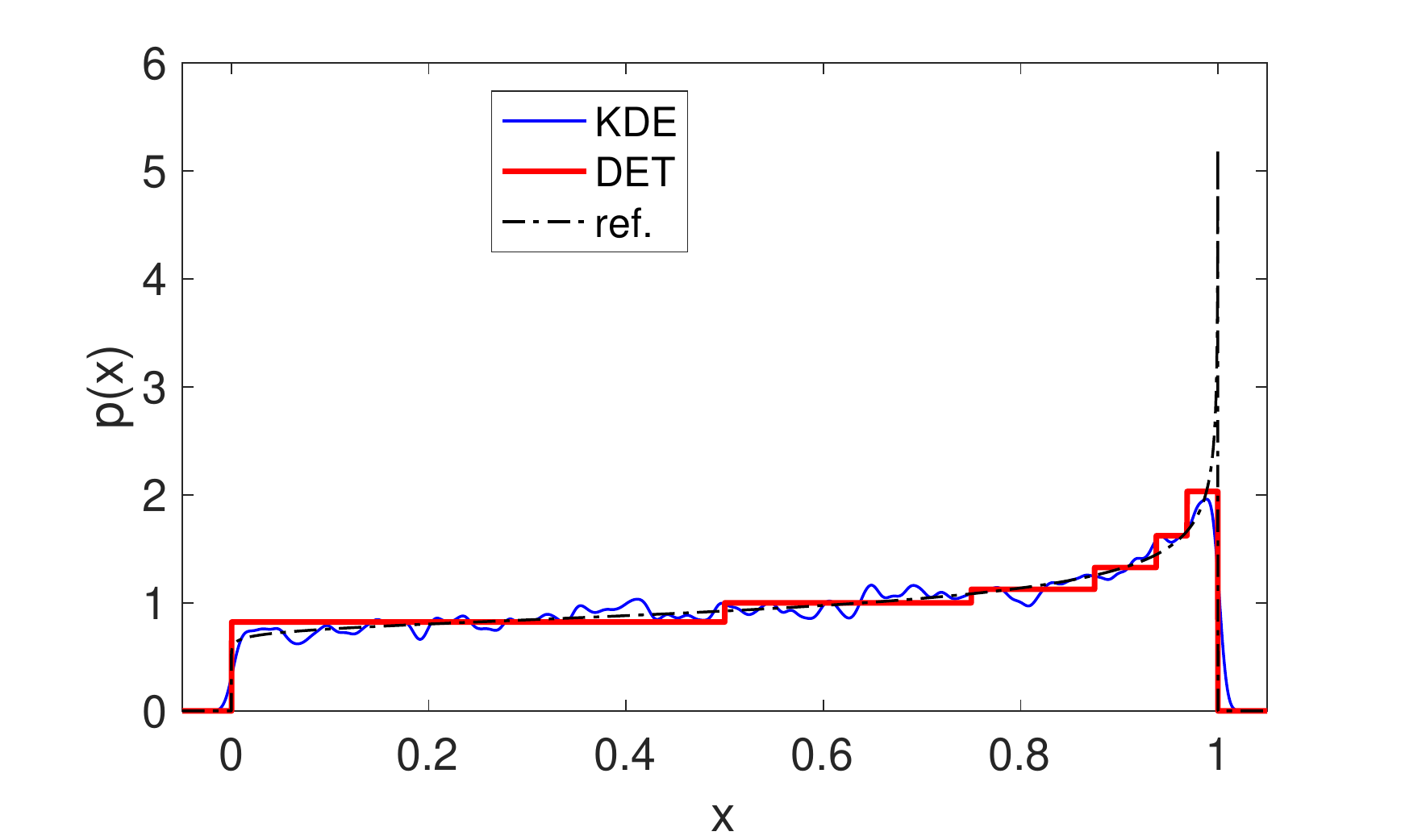}}
\put(0.01,0.61){\makebox(0,0){(a1)}}
\put(0.5,0.31){\includegraphics[width=0.5\textwidth]{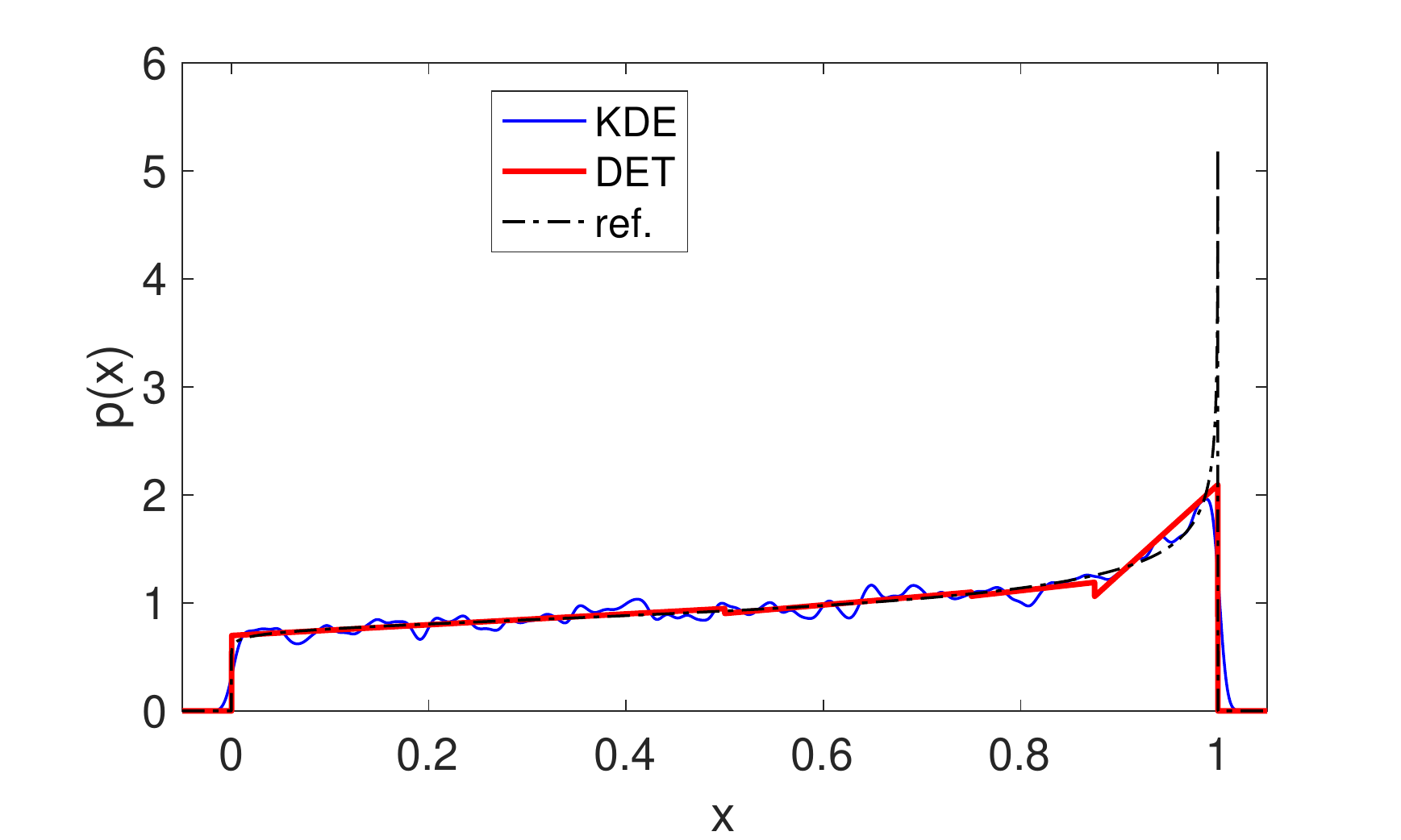}}
\put(0.51,0.61){\makebox(0,0){(b1)}}
\put(0.0,0.0){\includegraphics[width=0.5\textwidth]{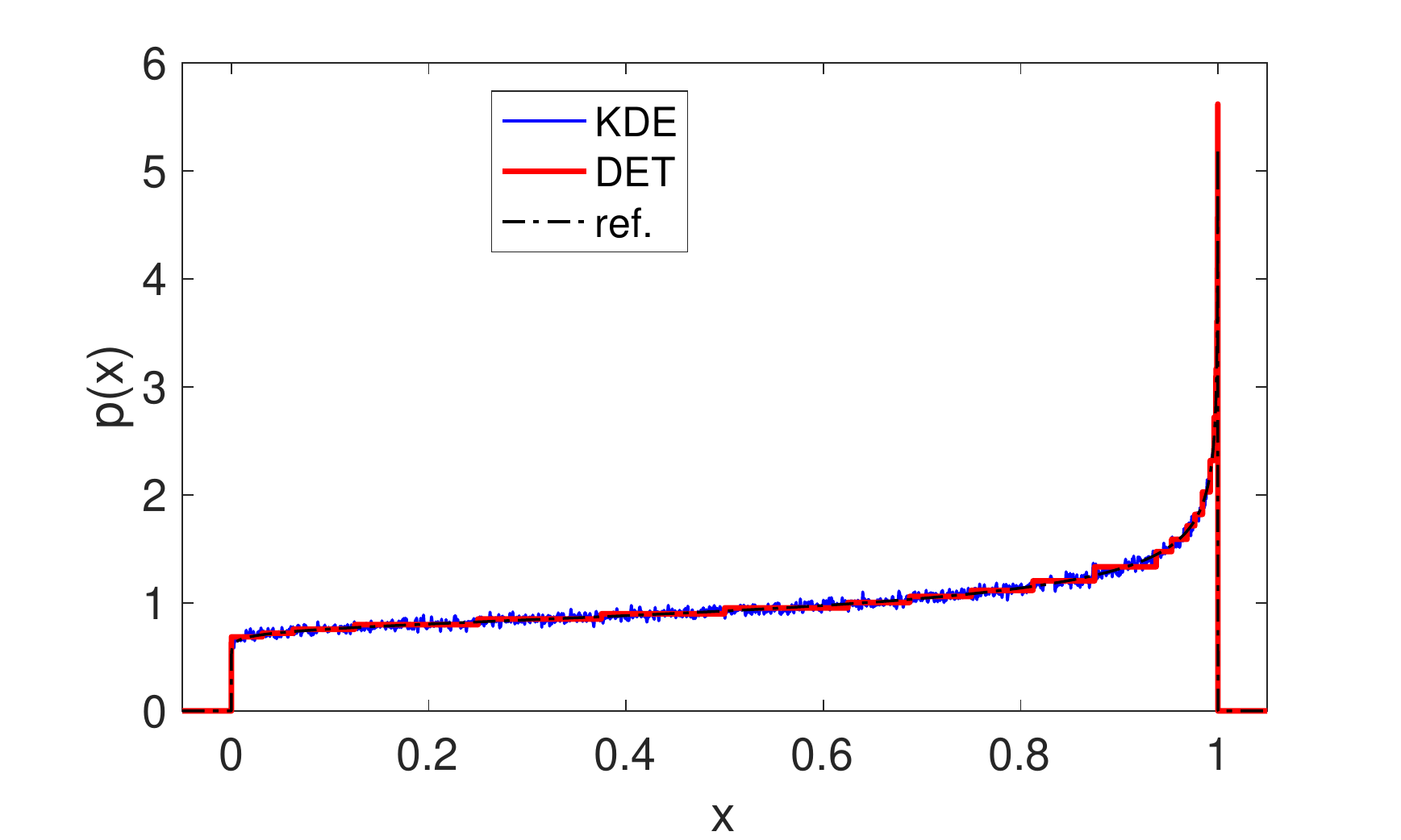}}
\put(0.01,0.3){\makebox(0,0){(a2)}}
\put(0.5,0.0){\includegraphics[width=0.5\textwidth]{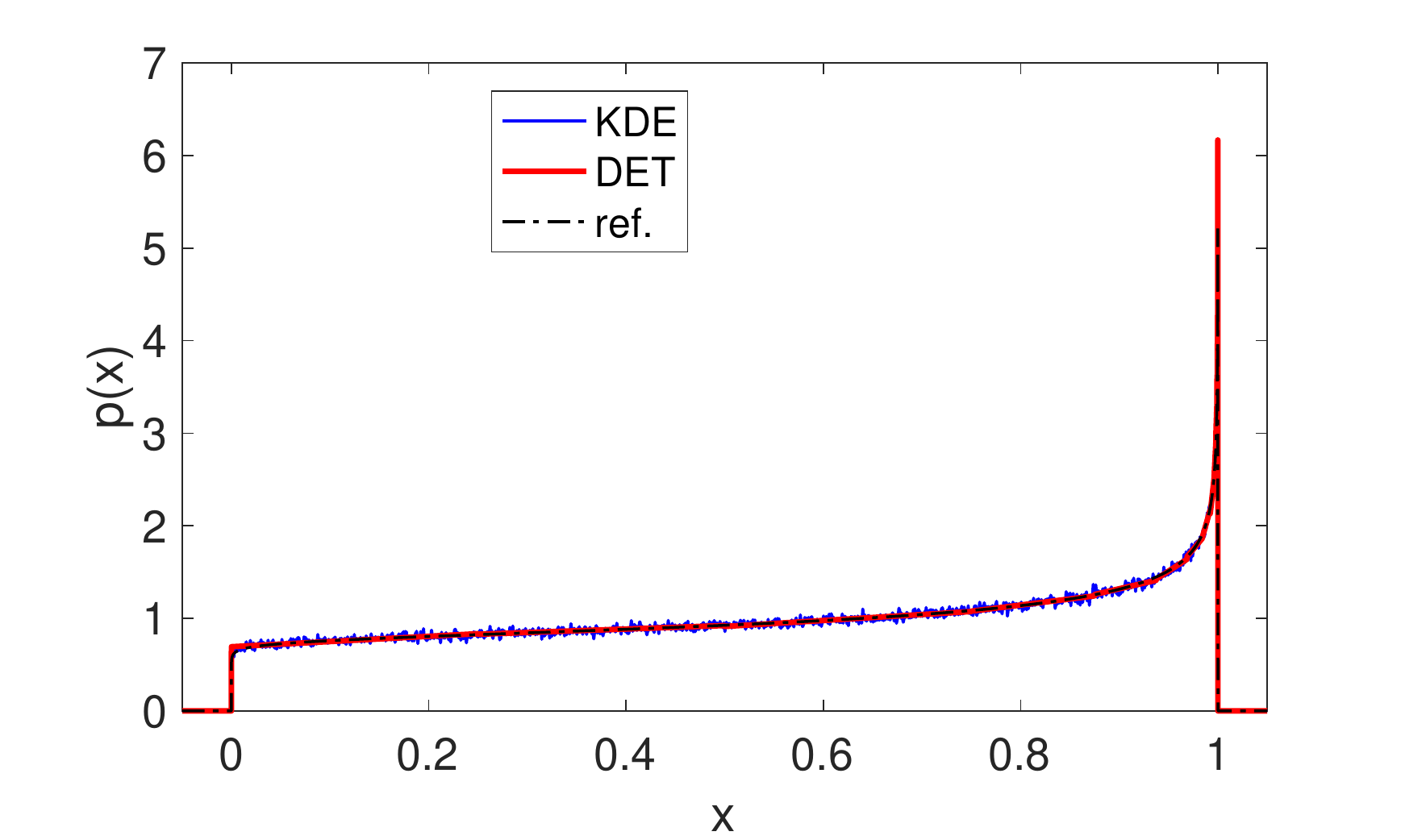}}
\put(0.51,0.3){\makebox(0,0){(b2)}}
\end{picture}
\caption{PDF estimates resulting from (blue thin solid) adaptive KDE and (red thick solid) the size-split DET method with particle ensembles including (1) $n = 10^4$ and (2) $10^6$ samples are compared with (black dash dot) the gamma PDF~\eq{eq1dC5PDF}. In panels (a) and (b), DET estimates with constant and linear elements are depicted, respectively.}\label{fig1dC5PDF}
\end{figure*}
\begin{figure*}
\unitlength\textwidth
\begin{picture}(1,0.375)
\put(0.0,0.0){\includegraphics[width=0.5\textwidth]{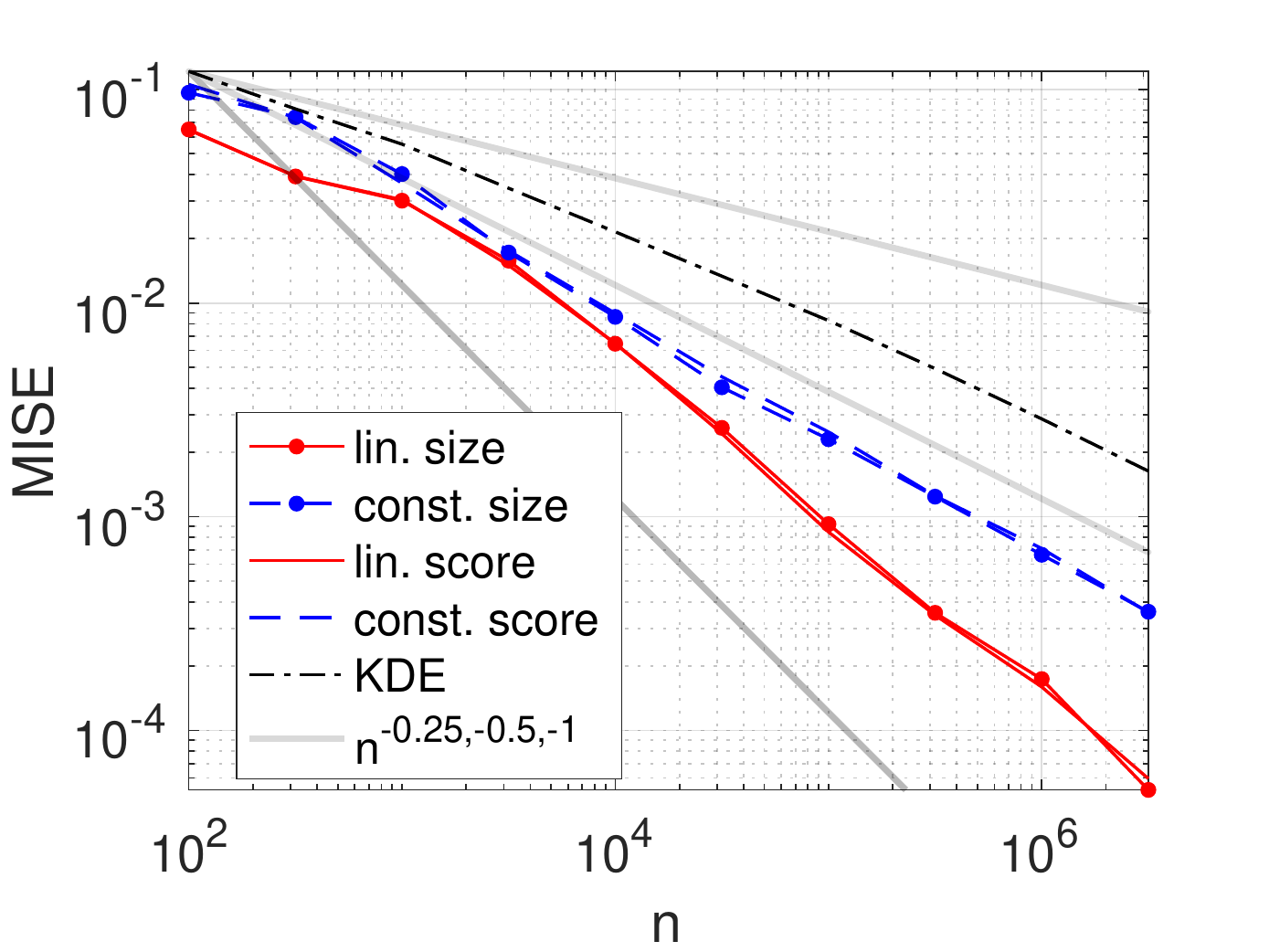}}
\put(0.01,0.365){\makebox(0,0){(a)}}
\put(0.5,0.0){\includegraphics[width=0.5\textwidth]{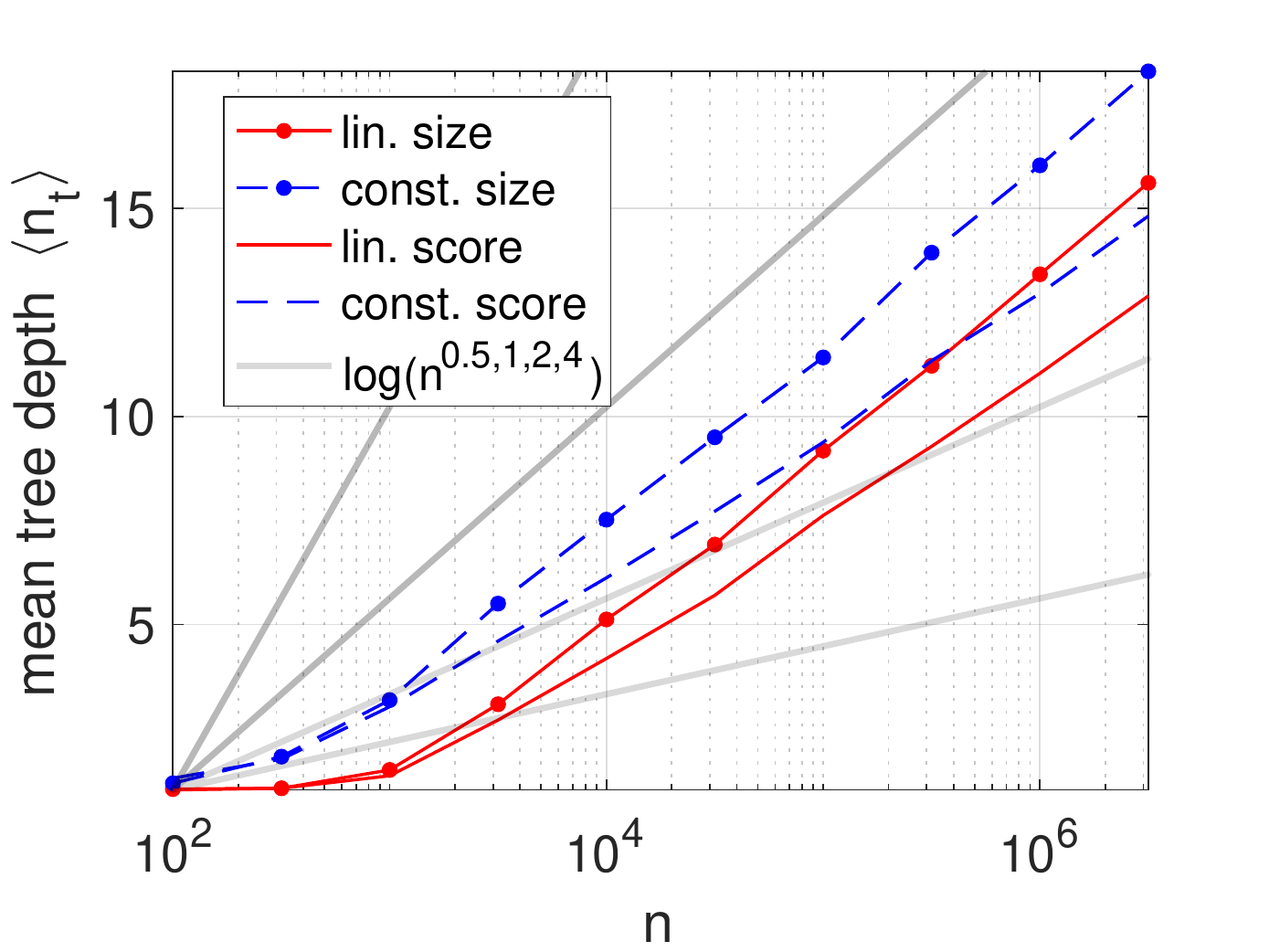}}
\put(0.51,0.365){\makebox(0,0){(b)}}
\end{picture}
\caption{Estimation of beta PDF~\eq{eq1dC5PDF}. See \figurename{}~\ref{fig1dC4MISE}.}\label{fig1dC5MISE}
\end{figure*}
In our fifth example, we perform density estimation based on data derived from the beta PDF
\begin{equation}\label{eq1dC5PDF}
p(x) = \frac{\Gamma(\alpha+\beta)}{\Gamma(\alpha)\Gamma(\beta)} x^{\alpha-1} (1-x)^{\beta-1}
\end{equation}
with $x\in[0,1]$, parameters $\alpha = 1.05$ and $\beta = 0.8$, and where $\Gamma(x)$ is the gamma function. Exemplary resulting estimates are depicted in \figurename{}~\ref{fig1dC5PDF}. While in the flat region of PDF~\eq{eq1dC5PDF} the constant DET estimator allocates few large elements, increasingly fine bins are placed towards the peak at $x = 1$. The linear DET estimator behaves similarly but requires fewer elements. In comparison with adaptive KDE, one can observe in \figurename{}~\ref{fig1dC5MISE}(a) that the DET methods converge faster and that the linear DET variants are more accurate. Unlike in the previous examples, it becomes apparent from \figurename{}~\ref{fig1dC5MISE}(b) that the score-split DET variants lead to smaller trees than their size-split counterparts.

\subsubsection{Gamma PDF}\label{subsubsec1dC6}

\begin{figure*}
\unitlength\textwidth
\begin{picture}(1,0.34)
\put(0.0,0.0){\includegraphics[width=0.56\textwidth]{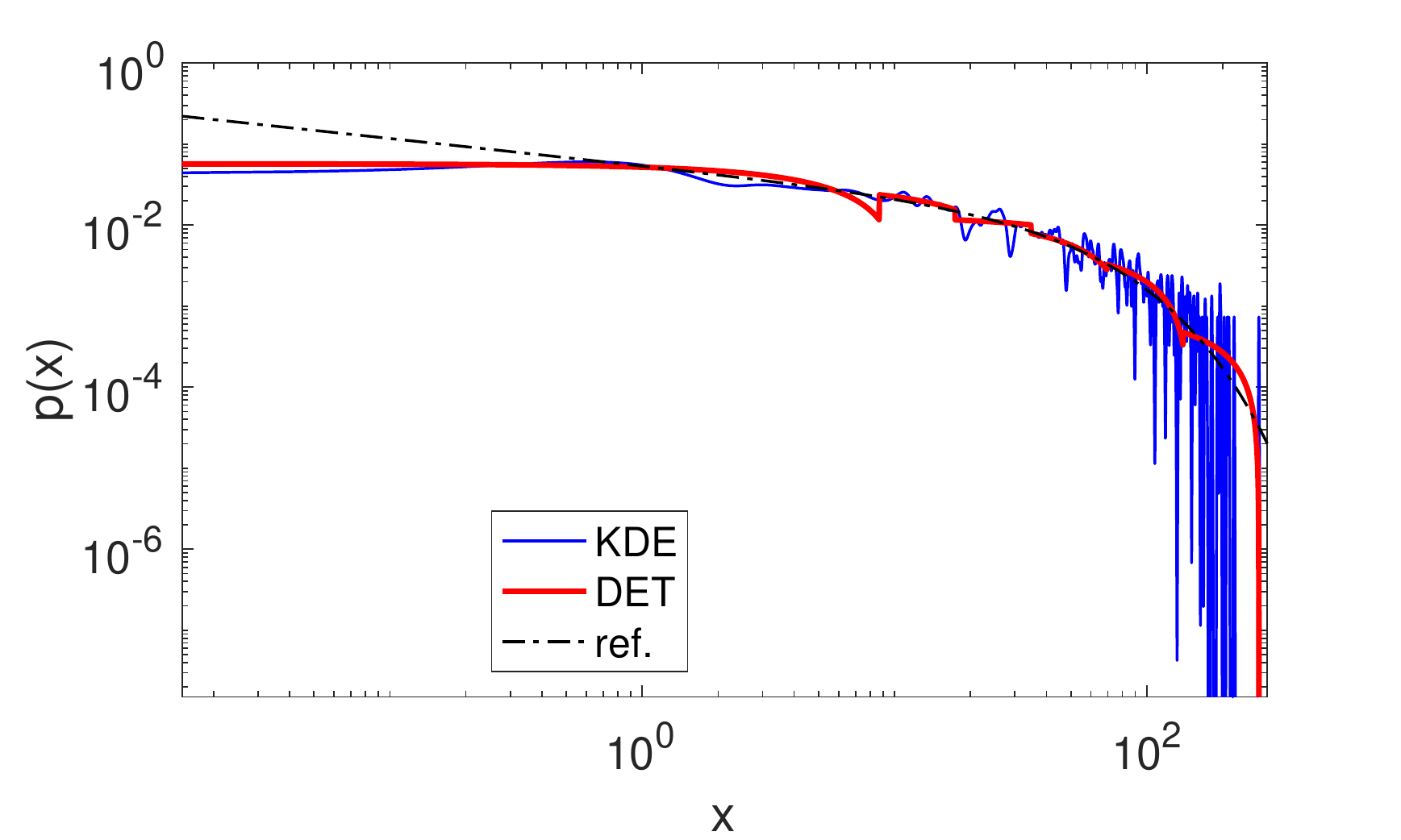}}
\put(0.01,0.33){\makebox(0,0){(a)}}
\put(0.56,0.0){\includegraphics[width=0.44\textwidth]{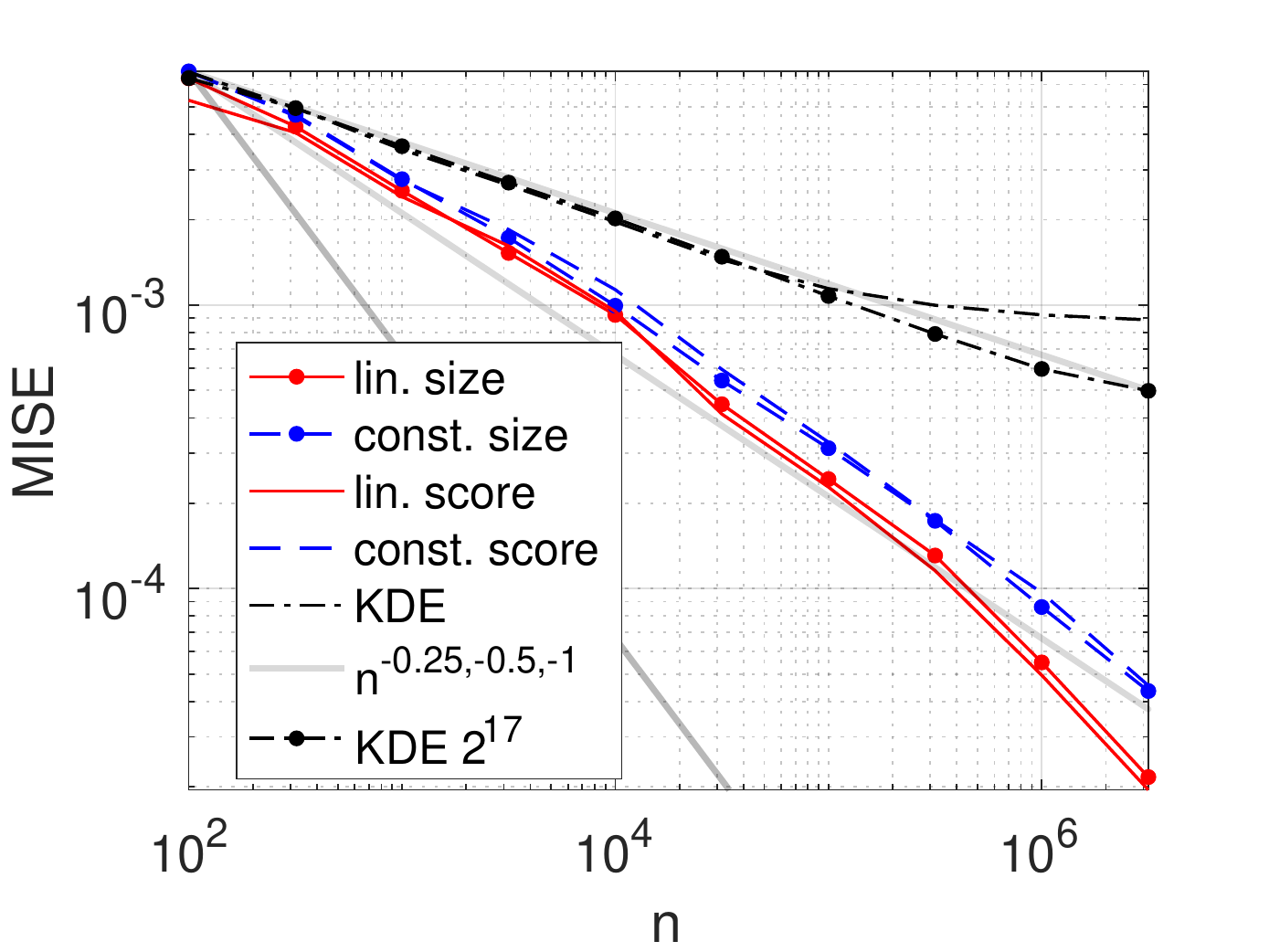}}
\put(0.57,0.33){\makebox(0,0){(b)}}
\end{picture}
\caption{(a) PDF estimates resulting from (blue thin solid) adaptive KDE and (red thick solid) the size-split linear DET method based on a particle ensembles with $n = 10^3$ samples are compared with (black dash dot) the reference PDF~\eq{eq1dC6PDF}. The corresponding MISE evolution as a function of the number of samples~$n$ is shown in panel (b) like in \figurename{}~\ref{fig1dC2MISE}(a). In addition to the adaptive KDE with $2^{14}$ grid points (black dash dot), MISE results based on $n^{17}$ points (black dash dot with symbol) are provided.}\label{fig1dC6MISEPDF}
\end{figure*}
The last one-dimensional example considered in this work is the gamma PDF
\begin{equation}\label{eq1dC6PDF}
p(x) = \frac{x^{a-1}e^{-x/b}}{b^a\Gamma(a)}
\end{equation}
with $x$ going from 0 to $\infty$ and parameters $a = 2/3$ and $b = 50$. As is visible from \figurename{}~\ref{fig1dC6MISEPDF}, this PDF is skewed and goes to infinity for $x \to 0$. The MISE results included in the figure document the accuracy of the DET estimators. The ability of the adaptive KDE to resolve the singularity of PDF~\eq{eq1dC6PDF} becomes for $n\approx 10^{4.5}$ limited by the number of grid points or cosine modes applied for its calculation. Increasing this number from $2^{14}$ to $2^{17} = 131072$ shifts the resolution limit to larger~$n$ as is shown in \figurename{}~\ref{fig1dC6MISEPDF}(b). For the gamma PDF as for the beta PDF, score-based splitting is slightly more effective than size-based splitting leading to smaller trees with fewer DEs.

\subsection{Two-Dimensional Examples}\label{subsec2d}

For the adaptive KDEs in the following two-dimensional cases, a grid with $2^{10}\times 2^{10} = 1048576$ nodes was deployed \citep{Botev:2007a}.

\subsubsection{Bi-Variate Gaussian PDF}

\begin{figure*}
\unitlength\textwidth
\begin{picture}(1,0.65)
\put(0.05,-0.07){\includegraphics[width=0.4\textwidth]{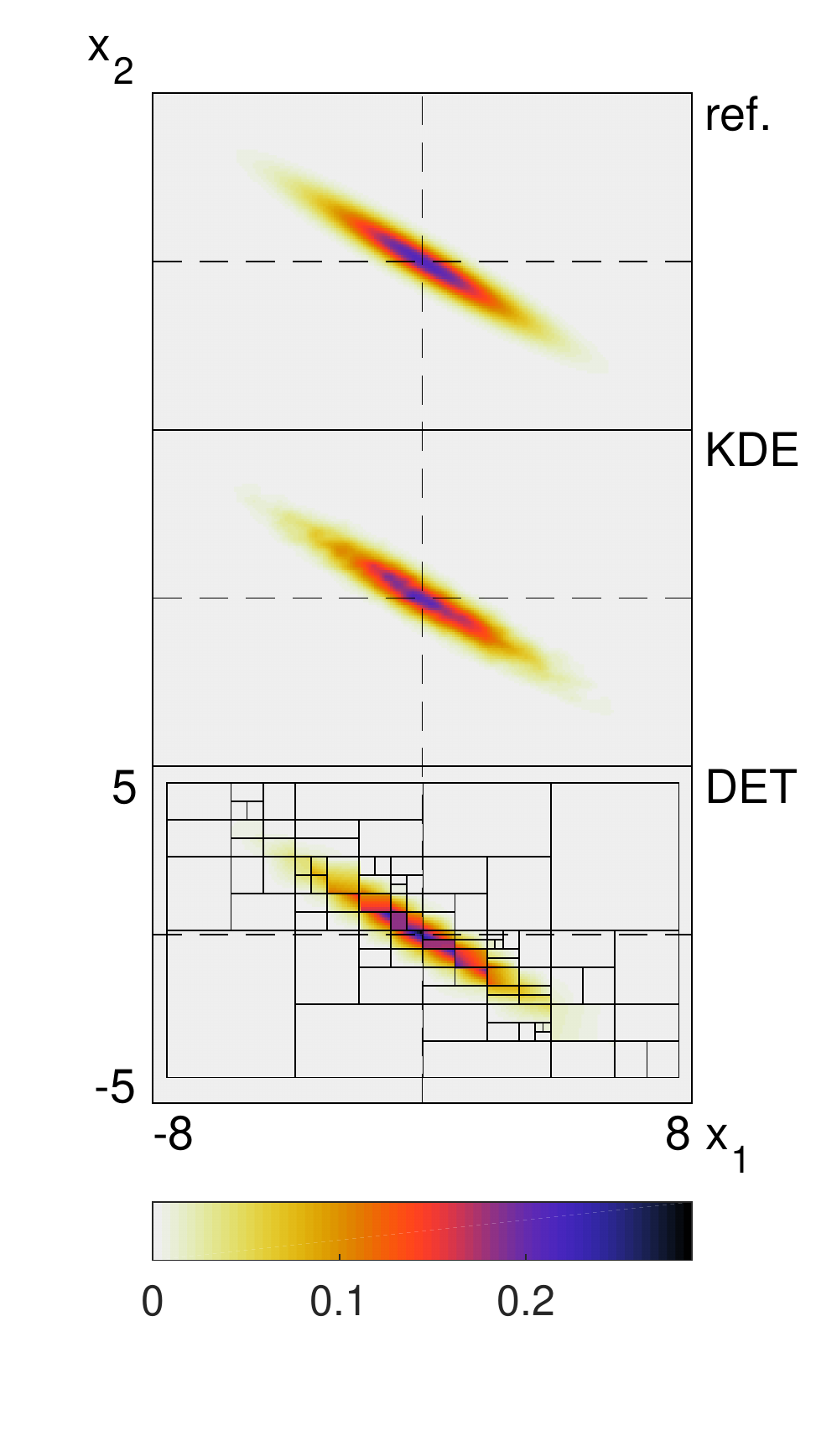}}
\put(0.01,0.64){\makebox(0,0){(a)}}
\put(0.55,-0.07){\includegraphics[width=0.4\textwidth]{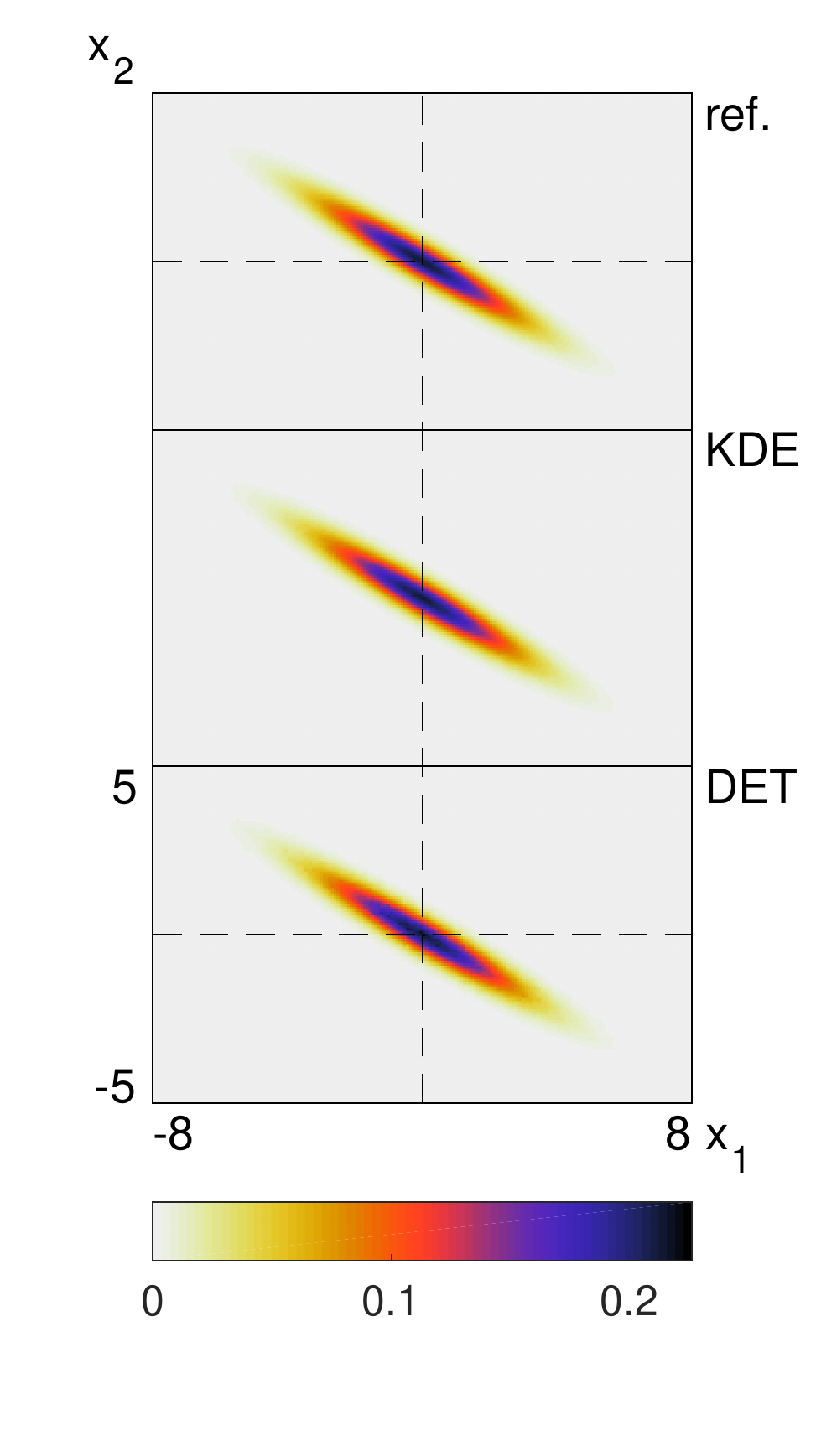}}
\put(0.51,0.64){\makebox(0,0){(b)}}
\end{picture}
\caption{PDF estimates resulting from adaptive KDE and the size-split linear DET method based on particle ensembles with (a) $n = 10^4$ and (b) $n \approx 10^{6.5}$ samples are compared with the reference PDF~\eq{eq2dC2PDF}}.\label{fig2dC2PDF}
\end{figure*}
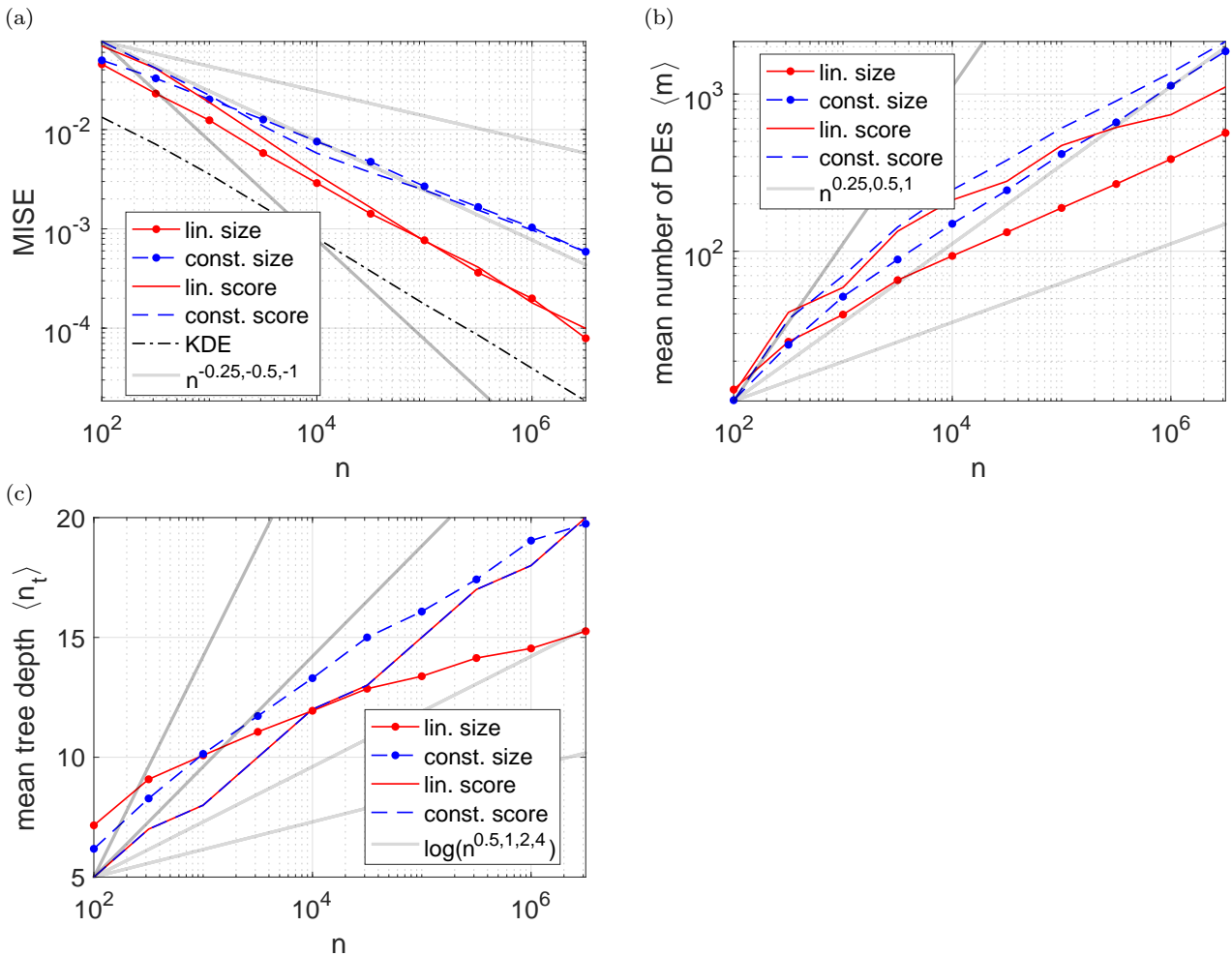
\begin{figure*}
\unitlength\textwidth
\begin{picture}(1,0.75)
\put(0.0,0.375){\includegraphics[width=0.5\textwidth]{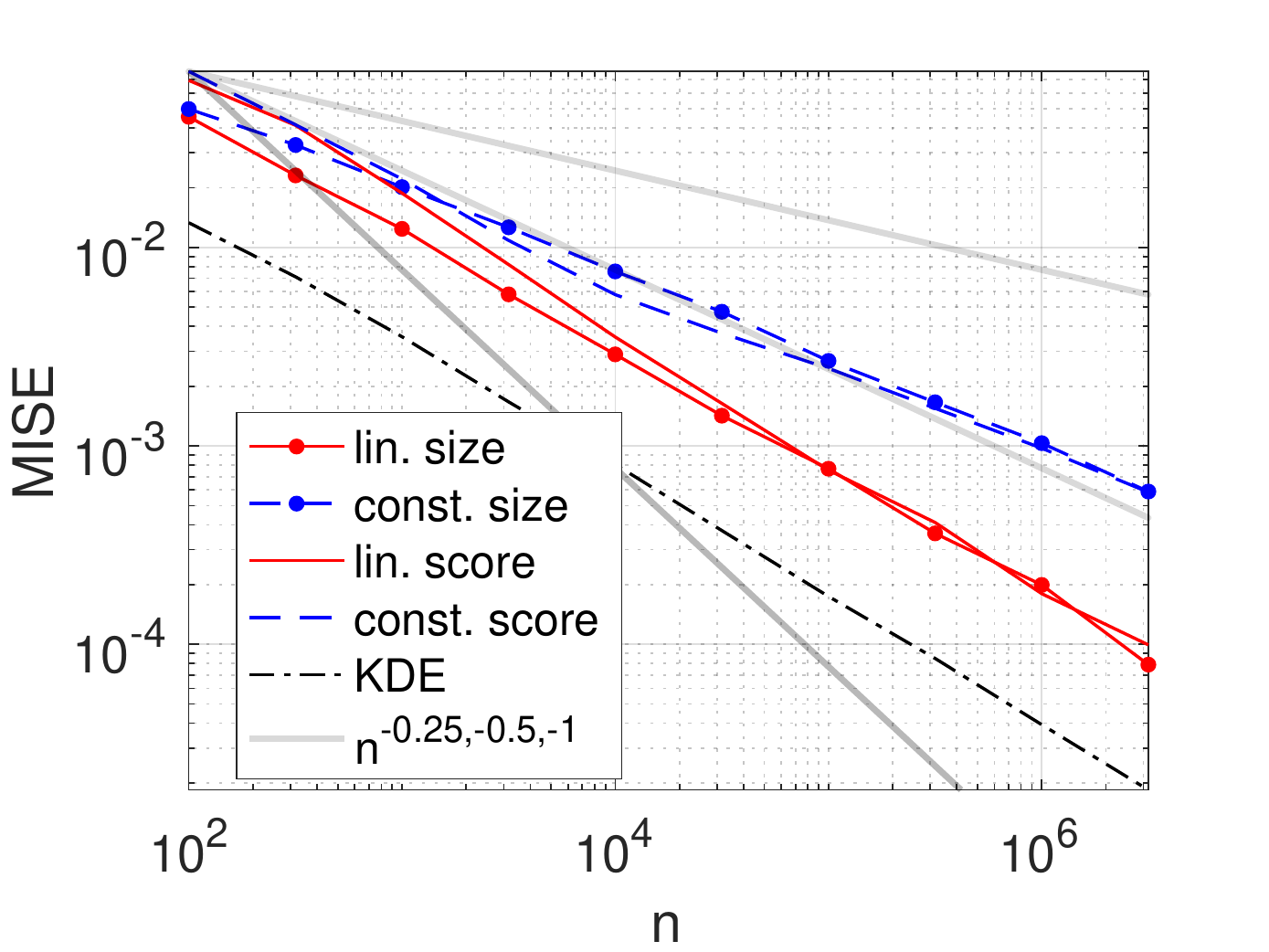}}
\put(0.01,0.74){\makebox(0,0){(a)}}
\put(0.5,0.375){\includegraphics[width=0.5\textwidth]{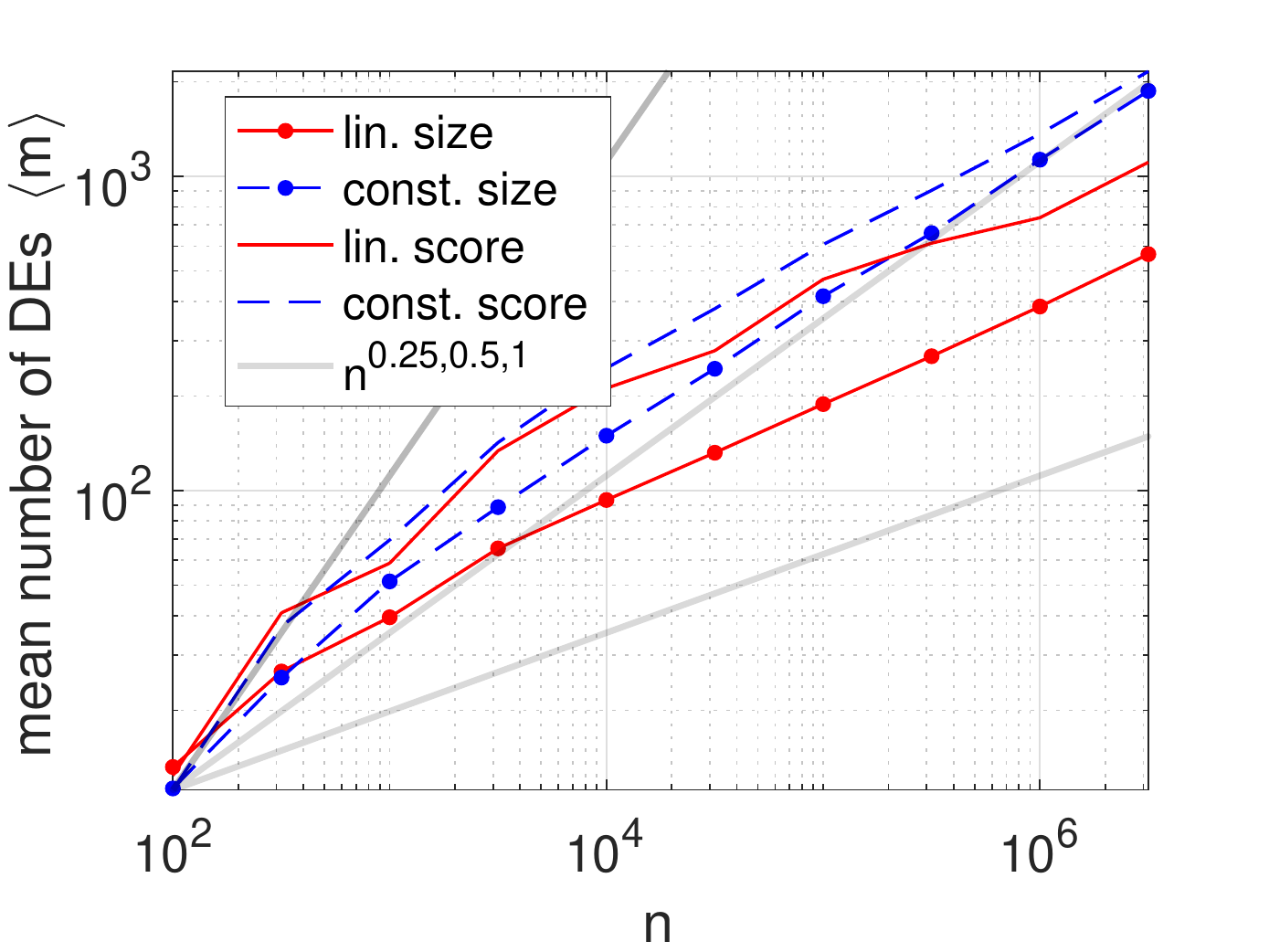}}
\put(0.51,0.74){\makebox(0,0){(b)}}
\put(0.0,0.0){\includegraphics[width=0.5\textwidth]{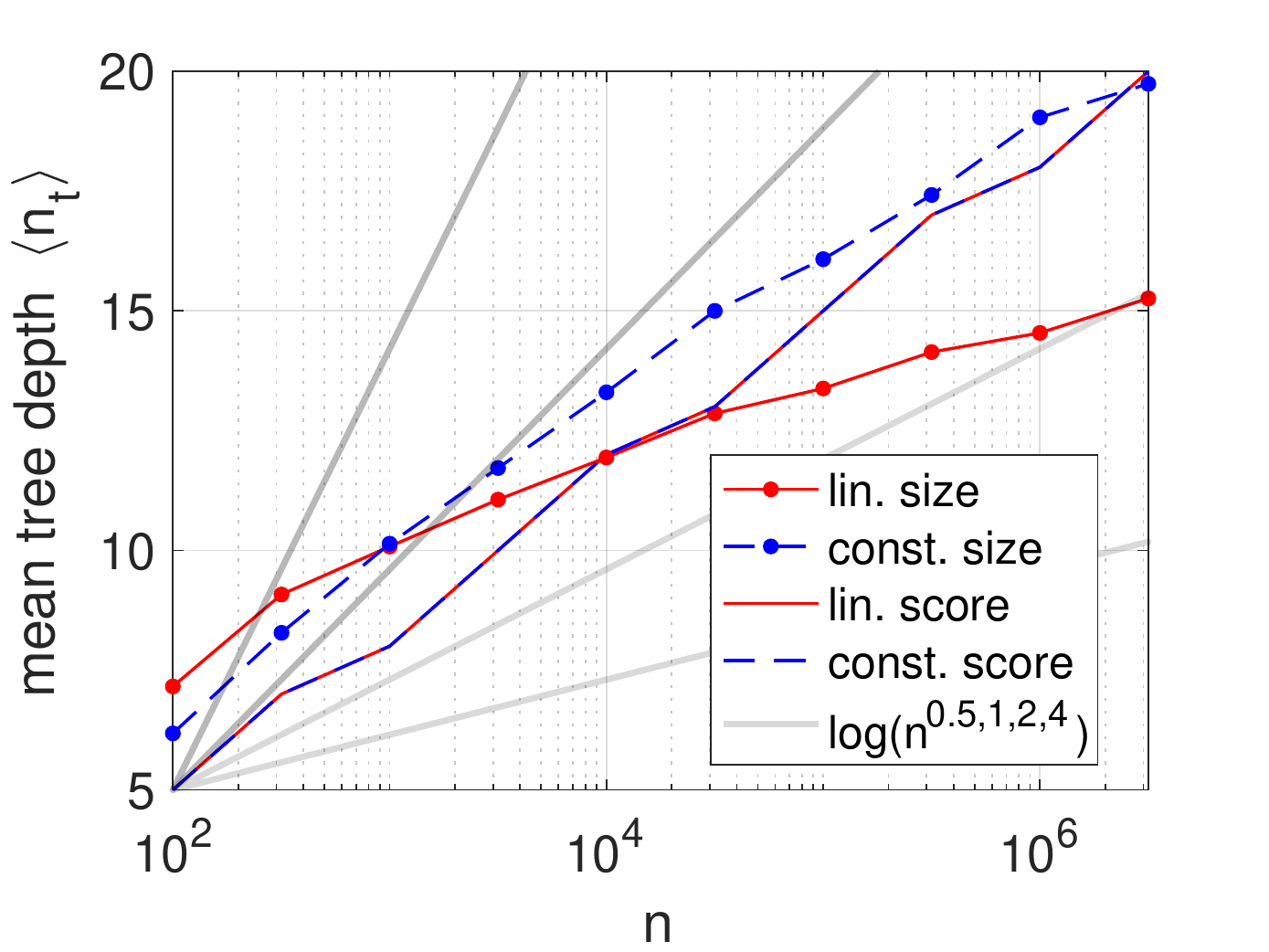}}
\put(0.01,0.365){\makebox(0,0){(c)}}
\end{picture}
\caption{Estimation of bi-variate Gaussian PDF~\eq{eq2dC2PDF}. See \figurename{}~\ref{fig1dC2MISE}. In panel (c), the data series from the score-split DETs coincide.}\label{fig2dC2MISE}
\end{figure*}
Before looking at more complex bi-variate distributions, we consider the joint Gaussian PDF
\begin{equation}\label{eq2dC2PDF}
p(\mathbf{x}) = \frac{\exp\left[-{\textstyle\frac{1}{2}}(\mathbf{x}-\mbf{\mu})^\top \mathbf{C}^{-1} (\mathbf{x}-\mbf{\mu})\right]}{\sqrt{(2\pi)^2 \det(\mathbf{C})}}
\end{equation}
with unbound probability space $\mathbf{x} = (x_1,x_2)^\top$, mean vector $\mbf{\mu} = (0,0)^\top$, and covariance matrix
\begin{displaymath}
\mathbf{C} = \left(\begin{array}{cc}4.0 & -2.28 \\ -2.28 & 1.44\end{array}\right).
\end{displaymath}
Gaussian PDFs are important in many applications, which emphasizes the present case. Resulting size-split linear DET estimates are compared with KDE and joint PDF~\eq{eq2dC2PDF} in \figurename{}~\ref{fig2dC2PDF}. The adaptivity of the DET method is illustrated in panel~(a), where regions of small density variation are represented by large DEs and sections with high variation were subdivided into several smaller elements.\footnote{In the singular case of linearly dependent components $x_1$ and $x_2$, many small DEs, resolving the probability peak along the diagonal of the $x_1$-$x_2$-space, result from a DET estimator.} The decay of the MISE depicted in \figurename{}~\ref{fig2dC2MISE}(a) is similar to the one-dimensional Gaussian mixture cases, but the convergence rates are slightly smaller (compare for example with \figurename{}~\ref{fig1dC2MISE}). Again, adaptive KDE is more accurate than the DET variants and the DE tree growth is logarithmic as seen in \figurename{}~\ref{fig2dC2MISE}(c). Moreover for large $n$, the mean number of DEs increases sublinearly with approximately $n^{1/4}$ (see \figurename{}~\ref{fig2dC2MISE}(b)).

\begin{figure*}
\unitlength\textwidth
\begin{picture}(1,0.375)
\put(0.04,0.025){\includegraphics[width=0.425\textwidth]{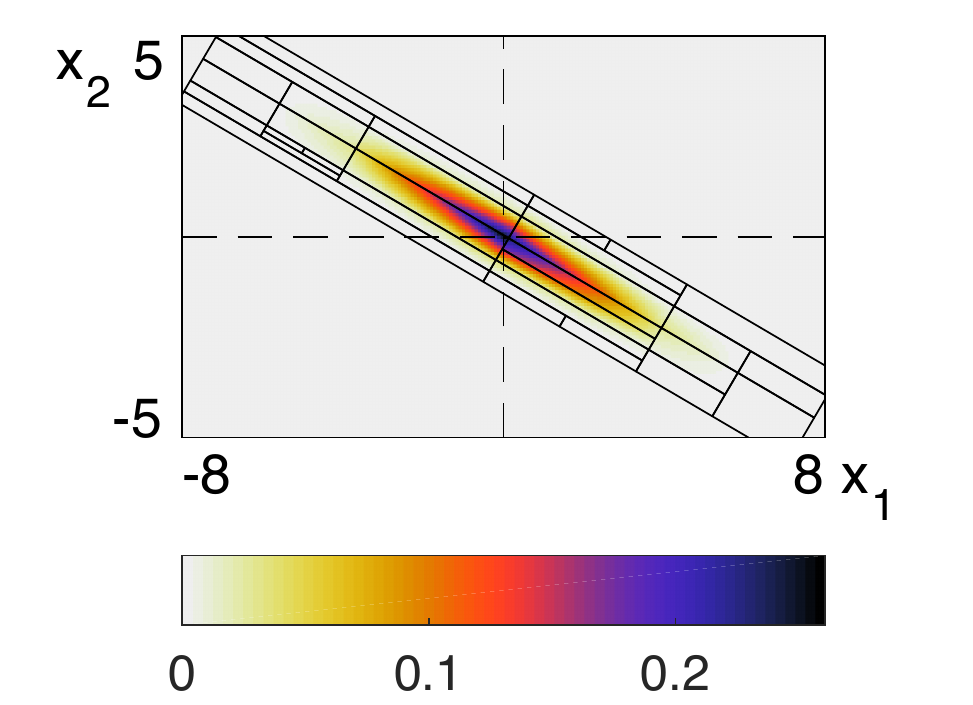}}
\put(0.01,0.365){\makebox(0,0){(a)}}
\put(0.5,0.0){\includegraphics[width=0.5\textwidth]{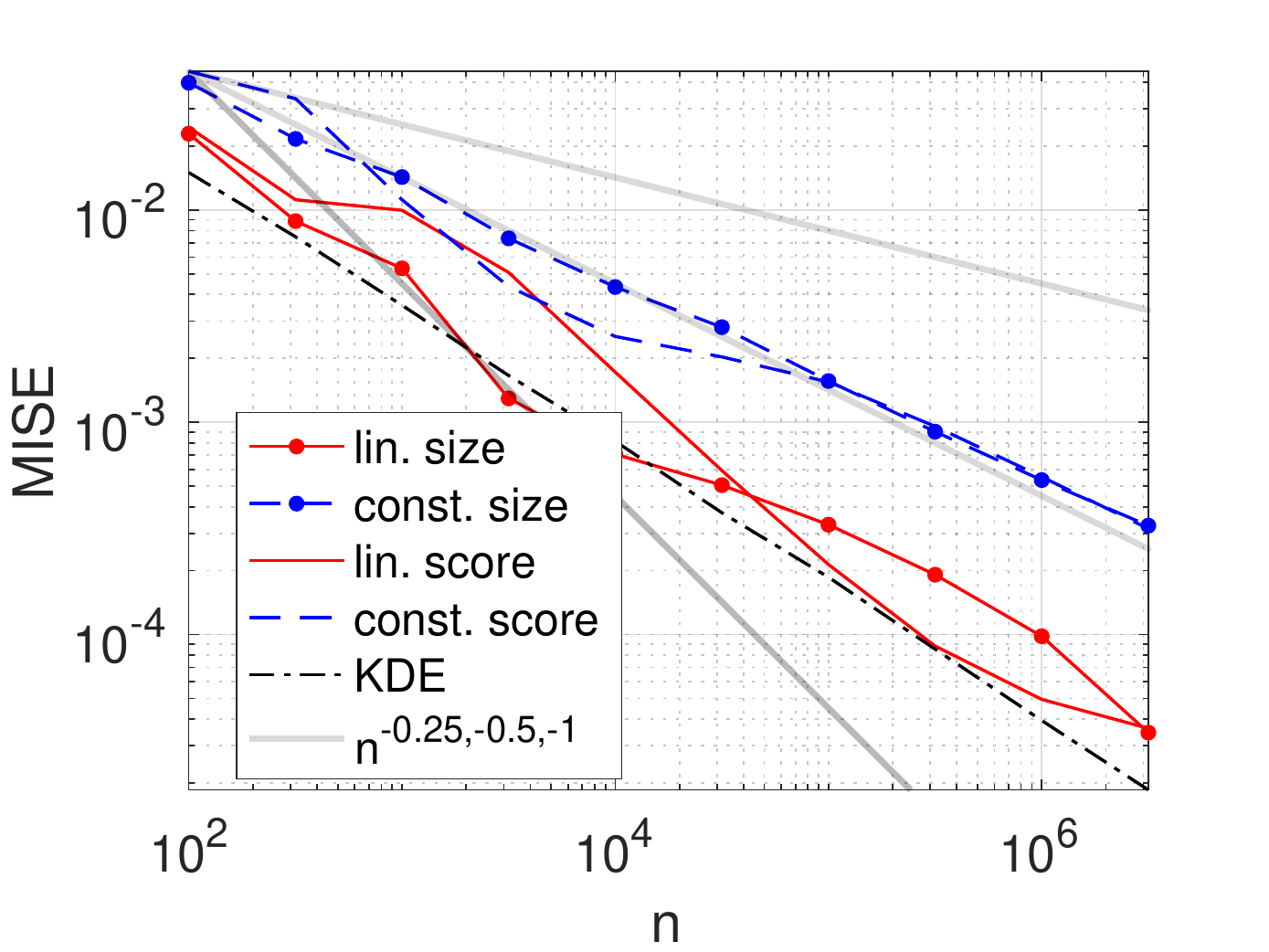}}
\put(0.51,0.365){\makebox(0,0){(b)}}
\end{picture}
\caption{Estimation of bi-variate Gaussian PDF~\eq{eq2dC2PDF}. (a) PDF estimate resulting from the linear size-split DET method combined with a principle axes transform and based on a particle ensemble with $n = 10^4$ samples. (b) Evolutions of the corresponding MISE for the indicated DET estimators and KDE as a function of the number of samples~$n$. All DET variants were combined with a principle axes transform.}\label{fig2dC2pat}
\end{figure*}
\figurename{}~\ref{fig2dC2PDF}(a) reveals a weakness of the present DET method in comparison to KDE, as the DET estimators or DET parametrizations will depend on the orientation of the coordinate system. In case of alignment of the joint Gaussian PDF~\eq{eq2dC2PDF} with the $x_1$-$x_2$-coordinate system, the DET estimators will require fewer DEs and will be more accurate. This implies the use of a covariance-matrix-based principle axes transform \citep[e.g.,][equation~(4.7)]{Silverman:1998a}, to achieve parametrization invariance and increased accuracy and computational efficiency. An illustration is given in \figurename{}~\ref{fig2dC2pat}, where results from DET estimators combined with principle axes transforms are provided. However, in order to demonstrate the versatility of the DET method, this approach is not pursued further in the present work.

\subsubsection{Uniform on an Ellipse}

\begin{figure*}
\unitlength\textwidth
\begin{picture}(1,1)
\put(0.0,0.5){\includegraphics[width=0.5\textwidth]{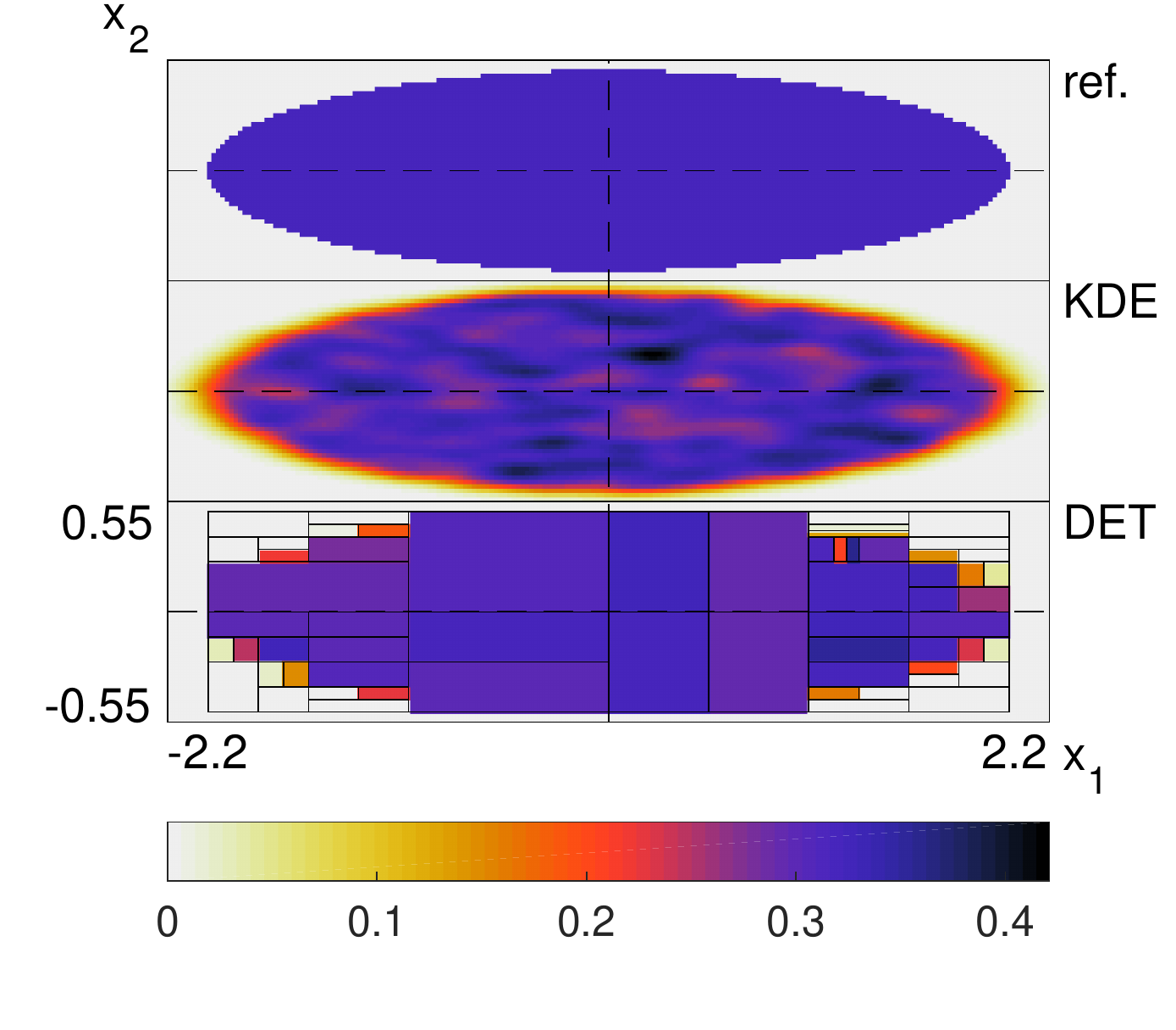}}
\put(0.01,0.99){\makebox(0,0){(a1)}}
\put(0.5,0.5){\includegraphics[width=0.5\textwidth]{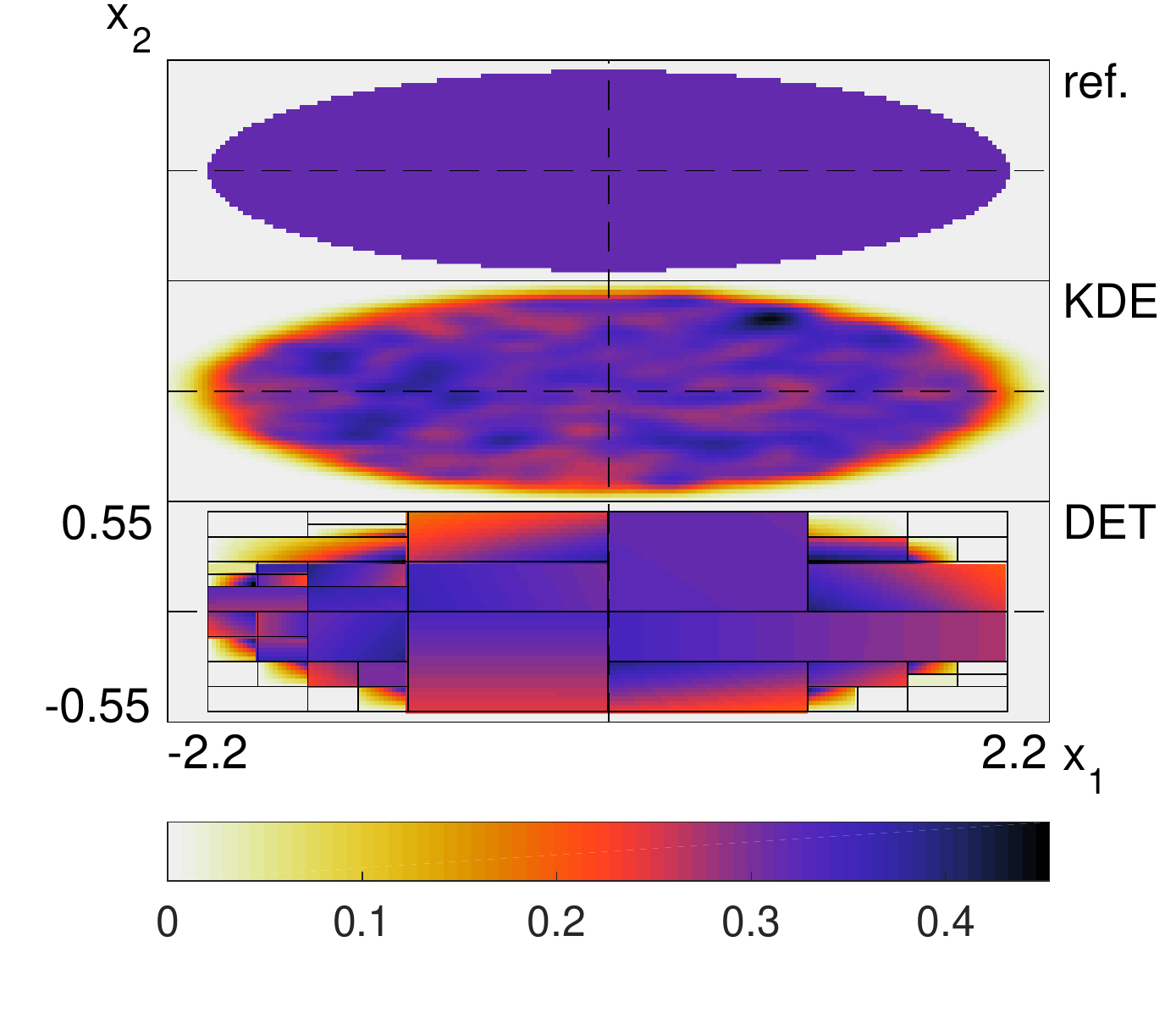}}
\put(0.51,0.99){\makebox(0,0){(b1)}}
\put(0.0,0.0){\includegraphics[width=0.5\textwidth]{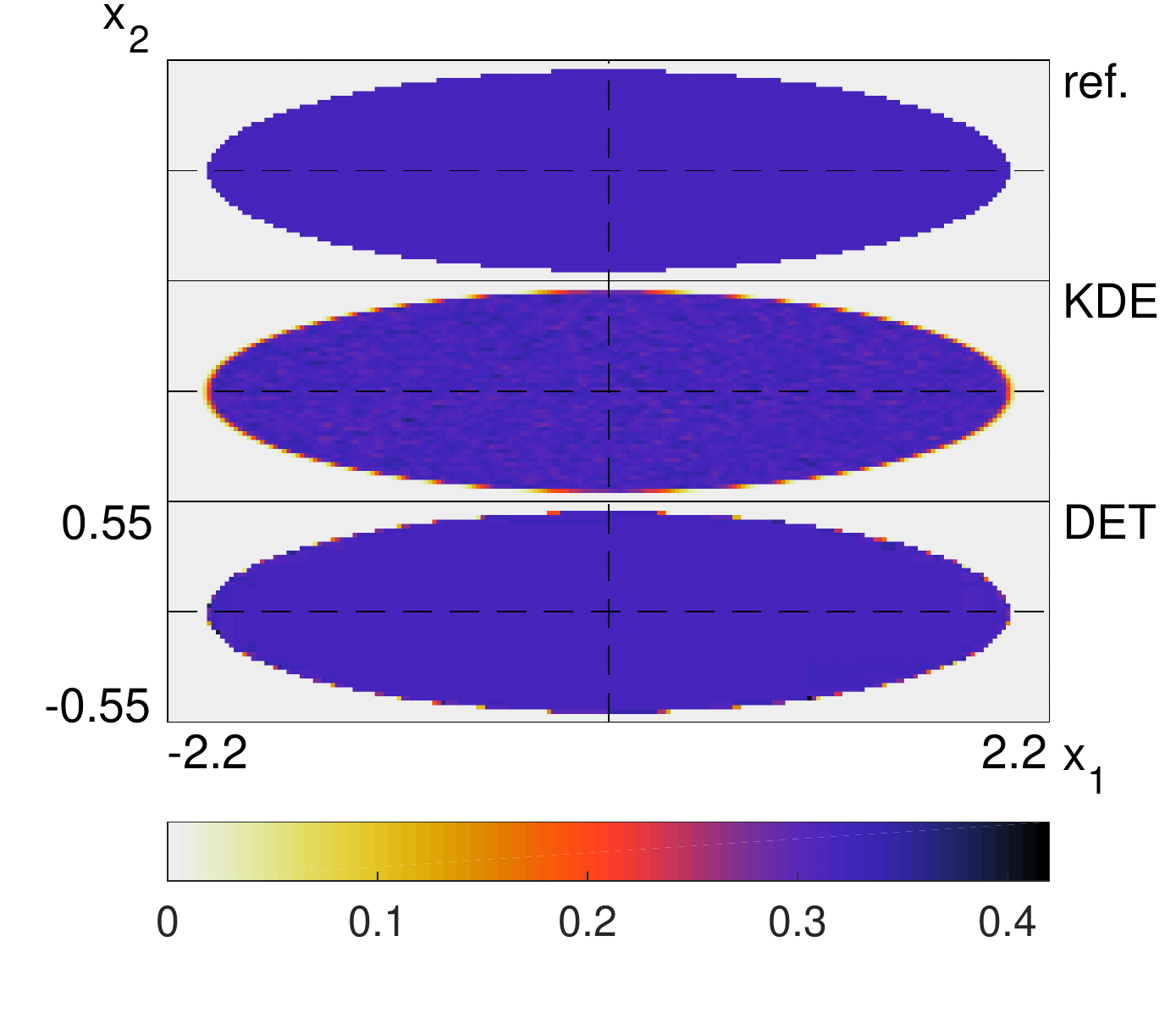}}
\put(0.01,0.49){\makebox(0,0){(a2)}}
\put(0.5,0.0){\includegraphics[width=0.5\textwidth]{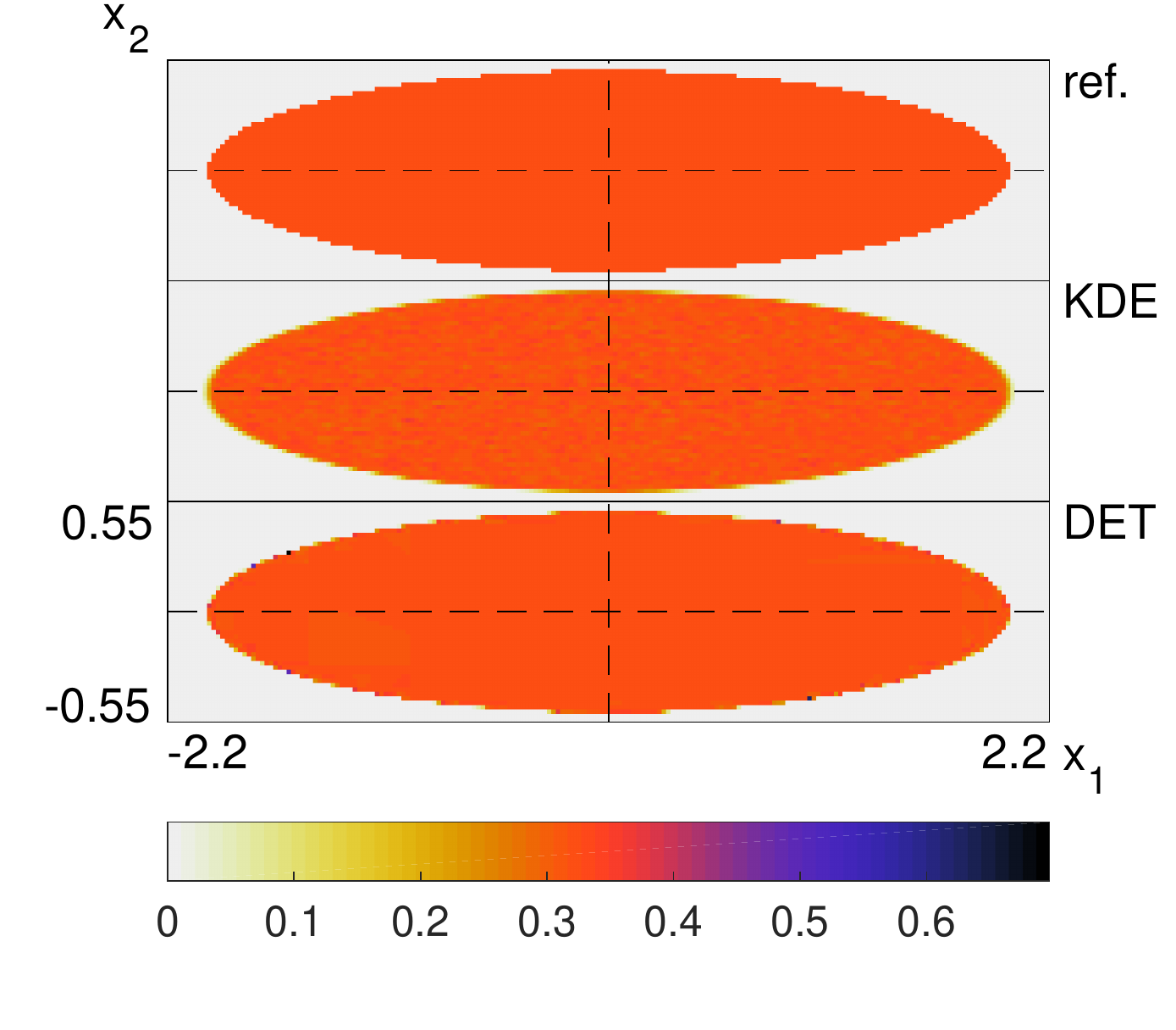}}
\put(0.51,0.49){\makebox(0,0){(b2)}}
\end{picture}
\caption{PDF estimates resulting from adaptive KDE and size-split DET methods with (a) constant and (b) linear DEs based on particle ensembles with (1) $n = 10^4$ and (2) $n \approx 10^{6.5}$ samples are compared with the reference PDF~\eq{eq2dC1PDF}.}\label{fig2dC1PDF}
\end{figure*}
\begin{figure*}
\unitlength\textwidth
\begin{picture}(1,0.375)
\put(0.0,0.0){\includegraphics[width=0.5\textwidth]{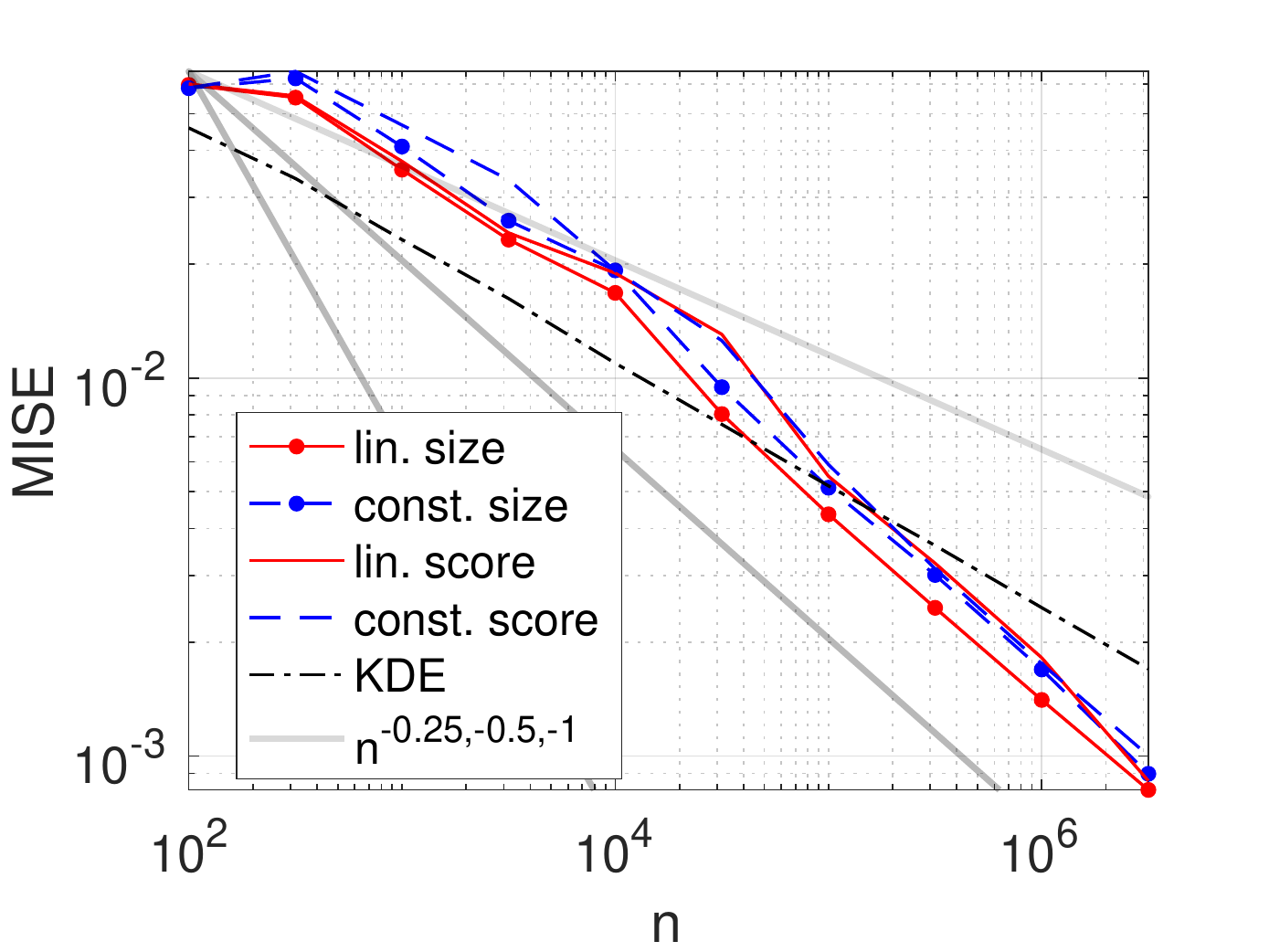}}
\put(0.01,0.365){\makebox(0,0){(a)}}
\put(0.5,0.0){\includegraphics[width=0.5\textwidth]{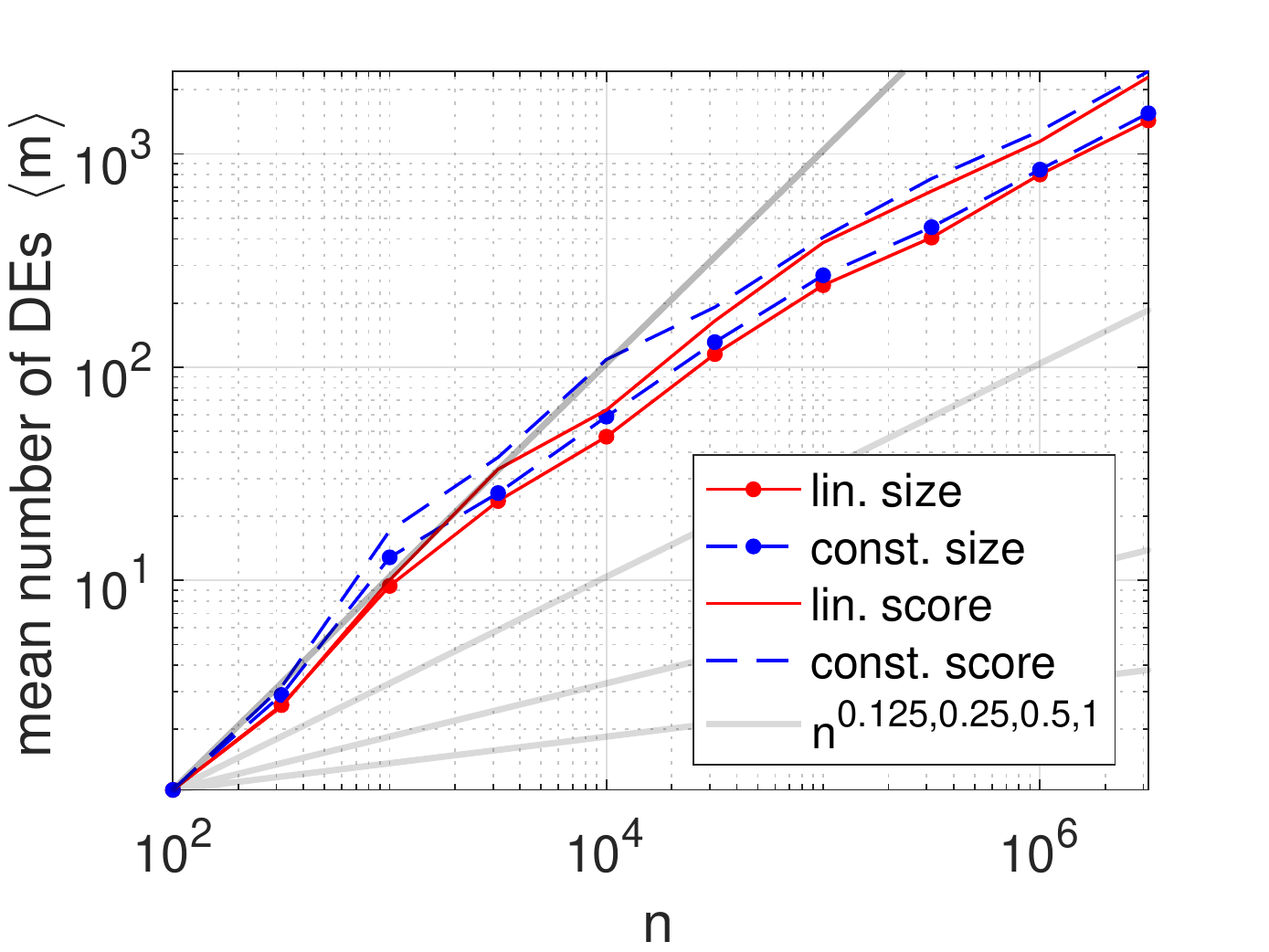}}
\put(0.51,0.365){\makebox(0,0){(b)}}
\end{picture}
\caption{Estimation of the bi-variate uniform PDF~\eq{eq2dC1PDF}. (a) Evolutions of the MISE as a function of the number of samples~$n$ for DET estimators (red solid and blue dashed) with equal size (symbols) and score splits (no symbols) are plotted. The MISE resulting from the adaptive KDE is included as well (black dash dot). For the DET estimators the resulting mean number of DEs~$\langle m\rangle$ are given in panel (b). Power law scalings with exponents indicated in the figure legends are depicted (gray thick solid).}\label{fig2dC1MISE}
\end{figure*}
Next, the two-dimensional case with uniform density on an ellipse outlined in \citep[p.~2942]{Botev:2010a} is inspected. The PDF is given by
\begin{equation}\label{eq2dC1PDF}
p(\mathbf{x}) = \left\{\begin{array}{ll}
1/\pi & \forall\;\mathbf{x}\in \{\mathbf{x}|x_1^2 + (4x_2)^2 \le 4\} \\
0 & \mbox{otherwise}
\end{array}\right.
\end{equation}
and in \figurename{}~\ref{fig2dC1PDF} KDE and DET-based estimation results based on ensembles stemming from this PDF are provided. Like in the one-dimensional case with uniform PDFs (see \figurename{}~\ref{fig1dC4PDF}(a2) and (b2)), constant DEs lead to smaller oscillations at the interface, where PDF~\eq{eq2dC1PDF} switches from 0 to $1/\pi$ (compare ranges of color bars in panels (1) and (2) of \figurename{}~\ref{fig2dC1PDF}). However, while the adaptive KDE displays a noisy density estimate within the ellipse, the DET variants capture the constant density quite accurately. This is reflected in the MISE results shown in \figurename{}~\ref{fig2dC1MISE}(a), where for $n > 10^5$ the DET estimators converge faster to the true PDF and become more accurate compared with adaptive KDE. Given the piecewise constant PDF~\eq{eq2dC1PDF}, linear DEs are as good as constant elements and like the MISE, the mean number of DEs $\langle m\rangle$ shown in \figurename{}~\ref{fig2dC1MISE}(b) and the mean tree depth $\langle n_t\rangle$ (not shown) behave similarly for both DE types. For large $n$, $\langle m\rangle$ increases approximately as $\sqrt{n}$ and $\langle n_t\rangle$ (not shown) grows similarly as in the one-dimensional spiky uniforms case (see \figurename{}~\ref{fig1dC4MISE}(b)).

One might argue that---in the absence of a principle axes transform---the present setup, with the ellipse aligned with the coordinate system, is in favor of the DET method. In an additional study the performance of the DET method was inspected for the ellipse rotated by an angle of $\pi/4$. With this modified setup, similar results were found, with the DET variants surpassing KDE at $n \approx 2\times 10^5$ and with $\langle m\rangle$ starting from~6 elements at $n = 100$ and growing to similar numbers as in the unrotated case.

\subsubsection{Bi-Variate Dirichlet PDF}

\begin{figure*}
\unitlength\textwidth
\begin{picture}(1,0.9)
\put(0.05,-0.1){\includegraphics[width=0.4\textwidth]{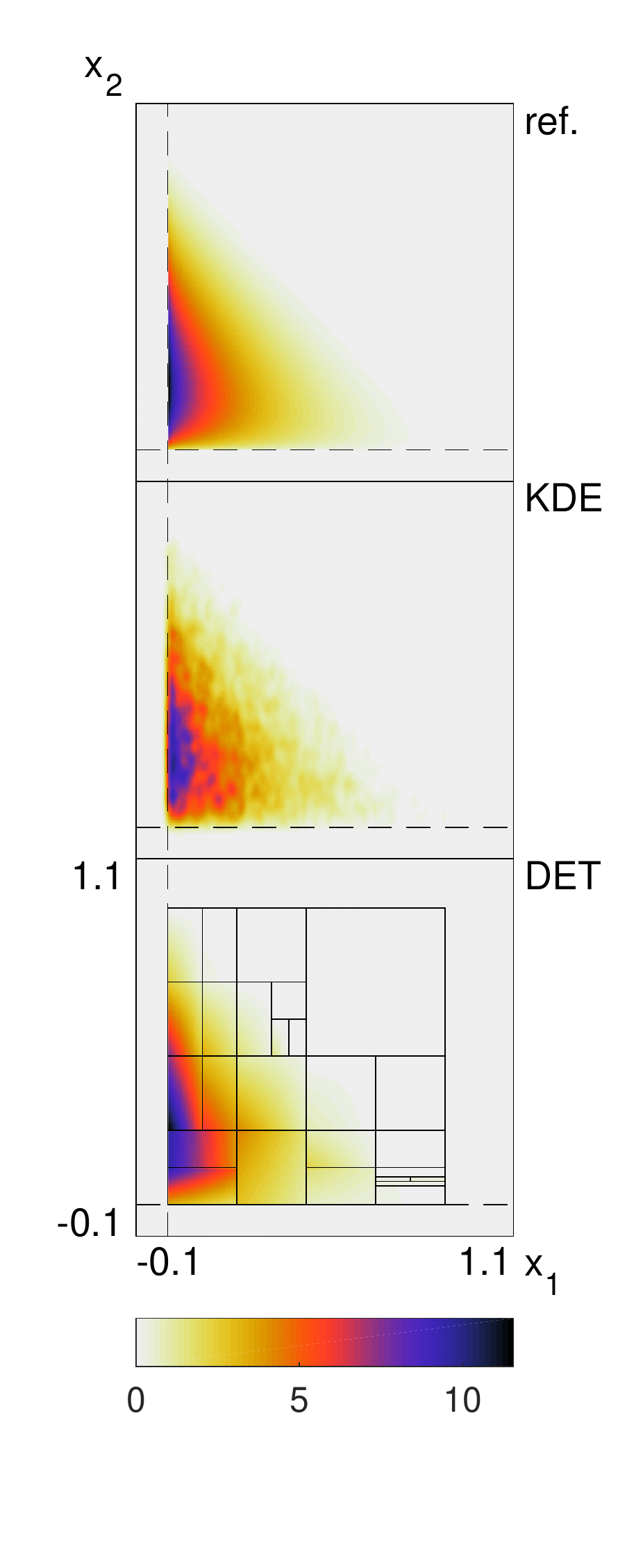}}
\put(0.01,0.89){\makebox(0,0){(a)}}
\put(0.55,-0.1){\includegraphics[width=0.4\textwidth]{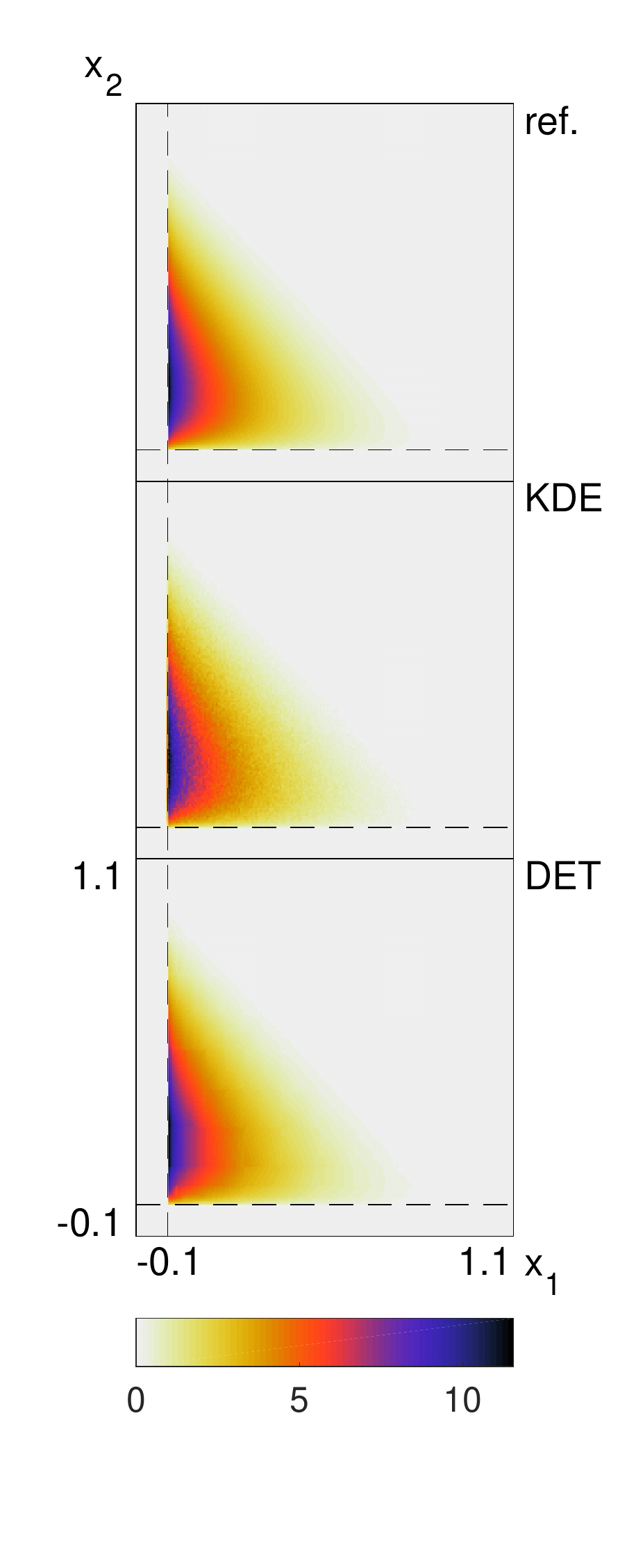}}
\put(0.51,0.89){\makebox(0,0){(b)}}
\end{picture}
\caption{PDF estimates resulting from adaptive KDE and the size-split linear DET method based on particle ensembles with (a) $n = 10^4$ and (b) $n \approx 10^{6.5}$ samples are compared with the reference PDF~\eq{eq2dC3PDF}.}\label{fig2dC3PDF}
\end{figure*}
\begin{figure*}
\unitlength\textwidth
\begin{picture}(1,0.375)
\put(0.0,0.0){\includegraphics[width=0.5\textwidth]{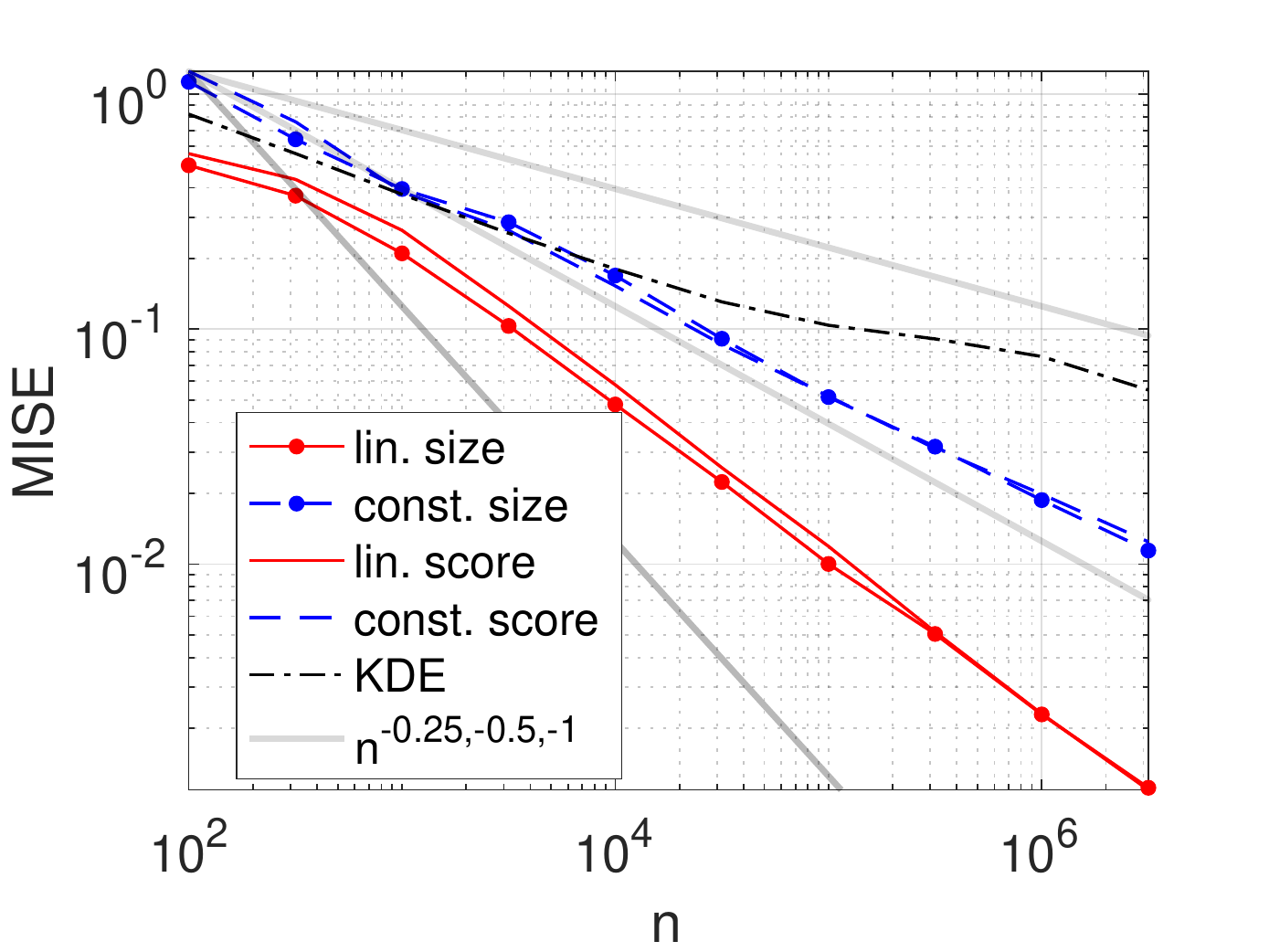}}
\put(0.01,0.365){\makebox(0,0){(a)}}
\put(0.5,0.0){\includegraphics[width=0.5\textwidth]{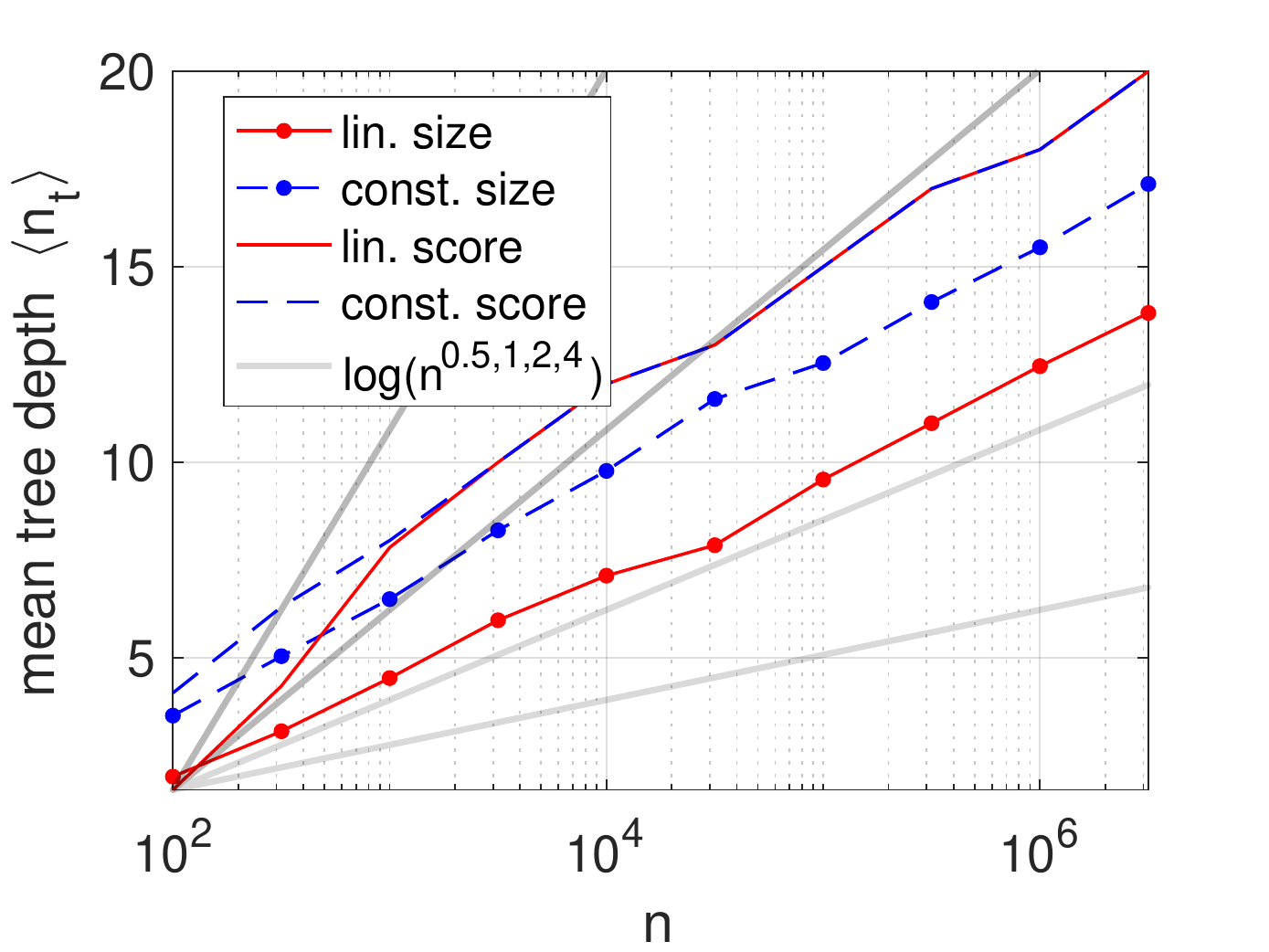}}
\put(0.51,0.365){\makebox(0,0){(b)}}
\end{picture}
\caption{Estimation of Dirichlet PDF~\eq{eq2dC3PDF}. See \figurename{}~\ref{fig1dC4MISE}.}\label{fig2dC3MISE}
\end{figure*}
The third bi-variate case addressed in this work is the Dirichlet PDF given by
\begin{eqnarray}\label{eq2dC3PDF}
p(\mathbf{x}) = \frac{\Gamma(\alpha_1+\alpha_2+\alpha_3)}{\Gamma(\alpha_1)\Gamma(\alpha_2)\Gamma(\alpha_3)}x_1^{\alpha_1-1}x_2^{\alpha_2-1}(1-x_1-x_2)^{\alpha_3-1} \nonumber \\
\end{eqnarray}
$\forall\;\mathbf{x}\in \{\mathbf{x}|x_1 + x_2 \le 1 \vee x_1 \ge 0 \vee x_2 \ge 0\}$ and $ = 0$ otherwise, with parameters $\alpha_1 = 0.9$, $\alpha_2 = 1.5$, and $\alpha_3 = 3$. The Dirichlet PDF~\eq{eq2dC3PDF} is a bi-variate generalization of the uni-variate beta PDF~\eq{eq1dC5PDF}. PDF~\eq{eq2dC3PDF} is depicted together with KDE results and DET estimates for two differently-sized ensembles in \figurename{}~\ref{fig2dC3PDF}. The DET method provides for both ensembles estimates that are in good agreement with the true density. MISE results are compared in \figurename{}~\ref{fig2dC3MISE}(a) and it is seen that like in the beta PDF example, the linear DET variants are most accurate, followed by their constant counterparts and adaptive KDE. As seen in \figurename{}~\ref{fig2dC3MISE}(b), the size-split DET methods produce the smallest trees with the linear DET approach being the most efficient. The mean number of DEs $\langle m\rangle$ (not shown) increases with~$n$ like in the bi-variate Gaussian case. The linear DEs and size-based splitting have the smallest $\langle m\rangle$ and all DET variants scale approximately with $n^{1/4}$ for $n$ large (see \figurename{}~\ref{fig2dC2MISE}(b)).

Rotating the bi-variate Dirichlet PDF by $\pi/4$ has a bigger impact on the DET performance as in the previously discussed ellipse case. While the MISE error decay rates remain similar, the MISE of the DET variants are larger than KDE for small~$n$ and surpass KDE for $n$ between~$10^4$ and~$10^5$.

In addition to the previous MISE-based performance analyses, we have evaluated the accuracy of the DET estimator with Hellinger \citep[e.g.,][]{Wang:2015a} and total variation distances \citep[p.~543]{Shorack:2000a}. These metrics are based on differences of square roots of densities and absolute density differences, respectively, and therefore penalize strong deviations less than the MISE. Since occasional outliers are more likely in the DET estimator and regular small noise is an issue in the adaptive KDE method (see for example \figurename{}~\ref{fig1dC4PDF}(b2)), the DET estimator was found to perform better than KDE in terms of Hellinger-distance metric as opposed to MISE (see exemplary comparison in \figurename{}~\ref{fig1dC2MISE}, panels~(a) and~(d)). In terms of total variation distance, the relative performance among the estimators and convergence trends were found to be overall similar to the MISE results.

\subsection{Comparisons Histogram and Tree-Based Methods}\label{subsec1d2d}

\begin{figure*}
\begin{center}
\includegraphics[width=0.85\textwidth]{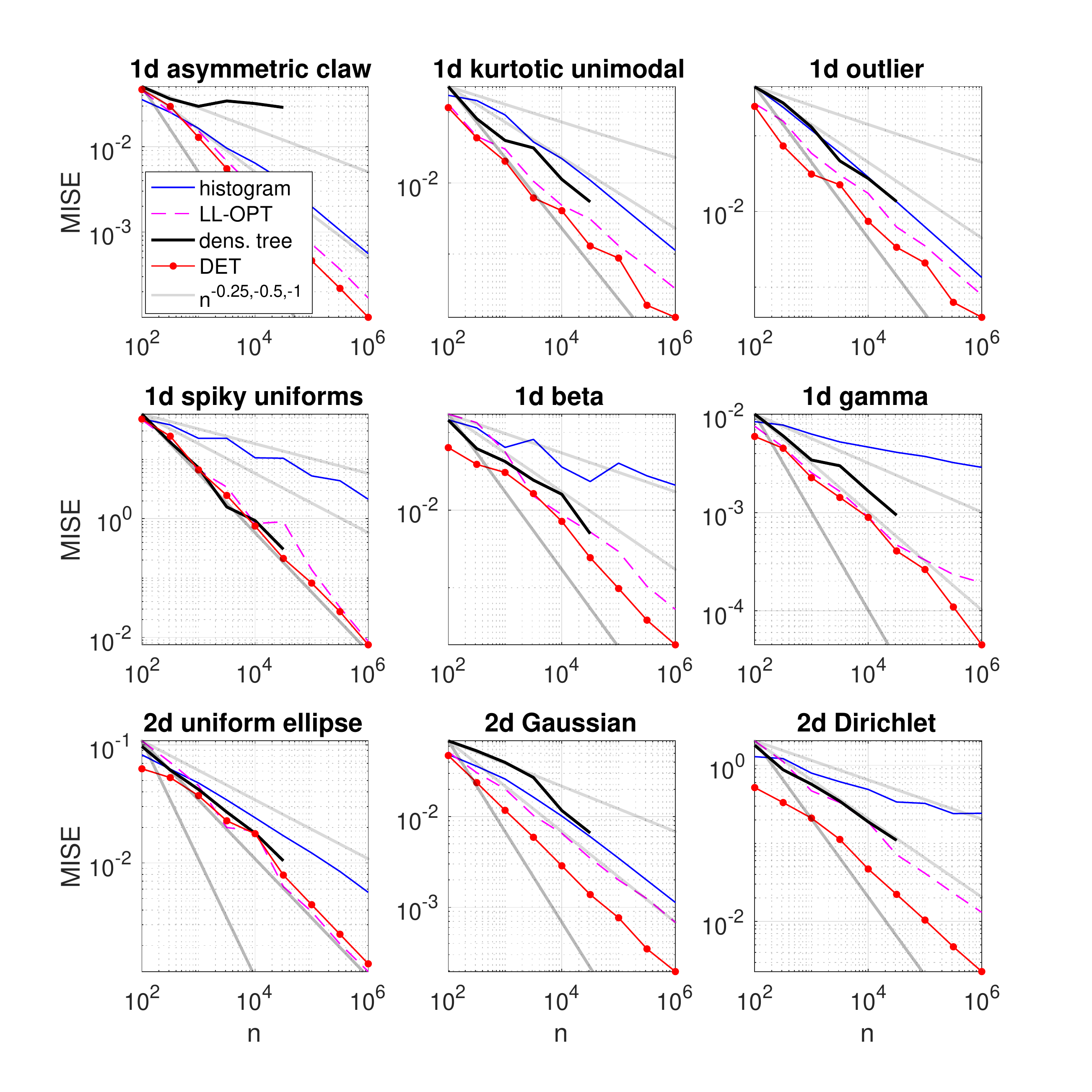}
\caption{MISE decay as a function of the number of samples~$n$ for the indicated one- and two-dimensional cases. Results from (blue solid) histogram, (magenta dashed) LL-OPT, (black thick solid) density tree, and (red symbols) size-split linear DET estimators are shown together with (gray thick solid) power-law scalings $n^{-0.25}$, $n^{-0.5}$, and $n^{-1}$.}\label{fig1d2dMISE}
\end{center}
\end{figure*}
\begin{figure*}
\unitlength\textwidth
\begin{picture}(1,0.31)
\put(0.0,0.0){\includegraphics[width=0.5\textwidth]{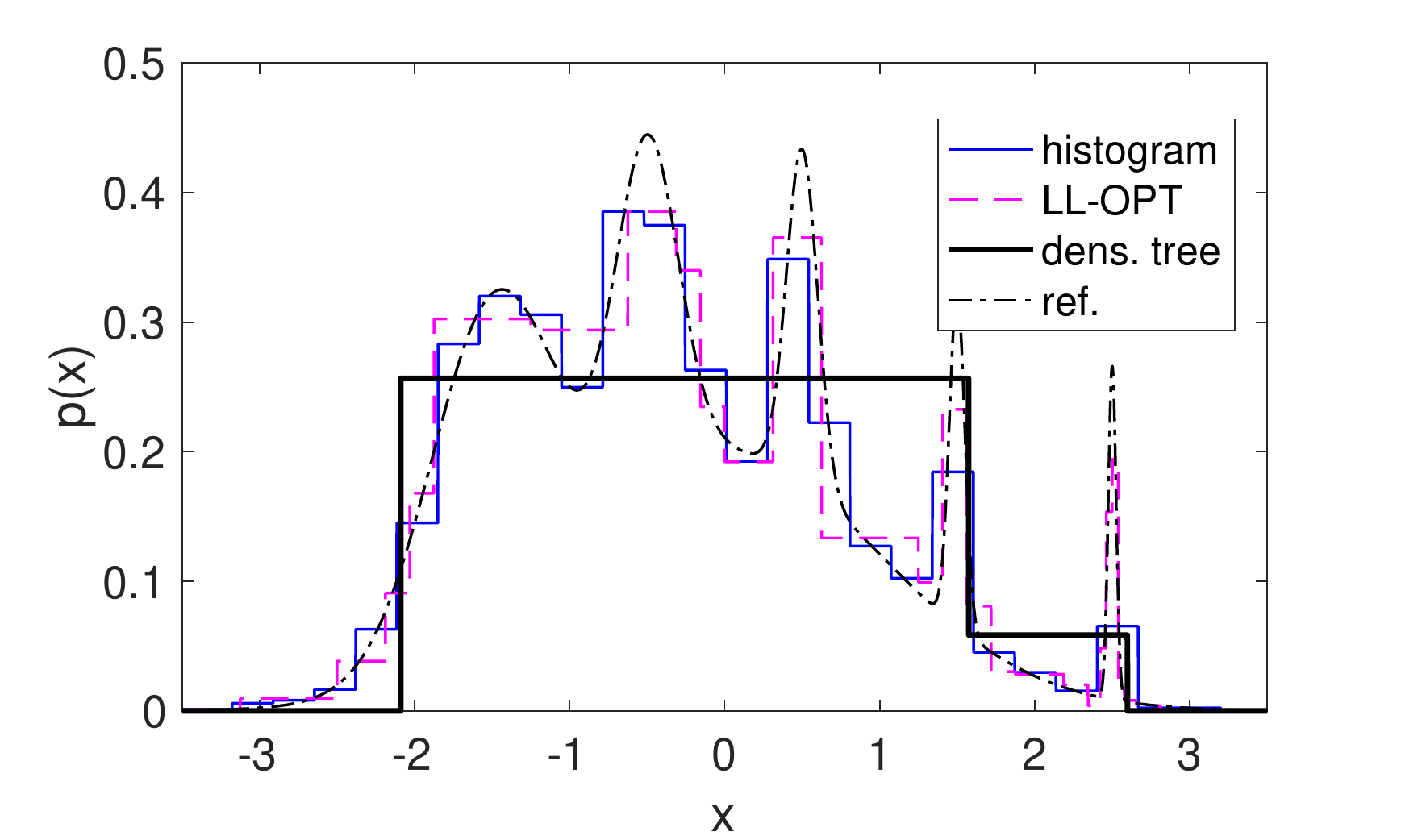}}
\put(0.01,0.3){\makebox(0,0){(a)}}
\put(0.5,0.0){\includegraphics[width=0.5\textwidth]{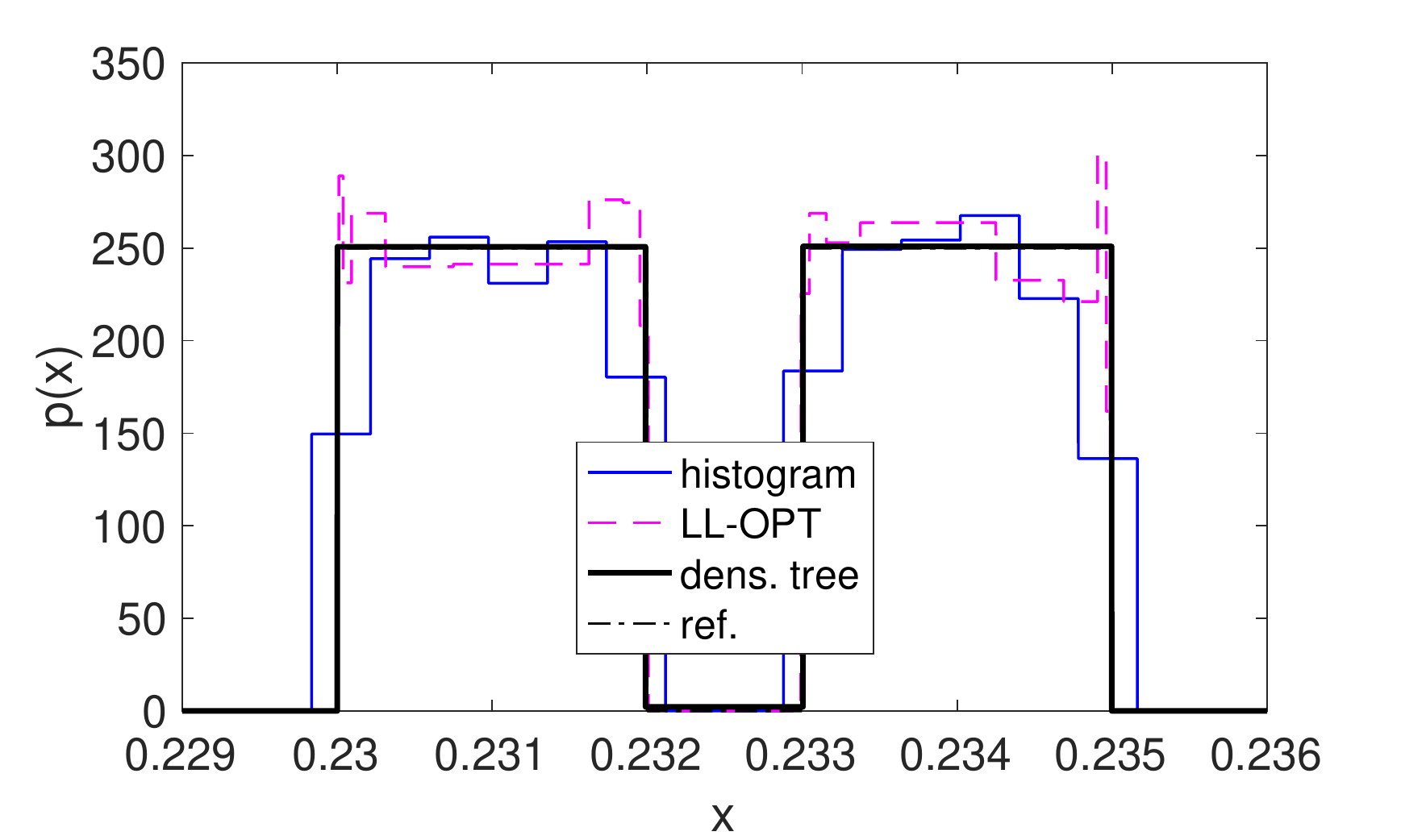}}
\put(0.51,0.3){\makebox(0,0){(b)}}
\end{picture}
\caption{PDF estimates resulting from (blue thin solid) histogram, (magenta dashed) LL-OPT, and (black thick solid) density tree estimators with $n = 3163$ samples are compared with (black dash dot) the reference PDFs~\eq{eq1dC1PDF} and~\eq{eq1dC4PDF}. In panels (a) and (b), the asymmetric claw~\eq{eq1dC1PDF} and spiky uniforms~\eq{eq1dC4PDF} PDFs, respectively, are depicted.}\label{fig1dC14PDF}
\end{figure*}
In the previous two sections, we found that the linear DET variant is most accurate with little difference between size- and score-based splitting. Accordingly, we proceed in this and the next sections by focusing on linear DETs with size-based splits. In \figurename{}~\ref{fig1d2dMISE}, a summary of MISE decay curves resulting from the histogram, LL-OPT, density tree, and linear DET estimators for all one- and two-dimensional examples is provided. Due to the rapidly growing computational costs associated with the density-tree estimator, we stop after $n = 31623 \approx 10^{4.5}$. The computing time for one density-tree estimate of, e.g., the beta PDF example with $10^5$ samples took around 10~minutes. \figurename{}~\ref{fig1d2dMISE} is accompanied by exemplary PDF estimates included in \figurename{}~\ref{fig1dC14PDF}. These estimates resulted from ensembles with $n = 3163$ samples.

One general observation from the MISE curves in \figurename{}~\ref{fig1d2dMISE} is the good accuracy of the DET estimator. While histograms with bin widths determined by the normal reference rule are reasonably accurate in the Gaussian mixture cases, they are inaccurate in the other cases. The density-tree estimator leads to mixed results and is most accurate for the examples involving uniform distributions (see \figurename{}~\ref{fig1dC14PDF}(b)). A MISE convergence that comes quite close to the DET estimator is resulting from the LL-OPT method. Notable is finally a reduction in MISE decay rates when going from one to two dimensions as seen in \figurename{}~\ref{fig1d2dMISE}. This is an indication of the curse of dimensionality mentioned in the introduction.

\subsection{Four- and Seven-Dimensional Examples}\label{subsec4d7d}

Next, we assess the performance of the linear DET method with equal-size splits together with the other estimators in examples involving four and seven sample-space dimensions.

\subsubsection{High-Dimensional Gaussian PDF}

\begin{figure*}
\unitlength\textwidth
\begin{picture}(1,0.375)
\put(0.0,0.0){\includegraphics[width=0.5\textwidth]{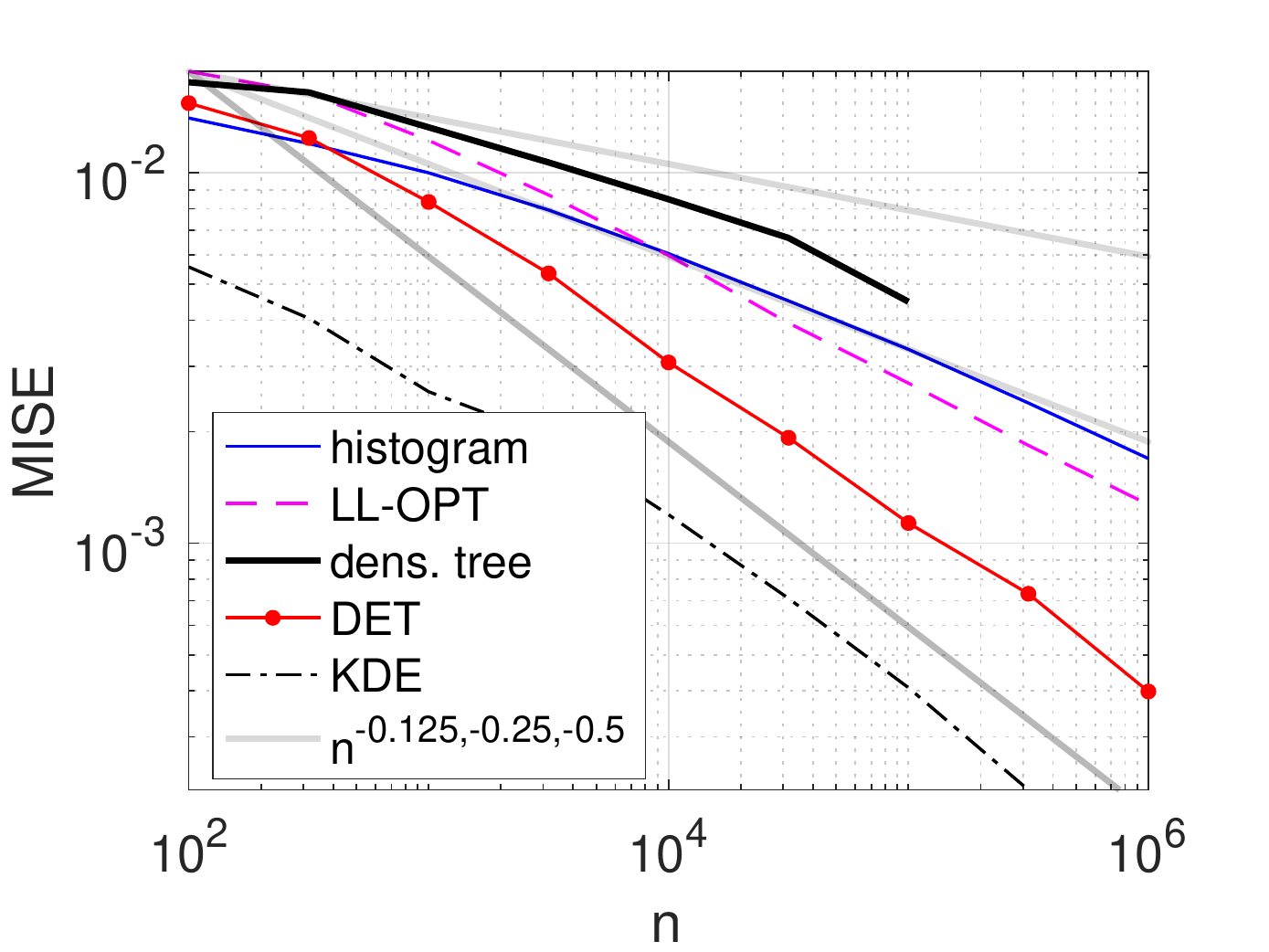}}
\put(0.01,0.365){\makebox(0,0){(a)}}
\put(0.5,0.0){\includegraphics[width=0.5\textwidth]{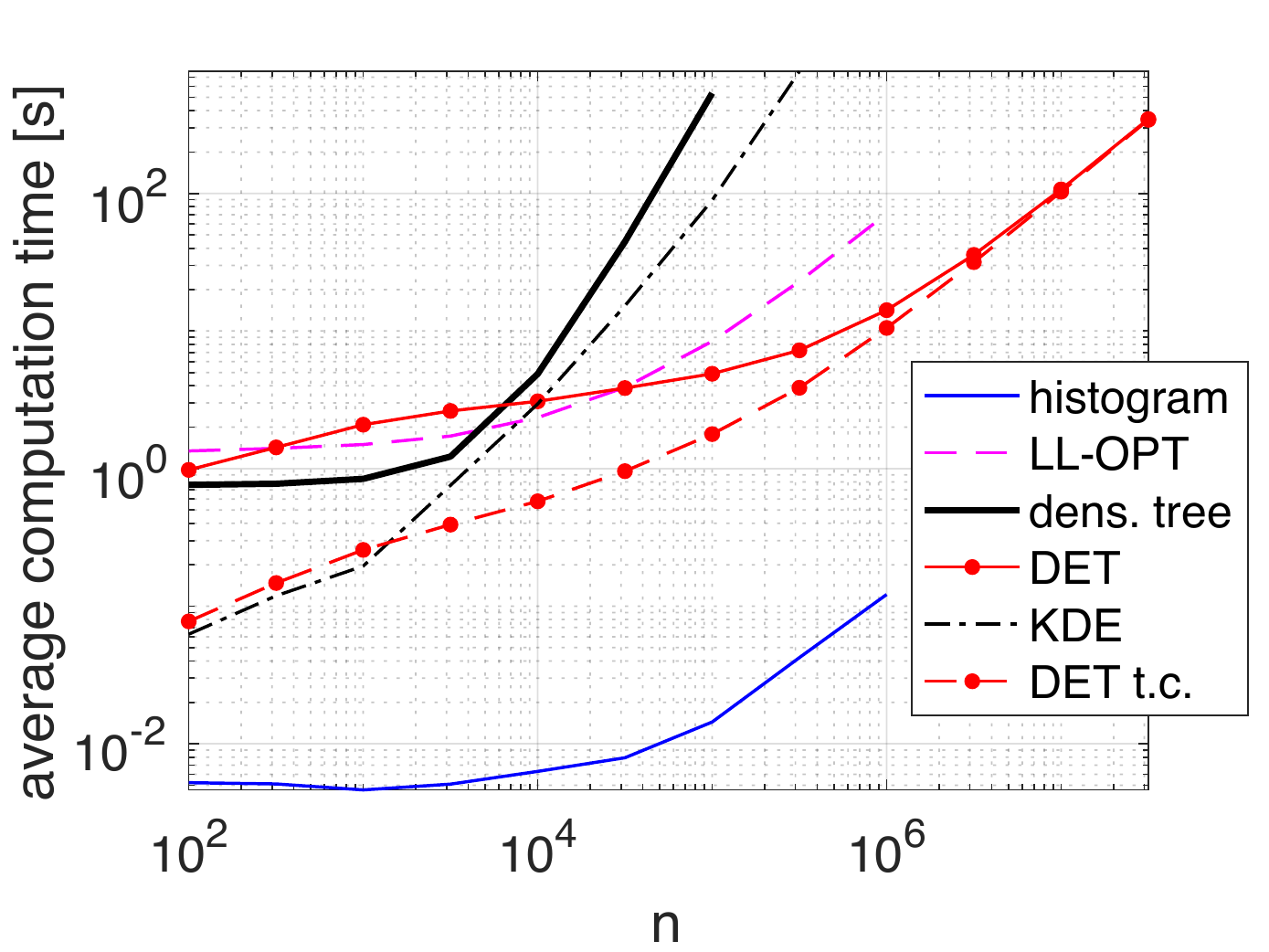}}
\put(0.51,0.365){\makebox(0,0){(b)}}
\end{picture}
\caption{Estimation of the four-dimensional joint Gaussian PDF. (a) Evolutions of the MISE and (b) computing time as a function of the number of samples~$n$ for histograms (blue solid), LL-OPT (dashed magenta), density tree (black thick solid), and DET (red solid symbols) estimators, and adaptive KDE (black dash dot) are plotted. In panel (b), the tree construction time for the DET estimator (red dashed symbol) is plotted as well.}\label{fig4dC1MISE}
\end{figure*}
\begin{figure*}
\unitlength\textwidth
\begin{picture}(1,0.375)
\put(0.0,0.0){\includegraphics[width=0.5\textwidth]{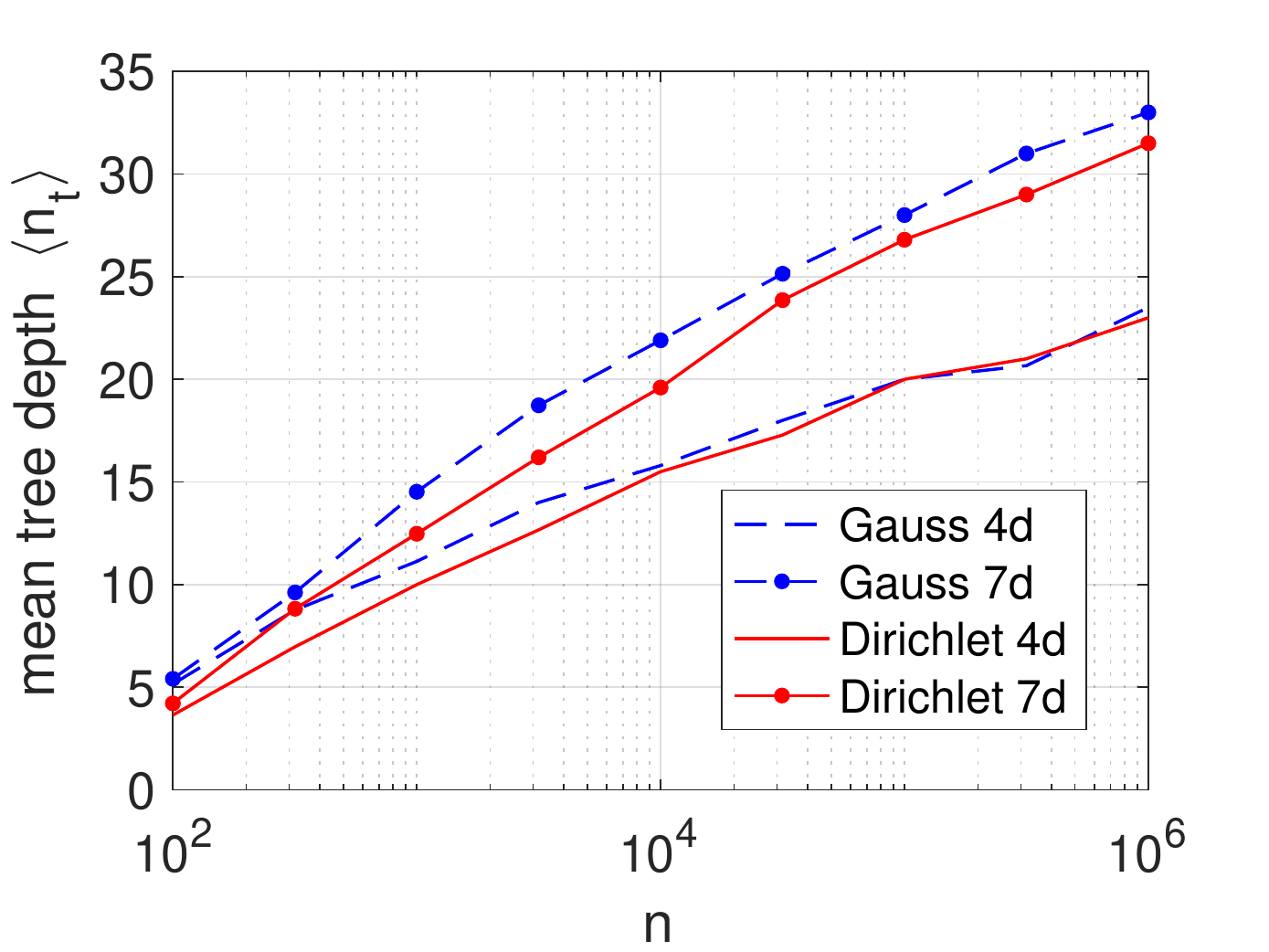}}
\put(0.01,0.365){\makebox(0,0){(a)}}
\put(0.5,0.0){\includegraphics[width=0.5\textwidth]{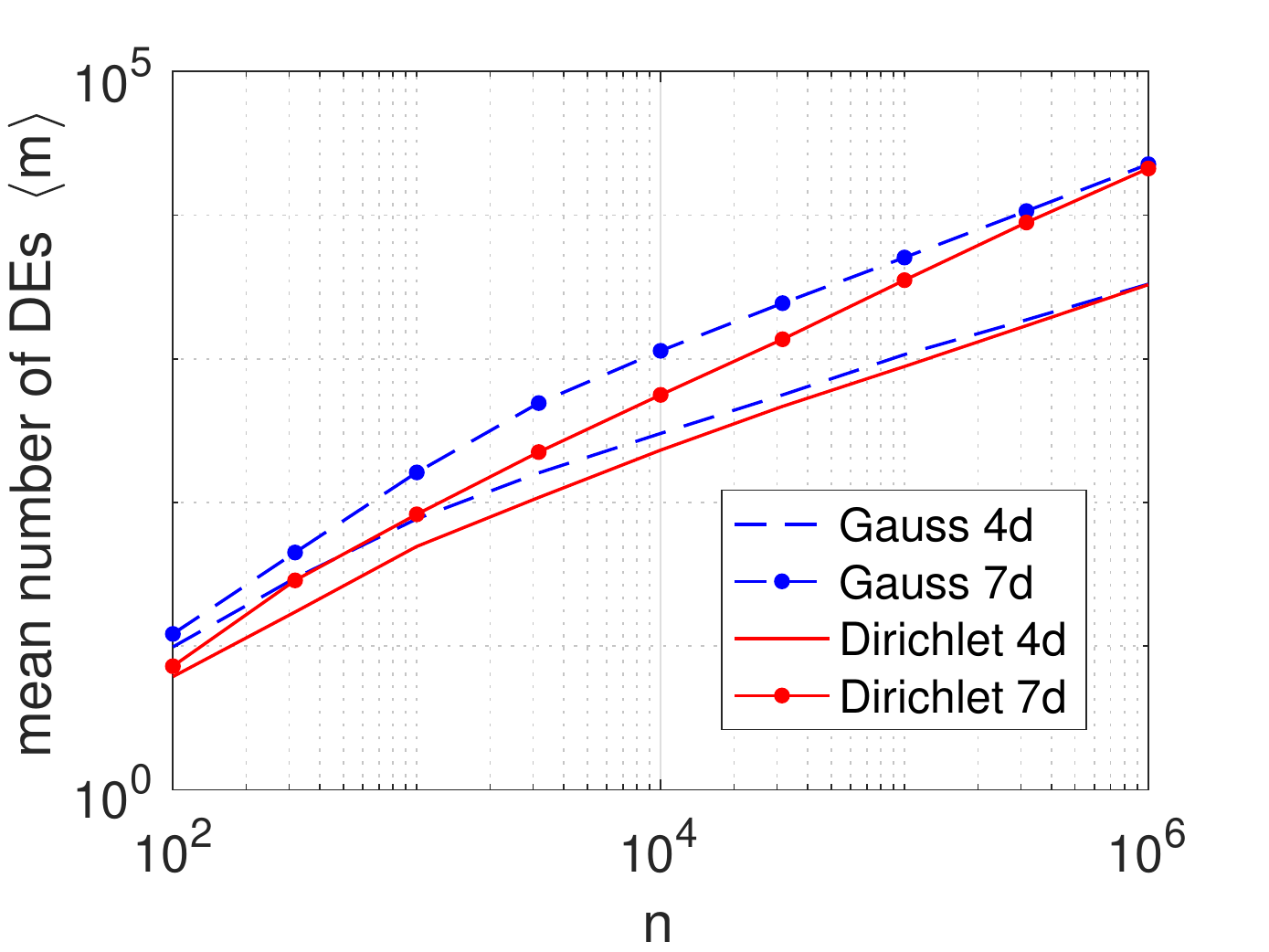}}
\put(0.51,0.365){\makebox(0,0){(b)}}
\end{picture}
\caption{Mean tree depths $\langle n_t\rangle$ (a) and mean number of DEs $\langle m\rangle$ (b) of the DET estimator in the examples involving the (lines) four- and (symbols) seven-dimensional (blue dashed) Gaussian and (red solid) Dirichlet PDFs.}\label{fig4d7dC1C2}
\end{figure*}
In a first multi-dimensional example, we reuse the joint Gaussian PDF~\eq{eq2dC2PDF} with $d$-dimensional probability space $\mathbf{x} = (x_1,x_2,\ldots,x_d)^\top$, mean vector $\mbf{\mu}$ being zero, and $d\times d$ covariance matrix~$\mathbf{C}$. In a first step, we inspect the performance of the different methods for $d = 4$ with the randomly chosen covariance matrix
\begin{displaymath}
\mathbf{C} = \left(\begin{array}{cccc}
1 & -0.344 & 0.141 & -0.486 \\
-0.344 & 1 & 0.586 & 0.244 \\
0.141 & 0.586 & 1 & -0.544 \\
-0.486 & 0.244 & -0.544 & 1
\end{array}\right).
\end{displaymath}
In \figurename{}~\ref{fig4dC1MISE}(a), MISE decays from the different estimators are compared. KDE performs best, followed by the DET estimator, which shares the same empirically determined convergence rate of approximately $1/\sqrt{n}$. The other estimators perform similarly to histograms. The data series of the density tree and KDE methods stop at $n \approx 10^{4.5}$ and $10^{5.5}$, respectively, due to the rapidly growing computing times of these estimators (see discussion in the introduction~\ref{secIntro}).

An analysis of computing times per density estimate is provided in \figurename{}~\ref{fig4dC1MISE}(b). The reported times are comprised of the construction of the estimator and density queries based on the estimator. The number of query points in the MC integration is kept constant for different estimators and number of samples~$n$. This explains the plateaus for small~$n$ in the LL-OPT, density tree, and histogram estimators, where the computing times for small~$n$ are governed by the query effort. For increasing~$n$, the estimator construction becomes noticeable, leading to growing times. In the KDE, the query cost is connected to the number of samples, which leads to a continuous increase in computing time. The relatively small growth in computing time of the DET estimator is based on two factors document in \figurename{}~\ref{fig4d7dC1C2}: (panel~a) the query time, which is driven by the logarithmically increasing tree depth, grows slowly compared to (b) the tree construction time, which depends on an exponential but sublinear growth in DEs $\langle m\rangle$. The construction time is plotted in \figurename{}~\ref{fig4dC1MISE}(b) and shown to converge to the total DET computing time, as the query time becomes comparably small for large~$n$.

\begin{figure*}
\unitlength\textwidth
\begin{picture}(1,0.375)
\put(0.0,0.0){\includegraphics[width=0.5\textwidth]{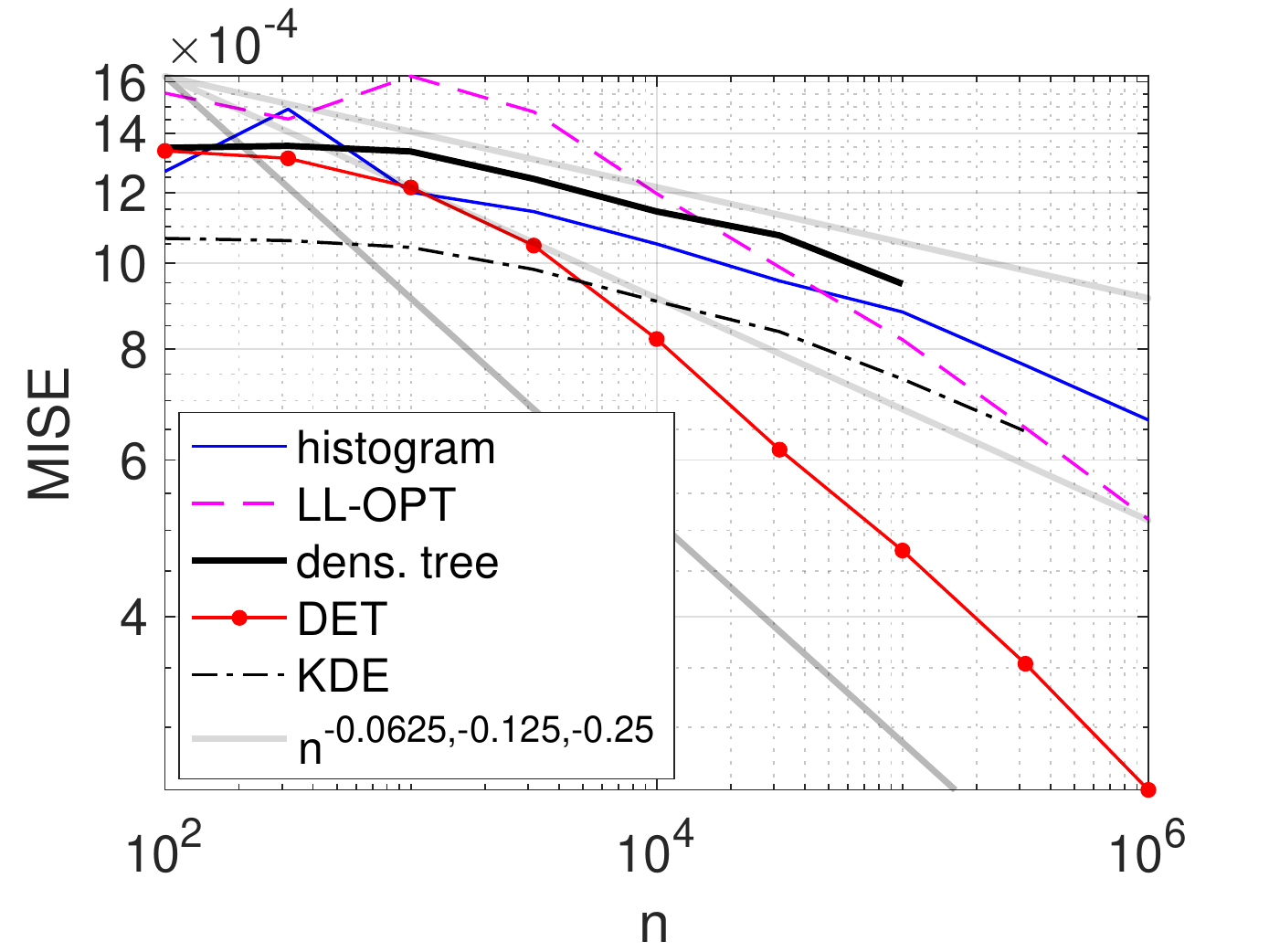}}
\put(0.01,0.365){\makebox(0,0){(a)}}
\put(0.5,0.0){\includegraphics[width=0.5\textwidth]{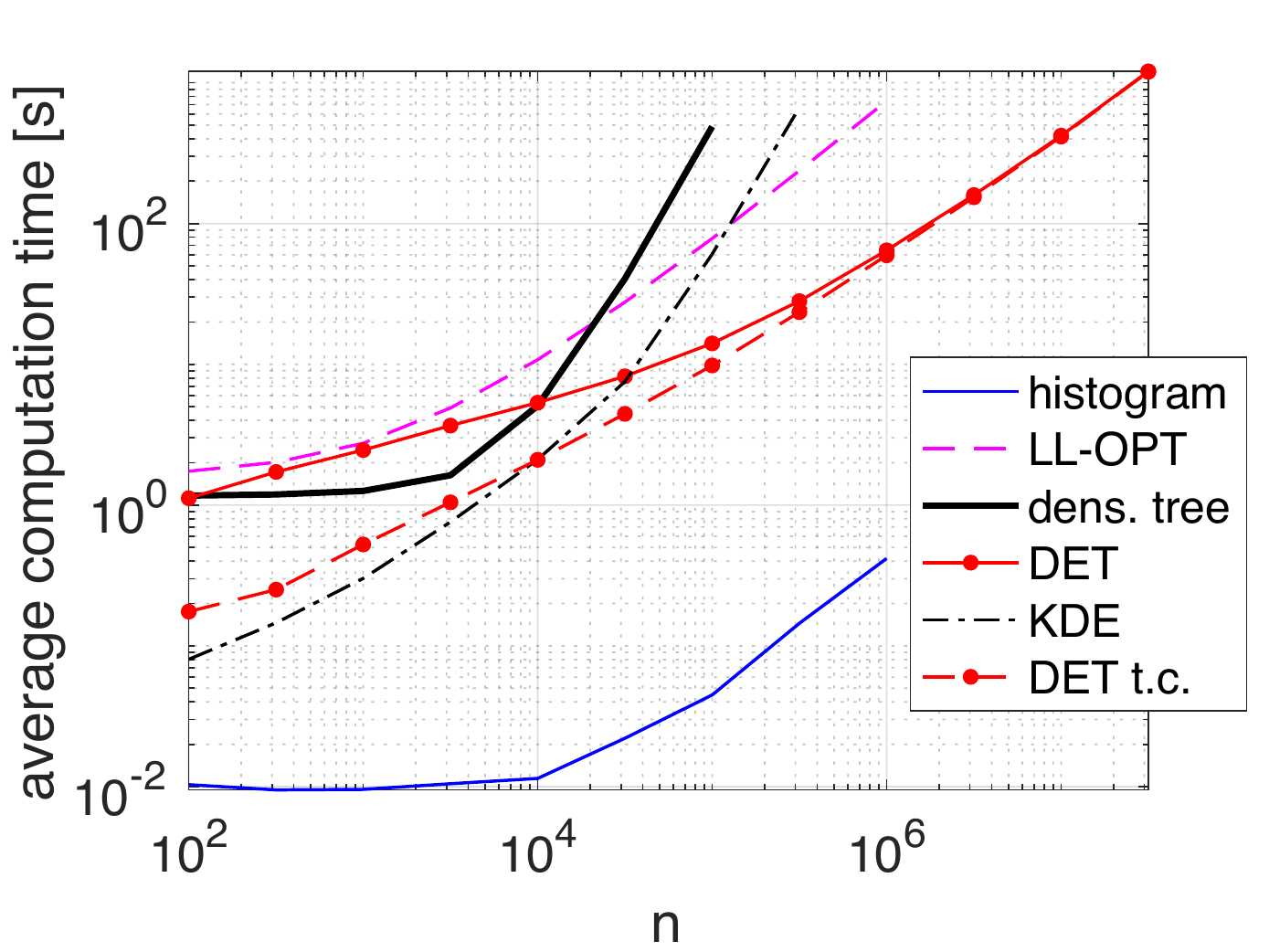}}
\put(0.51,0.365){\makebox(0,0){(b)}}
\end{picture}
\caption{Estimation of the seven-dimensional joint Gaussian PDF. See \figurename{}~\ref{fig4dC1MISE}.}\label{fig7dC1MISE}
\end{figure*}
In a next step, the dimensionality is increased to seven with the covariance matrix
\begin{eqnarray*}
\lefteqn{\mathbf{C} =} \\
& & \small\left(\!\!\begin{array}{ccccccc}
1 & -0.216 & 0.161 & -0.0496 & 0.0342 & -0.116 & 0.749 \\
-0.216 & 1 & 0.301 & 0.0391 & -0.217 & 0.0189 & -0.381 \\
0.161 &  0.301 & 1 & 0.574 & -0.312 & 0.109 & 0.386 \\
-0.0496 & 0.0391 & 0.574 & 1 & -0.438 & 0.730 & -0.0572 \\
0.0342 & -0.217 & -0.312 & -0.438 & 1 & -0.475 & 0.258 \\
-0.116 & 0.0189 & 0.109 & 0.730 & -0.475 & 1 & -0.386 \\
0.749 & -0.381 & 0.386 & -0.0572 & 0.258 & -0.386 & 1
\end{array}\!\!\right)
\end{eqnarray*}
again chosen randomly. The corresponding MISE and computing time curves for this example are provided in \figurename{}~\ref{fig7dC1MISE}. Compared to the previous four-dimensional case, the MISE convergence rate shown in panel~(a) has reduced to $n^{-1/4}$ for the DET estimator (curse of dimensionality). Unlike in the previous case, the DET estimator becomes more accurate than adaptive KDE for $n > 5000$. While the LL-OPT estimator has a similar MISE convergence rate for large~$n$ like the DET estimator, the density tree and histogram MISE are similar, like in the previous four-dimensional case. The computing times reported in \figurename{}~\ref{fig7dC1MISE}(b) show similar trends as well. Unlike in the four-dimensional case, however, we can identify the growth originating from the tree construction process at large~$n$ more clearly. Moreover, the computing times have increased noticeably compared to $d = 4$, which is documented as well in \figurename{}~\ref{fig4d7dC1C2}, where mean tree depth and number of DEs are higher for larger dimensionality~$d$.

\subsubsection{High-Dimensional Dirichlet PDF}

\begin{figure*}
\unitlength\textwidth
\begin{picture}(1,0.375)
\put(0.0,0.0){\includegraphics[width=0.5\textwidth]{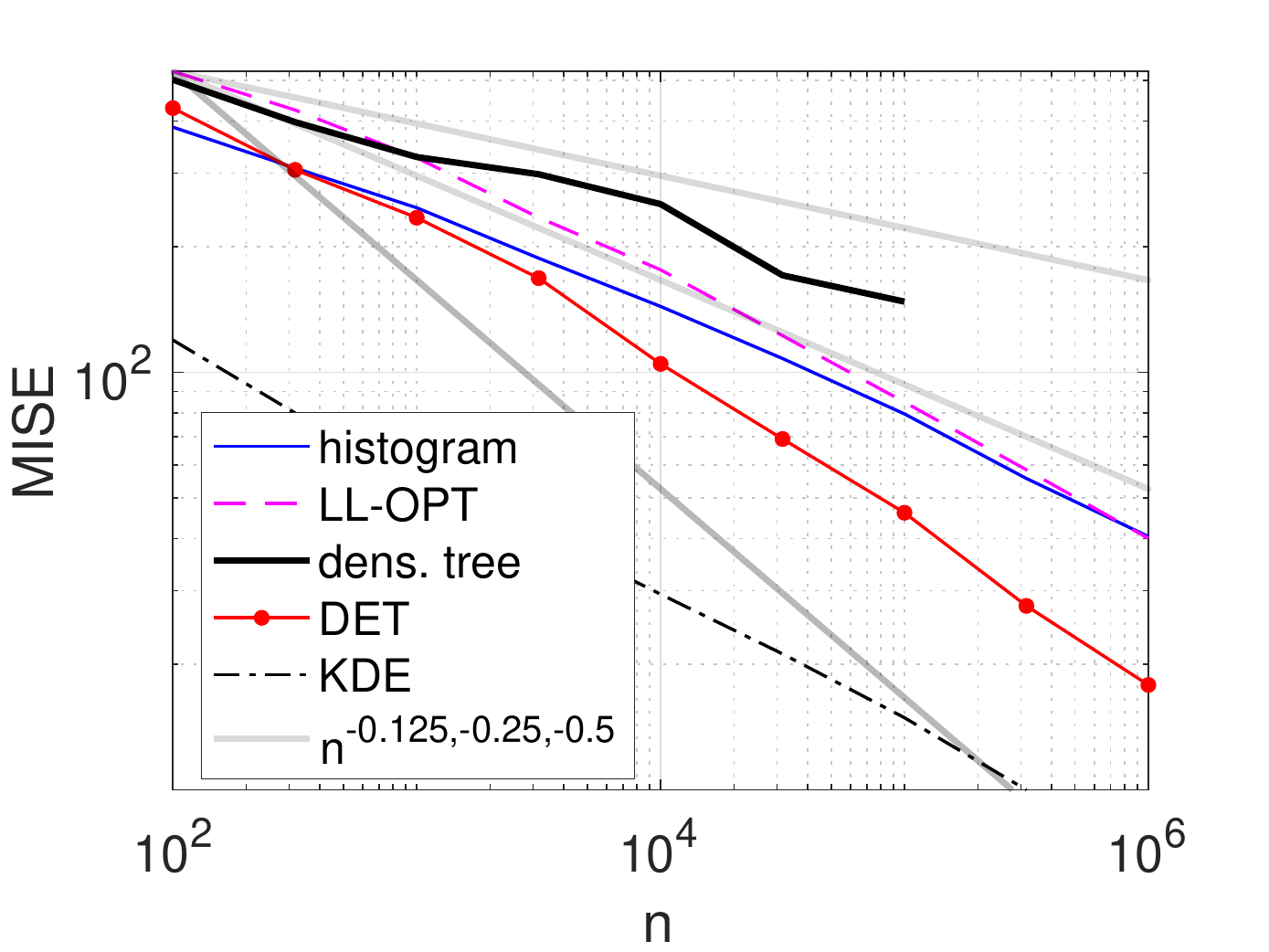}}
\put(0.01,0.365){\makebox(0,0){(a)}}
\put(0.5,0.0){\includegraphics[width=0.5\textwidth]{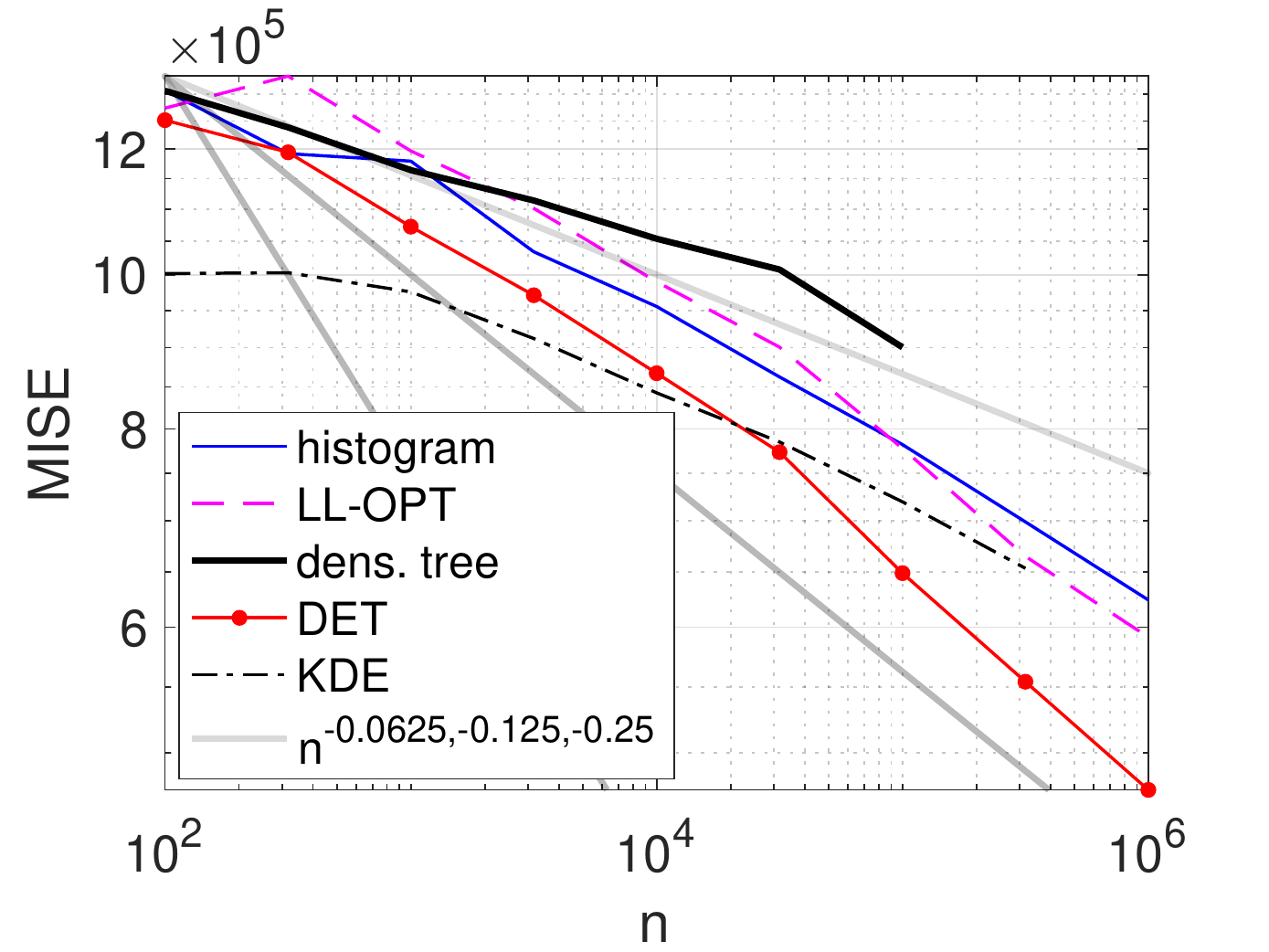}}
\put(0.51,0.365){\makebox(0,0){(b)}}
\end{picture}
\caption{Estimation of the (a) four- and (b) seven-dimensional Dirichlet PDFs. Evolutions of the MISE as a function of the number of samples~$n$ are plotted. See \figurename{}~\ref{fig4dC1MISE}.}\label{fig4d7dC2MISE}
\end{figure*}
A $d$-dimensional generalization of the Dirichlet PDF~\eq{eq2dC3PDF} is given by
\begin{eqnarray}\label{eq4d7dC2PDF}
p(\mathbf{x}) = \frac{\Gamma\left(\sum_{i = 1}^{d+1}\alpha_i\right)}{\prod_{i = 1}^{d+1}\Gamma(\alpha_i)} \prod_{i = 1}^{d} x_i^{\alpha_i-1}\left(1-\sum_{i = 1}^d x_i\right)^{\alpha_{d+1}-1} \nonumber \\
\end{eqnarray}
$\forall\;x_i \ge 0$ and $\sum_{i = 1}^d x_i \le 1$ with parameter vector $\mbf{\alpha} = (\alpha_1,\ldots,\alpha_{d+1})$ having components $\alpha_i > 0$. Again two examples of different dimensionality are considered, that is a four-dimensional case with
\begin{displaymath}
\mbf{\alpha} = (6.13,9.29,10.6,8.24,3.91)^\top
\end{displaymath}
and a seven-dimensional case with
\begin{displaymath}
\mbf{\alpha} = (9,5.71,8.96,4.51,5.81,4.06,10.7,1.51)^\top.
\end{displaymath}
Both parameter vectors were chosen randomly. MISE convergence curves for both cases are plotted in \figurename{}~\ref{fig4d7dC2MISE}. Except for somewhat smaller asymptotic decay rates, the observations from the previous Gaussian examples carry over to this case. This holds also true for the computing times of the different estimators (not shown) that display similar dependencies on~$n$ as seen in the previous section. Finally, by inspecting the mean tree depth and number of DEs in \figurename{}~\ref{fig4d7dC1C2}, we can observe that they grow again approximately logarithmically and sublinearly with a dependence on the dimension.

\subsection{Dependence on Test Parameters and Test Statistic}\label{subsecDepTestParams}

\begin{figure*}
\unitlength\textwidth
\begin{picture}(1,0.375)
\put(0.0,0.0){\includegraphics[width=0.5\textwidth]{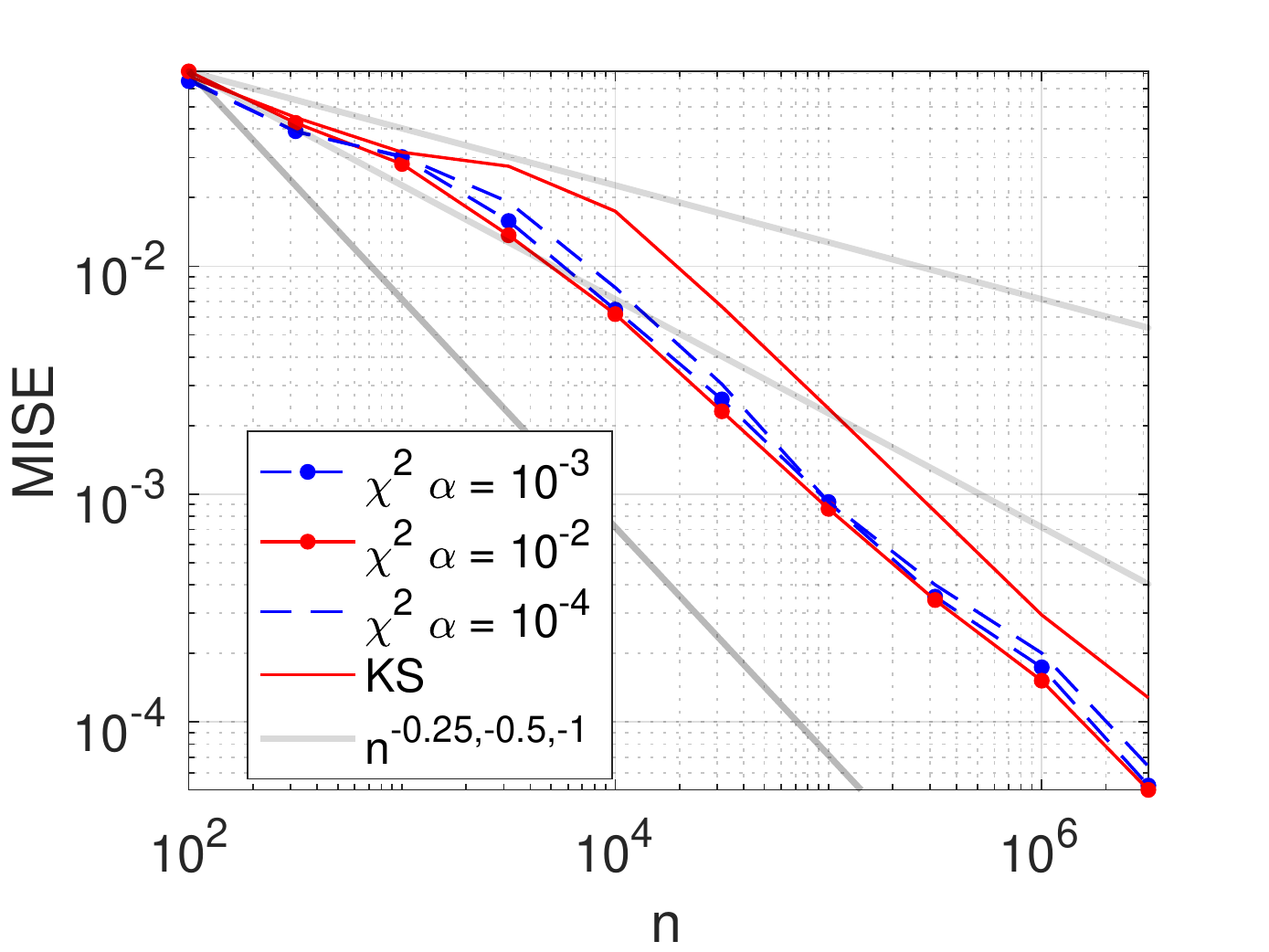}}
\put(0.01,0.365){\makebox(0,0){(a)}}
\put(0.5,0.0){\includegraphics[width=0.5\textwidth]{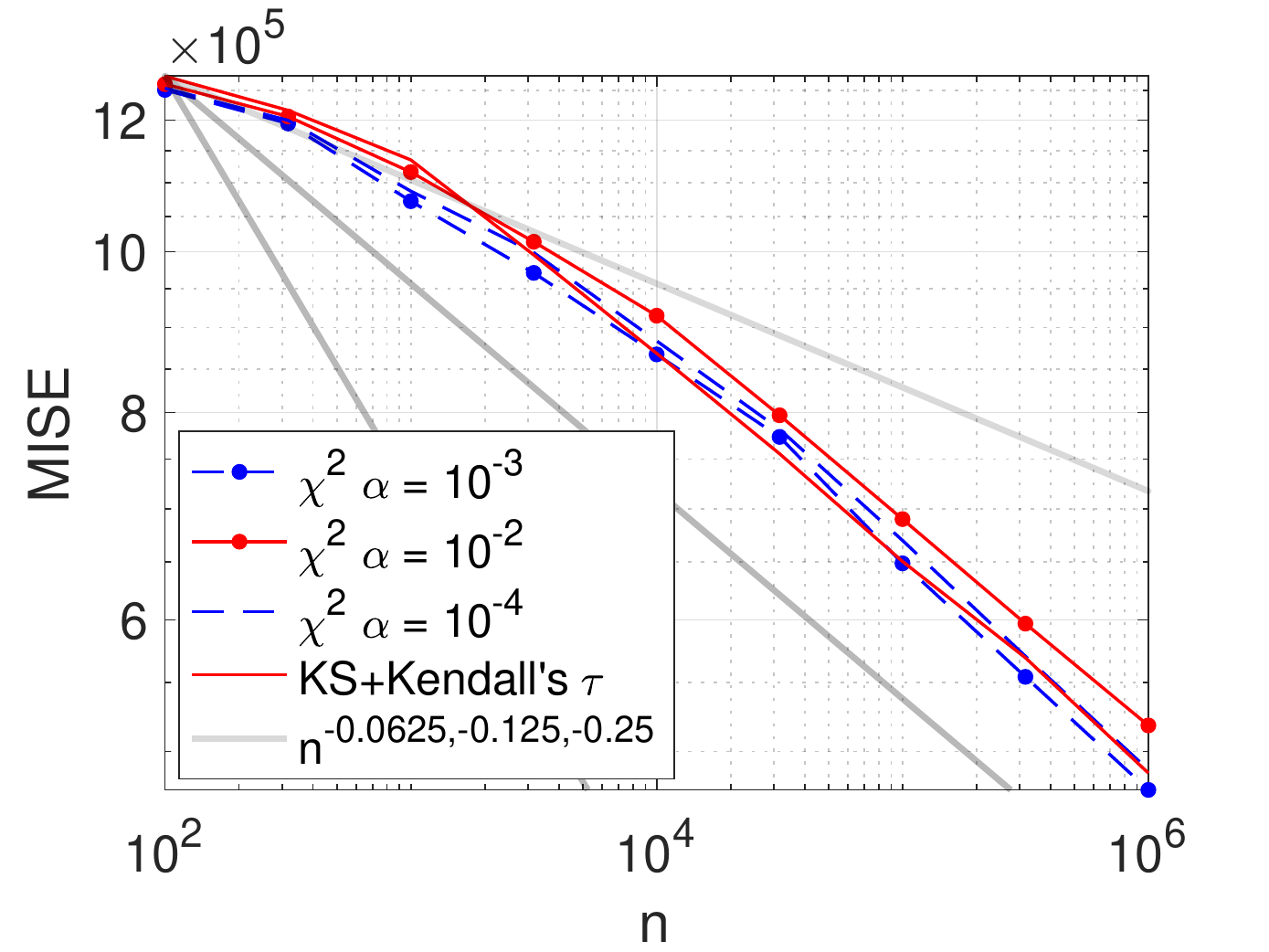}}
\put(0.51,0.365){\makebox(0,0){(b)}}
\end{picture}
\caption{Estimations of (a) beta PDF~\eq{eq1dC5PDF} and (b) seven-dimensional Dirichlet PDF~\eq{eq4d7dC2PDF}. Evolutions of the MISE as a function of the number of samples~$n$ for linear DET estimators with equal size splits and $\chi^2$ test statistics with significance level (red solid symbol) $\alpha = 0.001$, (blue dashed symbol) $\alpha = 0.01$, (blue dashed) $\alpha = 0.0001$, and (red solid) KS and Kendall's $\tau$ test statistics with $\alpha = 0.001$ are plotted. Power law and logarithmic scalings with exponents indicated in the figure legends are depicted (gray thick solid).}\label{figDepTestParams}
\end{figure*}
All previous results were obtained with DET methods that were based on $\chi^2$ test statistics both for goodness-of-fit and independence with identical significance levels $\alpha_g = \alpha_d = 0.001$, respectively. To inspect the influence of the choice of the test statistic on one hand, we performed DET computations with KS and Kendall's $\tau$ tests for goodness-of-fit and independence, respectively, while keeping the significance levels unchanged. On the other hand, we varied the significance level $\alpha = \alpha_g = \alpha_d$ by factors of 0.1 and 10, while maintaining the $\chi^2$ test statistics. No significant changes in the DET estimator performance was found with these variations as is for example shown in \figurename{}~\ref{figDepTestParams}, where MISE results for the one-dimensional beta and seven-dimensional Dirichlet cases are provided.

\begin{figure*}
\unitlength\textwidth
\begin{picture}(1,0.375)
\put(0.0,0.0){\includegraphics[width=0.5\textwidth]{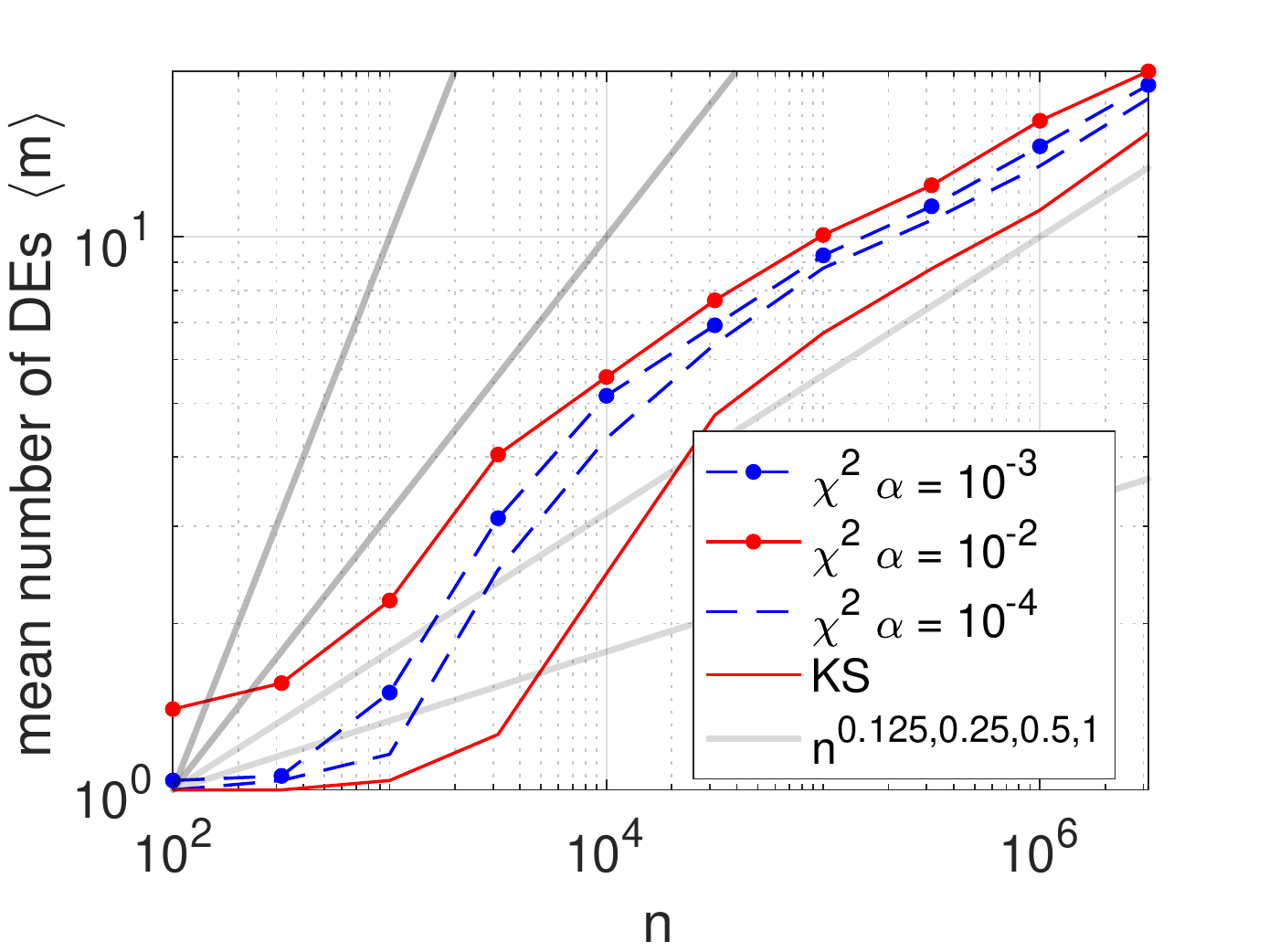}}
\put(0.01,0.365){\makebox(0,0){(a)}}
\put(0.5,0.0){\includegraphics[width=0.5\textwidth]{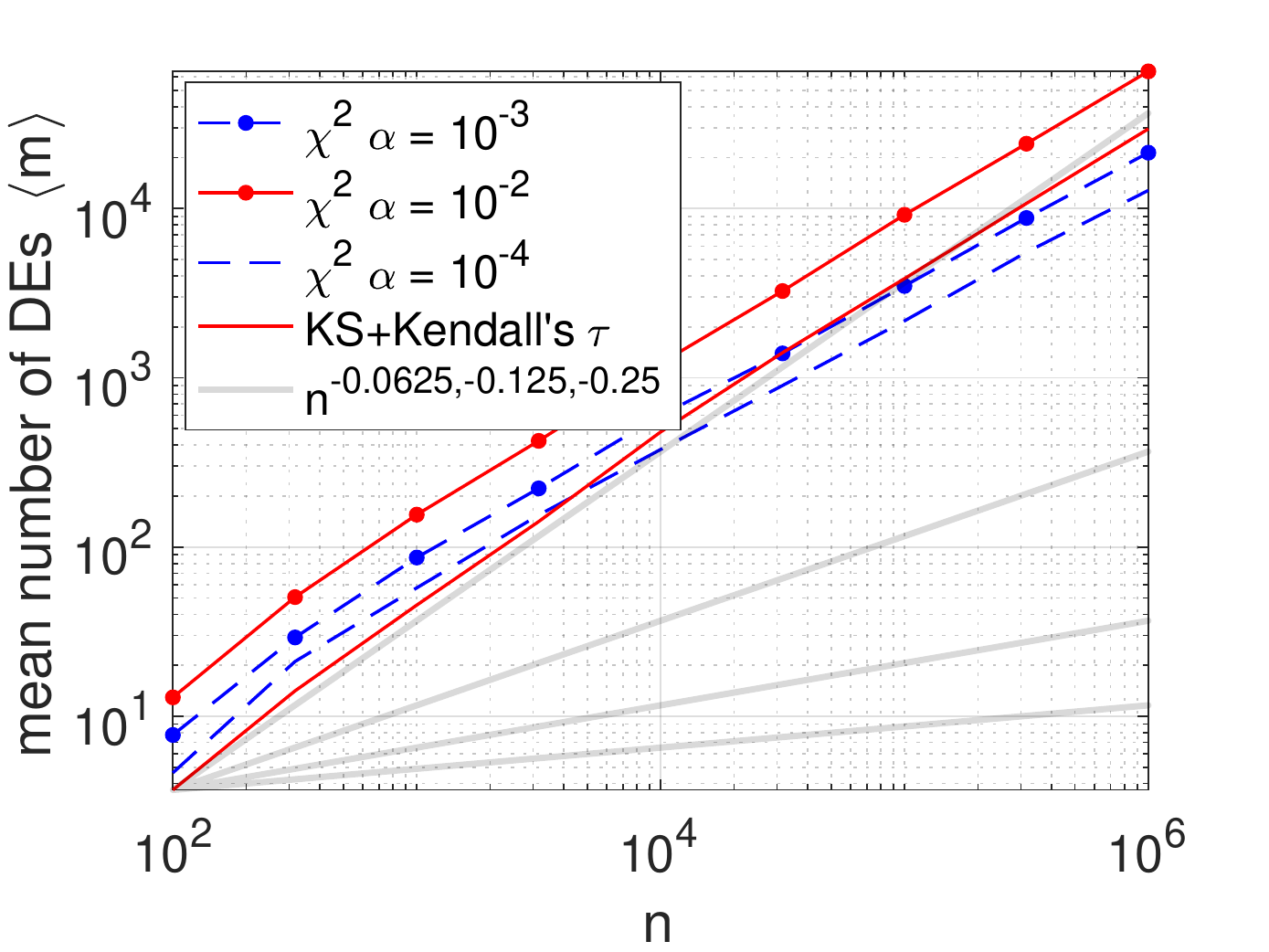}}
\put(0.51,0.365){\makebox(0,0){(b)}}
\end{picture}
\caption{Estimations of (a) beta PDF~\eq{eq1dC5PDF} and (b) seven-dimensional Dirichlet PDF~\eq{eq4d7dC2PDF}. Evolutions of the mean number of DEs $\langle m\rangle$ are depicted. See \figurename{}~\ref{figDepTestParams}.}\label{figDepTestParamsNe}
\end{figure*}
By varying, however, the significance level~$\alpha$, a dependence of the mean number of DEs is expected as is seen in \figurename{}~\ref{figDepTestParamsNe}. While an increase in $\alpha$ leads to an increase in the number of elements, more elements do not necessarily translate into higher accuracy or reduced MISE. This is due to a bias/variance trade-off, where as $\alpha$ is increased bias/statistical errors de-/increase, respectively, and vice versa. As a result, the total error or MISE does not change significantly.

\section{Concluding Remarks}\label{secConclusions}

The DET estimator provides an analytical density representation based on piecewise constant, linear, and possibly higher-order functions. This representation can be efficiently assembled and evaluated at arbitrary probability space positions. Therefore, the adaptivity of the DET method is not limited by a smallest scale that limits for example the performance of a grid-based estimation method. Unlike adaptive KDE, DET-based estimates provide limited differentiability, which may be a disadvantage in some applications. Moreover, unlike combined with a principal axes transform common in KDE \citep[e.g.,][equation~(4.7)]{Silverman:1998a}, the DET estimator is not invariant to the orientation of the coordinate system. A corresponding combination was found to enable an increase in the computational efficiency and accuracy of the DET method.

In terms of accuracy, the linear DET method showed, for the different examples considered, similar or better MISE convergence rates compared with adaptive KDE. The latter is, besides the DET method, the most accurate estimator in this study. Constant DEs are advantageous only in cases, where samples stem from piecewise constant PDFs, as they lead to smaller oscillations at PDF discontinuities. The LL-OPT estimator follows in terms of accuracy after the linear DET estimator. Conventional histograms are least accurate, but at the same time computationally inexpensive as long as the dimension is not too high. Density trees were found to provide mixed results in terms of accuracy.

Computing times of the DET method were found to scale in terms of the number of samples favorably compared to the other estimators. In the DET method, the number of DEs, which determines the time for the DET construction, was found to grow sublinearly. In the case of score-based splitting, there is a theoretical linear upper bound to the DE growth. The splitting method for the DET construction was found to have little effect on the number of DEs, but affects the tree depth. However, irrespective of the method applied, the mean tree depth scales approximately logarithmically with the number of samples. The computational cost of a density evaluation at a specific point is determined by the tree depth. By varying the test parameters that guide the splitting process, i.e., significance levels and test statistics, no significant change in the DET performance was found.

In conclusion, our new DET method is a good candidate for a computationally efficient general-purpose-density-estimator. The matlab implementation of the DET method that was applied in this study is available for download from the MathWorks File Exchange (tag `distribution element tree').

An important element that allowed us to break the exponential growth of the bin count with respect to the dimension~$d$ is the approximation of (mutual) independence by pairwise independence. The relation between pairwise and mutual independence is a subject of ongoing research \citep[e.g.,][]{Nelson:2012a}. Besides this important aspect, to further the DET method, it would be interesting to inspect the performance of higher order DEs and more advanced goodness-of-fit and statistical independence tests.

\appendix
\section{Derivation of MMSE Slope Estimator}

Writing without loss of generality the linear marginal PDF~\eq{eqDELinear} in a simpler form with $x_i\in[0,1]$ and the subscripts skipped, we obtain
\begin{displaymath}
p(x|\theta) = \left(x - {\textstyle\frac{1}{2}}\right)\theta + 1.
\end{displaymath}
By calculating the mean of random variable~$X$ based on this PDF we obtain $\langle X\rangle = \frac{1}{12}(6 + \theta)$ and therefore, can express the slope parameter in terms of this mean as
\begin{equation}
\theta = 6(2\langle X\rangle - 1).
\end{equation}
In the case of a finite ensemble, we estimate the mean with $\langle X\rangle_n = \frac{1}{n}\sum_{j = 1}^n x_j$ and the slope by
\begin{displaymath}
\hat{\theta} = 6c(2\langle X\rangle_n - 1).
\end{displaymath}
Here, $c$ is a correction factor that is determined by minimizing the mean square error (MSE) expressed as
\begin{eqnarray*}
\lefteqn{\langle(\hat{\theta} - \theta)^2\rangle = \left\langle\left[6c\left(\frac{2}{n}\sum_{j = 1}^n x_j - 1\right) - \theta\right]^2\right\rangle} \\
 & = & \left\langle 36c^2\left(\frac{2}{n}\sum_{j = 1}^n x_j - 1\right)^2 - 12c\left(\frac{2}{n}\sum_{j = 1}^n x_j - 1\right)\theta + \theta^2\right\rangle \\
 & = & 36c^2\left\langle\frac{4}{n^2}\sum_{j = 1}^n\sum_{k = 1}^n x_j x_k - \frac{4}{n}\sum_{j = 1}^n x_j + 1\right\rangle \\
 & & - 12c(2\langle X\rangle - 1)\theta + \theta^2 \\
 & = & 36c^2\left(\frac{4}{n^2}\left\langle\sum_{j = 1}^n\sum_{k = 1}^n x_j x_k\right\rangle - 4\langle X\rangle + 1\right) \\
 & & - 12c(2\langle X\rangle - 1)\theta + \theta^2 \\
 & = & 36c^2\left(\frac{4}{n^2}\left\langle\sum_{j = 1}^n\sum_{k = 1\atop k \ne j}^n x_j x_k + \sum_{j = 1}^n x_j^2\right\rangle - 4\langle X\rangle + 1\right) \\
 & & - 12c(2\langle X\rangle - 1)\theta + \theta^2 \\
 & = & 36c^2\left[\frac{4(n-1)}{n}\langle X\rangle^2 + \frac{4}{n}\langle X^2\rangle - 4\langle X\rangle + 1\right] \\
 & & - 12c(2\langle X\rangle - 1)\theta + \theta^2.
\end{eqnarray*}
To determine the minimum MSE, we set
\begin{eqnarray*}
\dder{}{c}\langle(\hat{\theta} - \theta)^2\rangle & = & 72c\left(\frac{4(n-1)}{n}\langle X\rangle^2 + \frac{4}{n}\langle X^2\rangle - 4\langle X\rangle + 1\right) \\
 & & - 12(2\langle X\rangle - 1)\theta = 0,
\end{eqnarray*}
which leads for the correction factor to
\begin{eqnarray}
c & = & \frac{(2\langle X\rangle - 1)n\theta}{6[4(n-1)\langle X\rangle^2 + 4\langle X^2\rangle - 4n\langle X\rangle + n]} \\
 & = & \frac{(2\langle X\rangle - 1)n\theta}{6(4n\langle X\rangle^2 - 4\langle X\rangle^2 + 4\langle X^2\rangle - 4n\langle X\rangle + n])} \nonumber \\
 & = & \frac{(2\langle X\rangle - 1)n\theta}{6[n(2\langle X\rangle - 1)^2 + 4\langle X^{\prime 2}\rangle]} = \frac{6n(2\langle X\rangle - 1)^2}{6[n(2\langle X\rangle - 1)^2 + 4\langle X^{\prime 2}\rangle]}. \nonumber
\end{eqnarray}
For $n\to\infty$ the correction factor~$c$ goes to one.


\begin{thebibliography}{46}
\bibitem[{Pearson(1900)}]{Pearson:1900a}
Pearson K (1900) On the criterion that a given system of deviations from the
  probable in the case of correlated system of variables is such that it can be
  reasonably supposed to have arisen from random sampling. Philosophical
  Magazine 50:157--175
\end{thebibliography}

%
%

\end{document}